\documentclass[11pt]{iopart}

\newcommand{\collop}[1][f_s]{\ensuremath{{C}\hspace{-0.5mm}\left[#1\right]}}
\newcommand{\vth}[1][s]{\ensuremath{v_{\mathrm{th}_#1}}}

\newcommand{\pd}[2]{\ensuremath{ \frac{\partial #1} {\partial #2} } }
\newcommand{\gyror}[1]{\ensuremath{ {\left< #1 \right>}_{\bm{r}}}}
\newcommand{\ensav}[1]{\ensuremath{ {\left< #1 \right>}_{\mathrm{turb}}}}  
\newcommand{\fav}[1]{\ensuremath{\left< #1 \right>_{\psi}}} 
\newcommand{\gyroR}[1]{\ensuremath{{\left< #1 \right>}_{\bm{R}}}}

\providecommand{\eqref}[1]{Eq.\ (\ref{#1})}

\providecommand{\Or}[1]{\mathcal{O}#1}
\providecommand{\Eref}[1]{Equation\ (\ref{#1})}
\providecommand{\eref}[1]{(\ref{#1})}

\newcommand{\dg}{\cdot\nabla}      
\newcommand{\dv}{\nabla\cdot}
\newcommand{\curl}{\nabla\times}
\newcommand{\pb}[2]{\left\{ {#1} , {#2} \right\}} 

\newcommand{\tor}{\phi} 
\newcommand{\gyr}{\vartheta} 
\newcommand{\pot}{\varphi} 
\newcommand{\fpot}{\Phi} 
\newcommand{\gkeps}{\epsilon} 
\newcommand{\energy}[1][s]{{{\varepsilon}_#1}} 

\newcommand{\source}[1][s]{\ensuremath{{{{S}}_{#1}}}} 
\newcommand{\gkupot}{\delta \pot'}
\newcommand{\gkpot}{\chi}
\newcommand{\psistar}[1][s]{{\psi^*_#1}}
\newcommand{\magmom}[1][s]{\mu_#1}

\newcommand{\ddR}[2][s]{\pd{#2}{\bm{R}_{#1}}}
\newcommand{\dgR}[2][s]{\cdot\ddR[#1]{#2}}

\providecommand{\tensor}[1]{ {\bm{\mathsf{#1}}}}
\newcommand{\viscosity}[1][s]{\tensor{\Pi}_#1}
\newcommand{\idmat}{\tensor{I}}
\newcommand{\ParticleFlux}[1][s]{\Gamma_{#1}}

\newcommand{\MomentumFlux}[1][s]{\pi^{(\psi\tor)}_{#1}}
\newcommand{\TotMomFlux}{{\pi}^{(\psi\tor)}_{\mathrm{tot}}}
\newcommand{\inertia}{J}

\newcommand{\daccel}[1][s]{\delta \bm{a}_{#1}}
\newcommand{\ddtpsi}{\left.\pd{}{t}\right|_{\psi}}
\newcommand{\oxford}{
Rudolf Peierls Centre for Theoretical Physics, University of Oxford, Oxford OX1 3NP, UK
}

\newcommand{\culham}{
EURATOM/CCFE Fusion Association, Culham Science Centre, Abingdon OX14 3DB, UK
}

\newcommand{\imperial}{
Blackett Laboratory, Imperial College, London SW7 2AZ, UK
}

\newcommand{\merton}{
Merton College, Oxford, OX1 4JD, UK
}

\newcommand{\wint}{\int\hspace{-1.25mm} d^3 \bm{w}}

\newcommand{\Esource}[1][s]{S^{({E})}_{#1}}
\newcommand{\Psource}[1][s]{S^{({n})}_{#1}}
\newcommand{\Msource}{S^{({\omega})}}

\newcommand{\infrac}[2]{ {#1}/{#2} }
\newcommand{\binfrac}[2]{\left(\infrac{#1}{#2}\right)}
\newcommand{\CollEnergy}[1][s]{C^{{(E)}}_#1}

\newcommand{\PotEng}[1][s]{P^{\mathrm{pot}}_#1}
\newcommand{\CompHeat}[1][s]{P^{\mathrm{comp}}_#1}

\newcommand{\vchi}{\bm{V}_\chi}
\newcommand{\vchiR}{\gyroR{\vchi}}
\newcommand{\vdrift}[1][s]{\bm{V}_{\mathrm{D}#1}}

\newcommand{\NotN}[1][s]{N_{#1}}
\newcommand{\massratio}{\delta}

\newcommand{\delB}{{\ensuremath{\delta\bm{B}}}}
\newcommand{\delBp}{{\ensuremath{\delta B_\parallel}}}

\newcommand{\delAp}{{\ensuremath{\delta A_\parallel}}}
\newcommand{\delApN}[1]{{\ensuremath{\delta A_\parallel^{(#1)}}}}
\newcommand{\delE}{{\ensuremath{\delta\bm{E}}}}
\newcommand{\delA}{{\ensuremath{\delta\bm{A}}}}

\newcommand{\delpot}{{\delta\pot}}
 \newcommand{\bhat}{{\widetilde{\bm{b}}}}

\newcommand{\MeanB}{{\bm{B}}}
\newcommand{\MeanE}{{\bm{E}}}
\newcommand{\MeanMagB}{{B}}
\newcommand{\Meanb}{{\bm{b}}}

\newcommand{\Efield}{{\widetilde{\bm{E}}}}
\newcommand{\Bfield}{{\widetilde{\bm{B}}}}
\newcommand{\MagBfield}{{\widetilde{B}}}

\newcommand{\MeanA}{{\bm{A}}}

\newcommand{\bav}[1]{\left<#1\right>_{\parallel}}
\newcommand{\veff}{\bm{u}_{\mathrm{eff}}}
\newcommand{\fluctfav}[1]{\left<#1\right>_{\tilde\psi}}

\newcommand{\upar}{\delta u_{\parallel e}}
\newcommand{\tpsi}{\tilde{\psi}}
\newcommand{\talpha}{\tilde{\alpha}}
\newcommand{\tl}{\tilde{l}}
\newcommand{\ddttwiddles}[1][]{\left.\pd{#1}{t}\right|_{\tilde{\psi},\tilde{\alpha},\tilde{l}}}

\newcommand{\angvel}{\omega}
\newcommand{\cycfreq}[1][s]{\Omega_{#1}}

\newcommand{\vA}{v_{\mathrm{A}}} 

\newcommand{\vpsi}{\bm{V}_\psi}

\newcommand{\tw}{\widetilde{\bm{w}}}
\newcommand{\twint}{\int d^{3}\widetilde{\bm{w}}}
\newcommand{\Magtw}{\widetilde{{w}}}

\newcommand{\dgt}{\cdot\widetilde{\nabla}_\perp}

\newcommand{\MaxVDrift}{\widehat{\bm{V}}_D}
\newcommand{\ddtpsitwiddles}{\left.\pd{ }{t}\right|_{\tpsi}}
\newcommand{\Pturb}{P^{\mathrm{turb}}_e}

\newcommand{\tvp}{\tilde{V}'}
\newcommand{\tee}{\tilde{\varepsilon}_e}
\newcommand{\delNe}{\overbar{\delta n}_e}
\newcommand{\delTe}{\overbar{\delta T}_e}

\newcommand{\overbar}[1]{\mkern 1.5mu\overline{\mkern-1.5mu#1\mkern-1.5mu}\mkern 1.5mu}
\newcommand{\nustar}[1][e]{\nu^{*}_{#1}}

\usepackage{iopams}
\usepackage{setstack}
\usepackage{graphicx}
\usepackage{bm,ulem,color}
\usepackage{xr}
\usepackage{xr-hyper}
\usepackage[colorlinks=true]{hyperref}
\begin{document}

\date{\today}

\title[The Electron Response]{Multiscale Gyrokinetics for Rotating Tokamak Plasmas II: Reduced Models for Electron Dynamics}

\author{I.\ G.\ Abel$^{1,2,3}$, S.\ C.\ Cowley$^{2,4}$}
\ead{i.abel1@physics.ox.ac.uk}
\address{$^1$\oxford}
\address{$^2$\culham}
\address{$^3$\merton}
\address{$^4$\imperial}

\begin{abstract}
In this paper, we extend the multiscale approach developed in [Abel et. al., Rep. Prog. Phys., submitted] by exploiting the scale separation between
ions and the electrons. The gyrokinetic equation is expanded in powers of the electron to ion mass ratio, which provides a rigorous
method for deriving the reduced electron model. We prove that 
ion-scale electromagnetic turbulence cannot change the magnetic topology, and argue that
to lowest order the magnetic field lies on fluctuating flux surfaces.
These flux surfaces are used to construct magnetic coordinates, and in these coordinates a closed system of equations for
the electron response to ion-scale turbulence is derived. All fast electron timescales have been eliminated from these equations.
We also use these magnetic surfaces to construct transport equations for electrons and for electron heat in terms of the reduced electron model.
\end{abstract}

\maketitle
\section{Introduction}
\label{Sintro}
The energy, particle, and momentum confinement of present-day fusion experiments and proposed future devices is limited by turbulent, rather than collisional transport.
The turbulence that causes this transport has an essentially multiscale character. It occurs on (perpendicular) spatial scales smaller than those associated with
the equilibrium and on timescales shorter than those associated with transport of mean quantities but much longer than the gyroperiod. This is the basis for the multiscale gyrokinetic approach presented in \cite{flowtome1,sugama1998neg}, characterised by the small parameter $\gkeps = \rho / a$, where $\rho$ is the thermal gyroradius and $a$ is a typical equilibrium length scale.
In these treatments, all species are considered equal. However, there is a second scale separation that is important, the scale separation between the
plasma ions and the electrons, characterised by the small parameter $\massratio = \sqrt{m_e / m_i}$, where $m_e$ and $m_i$ are the electron and ion masses respectively. For a deuterium plasma $\massratio \approx 1/60$.

Ion-scale fluctuations (e.g. Ion Temperature Gradient (ITG)~\cite{sagdeevITG,coppiITG,cowley1991considerations}, Parallel Velocity Gradient (PVG)~\cite{cattoPVG,newton2010understanding} or Trapped Electron Mode (TEM)~\cite{kadomtsevpogutse} driven turbulence) have length scales perpendicular to the magnetic field comparable to the ion gyroradius and have frequencies comparable to the ion transit or bounce frequencies.
In contrast, electron-scale fluctuations have perpendicular length scales comparable to the electron gyroradius, and frequencies comparable to the electron transit or bounce frequency.
Unless they are suppressed (e.g. by strongly sheared flows), ion-scale fluctuations with their larger ``mixing length'' usually dominate the transport -- and are therefore considered to be more important. In this paper we focus on the electron response to such fluctuations.

The electrons are much lighter than the ions, hence ion-scale fluctuations are long-wavelength compared to the electron gyroradius and low-frequency when compared to the electron transit or bounce frequencies. 
\begin{table}[h]
\centering
\begin{tabular}{|c||c|c|c|}
	\hline
Parameter & JET & D-IIID & ITER \\
\hline
\hline
$n_e$, m$^{-3}$ & 1-5$\times 10^{19}$ & $10^{18}$ & $10^{20}$ \\
$T_i$, keV & 5-15 & 1-5 & 15-25 \\
$B$, T & 2-4 & 1-2 & 3-5 \\
$\beta_i$, \% & 0.01-2.0 & 1.0--4.0 & 1.0--4.0 \\
	\hline

$\rho_i$, m &  $5.1\times10^{-3}$ & $6.0\times10^{-3}$ & $3.2\times10^{-3}$ \\
$\rho_e$, m & $8.2\times10^{-5}$ & $9.7\times10^{-5}$ & $5.2\times10^{-5}$ \\
$a$, m & 1 & .5 & 2 \\
	\hline
$\nu_{ee}$, Hz & $1.6\times10^3$ & $4.5\times10^3$ & $1.3\times10^3$ \\
$\omega\approx\vth[i]/a$, Hz & $2.0\times10^4$ & $2.8\times10^4$ & $5.5\times10^5$ \\
$\cycfreq[i]$, Hz & $2.6\times10^7$& $1.52\times10^7$ & $3.8\times10^7$\\
$\tau_E$, s & 0.5 & 0.1 & 3.5 \\
	\hline
$\gkeps= \rho_i / a$  & $5\times10^{-3}$ & $1.2\times10^{-2}$ & $1.6\times10^{-3}$\\
$\delta$ & 1.5-1.6\% & 1.5-1.6\% & 1.5-1.6\% \\
\hline
\end{tabular}
\label{tableNumbers}
\caption{Typical length and time scales in selected fusion devices.}
\end{table}
It is clear that in simulating ion-scale turbulence one does not wish to resolve the short electron timescales.
This has motivated the development of models for the electron response to ion-scale turbulence that somehow eliminate fast processes due to electrons~\cite{gang1990nonlinear,linelectrons01,chenparker2001,hinton:168}. 
In this paper, we derive for the first time a set of equations governing the electron behaviour in the presence of ion-scale fluctuations. These equations 
eliminate the fast electron timescale via a rigorous expansion of the gyrokinetic equations in $\massratio \ll 1$, subsidiary to the expansion in $\gkeps \ll 1$.

The fast processes arise from the streaming of electrons along magnetic field lines. In electrostatic turbulence simulations, this streaming is along the unperturbed mean field and easily handled. The electrostatic approximation only holds at very low $\beta$ (the ratio of thermal to magnetic pressure, and a measure of the importance of magnetic perturbations). However, any future reactor design will operate with as high a $\beta$  as possible. Indeed, spherical tokamaks (such as MAST~\cite{akers2003transport}) already operate at finite $\beta$. Thus, we require an electron response model that incorporates the fluctuations of the magnetic field in a fully general way.
As we shall see, if we consider only ion-scale fluctuations then the topology of the magnetic field is preserved by the fluctuations (see \Sref{Sflux}).
However, to determine the electron behaviour we will need to make assumptions about the initial structure of the magnetic field and the effect of other scales. 
In particular,
whether the field is regular, with nested flux surfaces, or stochastic.	
We show in \Sref{SClebsch}, that if the magnetic field were significantly stochastic (i.e. if the ion heat transport due to the stochastic field were comparable to the transport due to the fluctuating $\bm{E}\times\bm{B}$ drifts) then the electron heat transport would be larger than the ion transport by a factor of $\massratio^{-1} \gg 1$.
Such poor electron heat confinement is not observed in experiments, indeed it is common for the level of ion heat transport to exceed the level of electron heat transport in conventional tokamaks.
Thus, we assume that any stochastic component of the magnetic field is small enough to be ignored (see \Sref{SClebsch}) and therefore that the field lines lie on fluctuating magnetic surfaces.

The structure of the rest of the paper is as follows.
In \Sref{Sgk}, we introduce the electron gyrokinetic equation and attendant notation.
In \Sref{Sexpansion}, we develop the formal expansion in $\massratio \ll 1$ and order all physical quantities with respect to this small parameter. 
The key result of this paper -- the gyrokinetic system with reduced electron equations -- is derived via a systematic expansion in $\massratio$. This
derivation is performed in Sections \ref{Smagfield} and \ref{Sreduced} with the results concisely summarised in \Sref{Turb_evol}.

In \Sref{Sflux}, we use the expansion in $\massratio \ll 1$ to prove that long-wavelength low-frequency fluctuations preserve the topology of magnetic field lines, generalising the results of \cite{snyder:3199,snyderthesis,snyder:744,stevefreezing}. With the assumption of an initially regular magnetic field, this result enables us to introduce a coordinate system aligned to the exact magnetic field in \Sref{Smagfield}. It is this set of coordinates that allows us, in Sections \ref{Sbounce} and \ref{Sfluxav}, to average over the fast electron motion along exact magnetic field lines and so close our equations for the electron distribution function. 
These equations are particularly simple in the limit of strong collisions ($\nu_{e} \gg \omega$) which we derive in \Sref{Scollisional}.

The second major result of this paper is a simplified system of transport equations for electrons. This is presented in \Sref{STransp} and derived in \ref{magicalTransp}. These equations emphasise the particularly simple form that the transport fluxes take in the low-mass-ratio limit. We also present the transport equations for the collisional electron model of \Sref{Scollisional}, which take on an even simpler form.

We draw the results of the paper together in \Sref{Sconclusions}. We make a detailed comparison of our equations to previously-derived models in \Sref{ap-comp}, emphasising both the differences and the similarities. Finally, we conclude in \Sref{ASConcFinal}.

\section{Electron Gyrokinetics and the Mass-Ratio Expansion}
We study the response of electrons to ion-scale fluctuations within the framework of multiscale gyrokinetics~\cite{flowtome1}. First, we will introduce this framework in \Sref{Sgk} along with all attendant notation. Then, in \Sref{Sexpansion}, we formally order all quantities in the small parameter $\massratio = \sqrt{{m_e}/{m_i}}$. 

\subsection{Gyrokinetics in a Rotating Tokamak Plasma}
\label{Sgk}
In this section, we introduce the assumptions and resulting equations of this formalism. The detailed derivation of these results can
be found in \cite{flowtome1}.

First, we split all physical quantities into mean and fluctuating parts:
\begin{eqnarray}
\Bfield &= \MeanB + \delB, \qquad \MeanB &= \ensav{\Bfield},\\
\Efield &= \MeanE + \delE, \qquad \MeanE &= \ensav{\Efield},\\
f_s &= F_s + \delta f_s, \qquad F_s &= \ensav{f_s},
\end{eqnarray}
where $\Bfield$ is the magnetic field, $\Efield$ the electric field, $f_s$ the distribution function of species $s$ and $\ensav{\cdot}$ is some average over the fluctuations.

We assume that the fluctuations obey the standard gyrokinetic ordering\footnote{In this paper, as in \cite{flowtome1}, the symbol $\sim$ is used to mean ``is the same order as'' (with respect to the appropriate expansion) rather than the more usual ``is asymptotically equivalent to''.} :
\begin{equation}
\frac{\omega}{\Omega_s} \sim \frac{k_\parallel}{k_\perp} \sim \frac{|\delE|}{|\MeanE|} \sim \frac{|\delB|}{|\MeanB|} \sim\frac{\delta f_s}{F_s} \sim \frac{\nu_s}{\Omega_s} \sim \frac{\rho_s}{a} = \gkeps \ll 1,
\end{equation}
where $\omega$ is the typical fluctuation frequency in the frame rotating with the plasma, $\Omega_s = Z_s e \MeanMagB / m_s c$ is the cyclotron frequency of species $s$ with charge $Z_s e$ and mass $m_s$ in the mean magnetic field $\MeanMagB=|\MeanB|$, $k_\parallel$ and $k_\perp$ are typical wavenumbers parallel and perpendicular to the mean magnetic field, $\nu_s$ is a typical collision frequency,
$\rho_s = \vth[s] / \Omega_s$ is the thermal Larmor radius, $\vth[s] = \sqrt{2T_s / m_s}$ is the thermal speed of species $s$ with mean temperature $T_s$ and $a$ is a typical equilibrium length scale.
The mean quantities ($\MeanB,\;\MeanE,\;F_s$) are assumed to vary on the long length scale $a$ and the long (transport) timescale $\tau_E$, the energy confinement time. This long timescale is ordered so that
\begin{equation}
\tau_E^{-1} \sim \gkeps^{3} \Omega_s.
\end{equation}
Typical values for the timescales and length scales in a tokamak are given in Table~(\ref{tableNumbers}), which shows the very large scale separation, characterised by the small parameter $\gkeps$.
This scale separation in time and space of the mean and fluctuating quantities motivates us to define the average $\ensav{\cdot}$ in terms of temporal and spatial averages over intermediate time and length scales. This is done precisely in equations (\ref{P1-LambdaScale})--(\ref{P1-avdef}) of \cite{flowtome1}.

It is shown in \Sref{P1-maggeom} of \cite{flowtome1} that if we introduce the cylindrical coordinate system $(R,z,\tor)$ and assume axisymmetry with respect to $\tor$, the mean magnetic field can be written (to lowest order in $\gkeps$) as
\begin{equation}
\label{magfield}
\MeanB = I(\psi,t) \nabla \tor + \nabla\psi \times\nabla\tor,
\end{equation}
where the poloidal flux function $\psi$ is given in terms of the mean vector potential $\MeanA$ by
\begin{equation}
\psi(R,z,t) = A_\tor = R^2\MeanA\dg\tor,
\label{psiDef}
\end{equation}
and $I(\psi,t) = R^2\MeanB \dg\tor$ is the toroidal component of the mean magnetic field. We will also need the following alternate 
form of the magnetic field~\cite{kruskal1958magnetic,beer1995field}:
\begin{equation}
\MeanB = \nabla\psi\times\nabla \alpha,
\end{equation}
where $\alpha$ is a variable that labels different field lines within a flux surface. We cut the toroidal surface on the inboard side to force $\alpha$ to be single-valued. This cut forms an axisymmetric ribbon with one edge the magnetic axis and the other the plasma boundary or separatrix.  Letting $l$ be the distance along a field line, ($\psi$, $\alpha$, $l$) is a set of field-aligned coordinates for $\MeanB$.  There are many different forms of field aligned coordinates and none of our results depend on a particular choice. Nevertheless, we will use ($\psi$, $\alpha$, $l$) for specificity.  Axisymmetric functions are functions of $\psi$ and $l$ but not $\alpha$.  Note that both $\psi(R,z)$ and $I(\psi)$ are mean quantities and therefore vary on the transport timescale.

We assume that the mean electric field is large, $\MeanE\sim \binfrac{\vth[s]}{c} \MeanB$, which can be shown to imply that the plasma rotates toroidally with a velocity 
that is species-independent and whose angular velocity only depends on the flux label $\psi$~\cite{catto:2784} and the slow transport timescale:
\begin{equation}
\bm{u} = \angvel(\psi,t) R^2\nabla\tor.
\label{torrot}
\end{equation}
If we express the mean electric field in terms of potentials using Gaussian units and Coulomb gauge,
\begin{equation}
\MeanE = -\nabla\pot - \frac{1}{c}\pd{\MeanA}{t},
\end{equation}
then the scalar potential takes the form
\begin{equation}
\pot = \fpot(\psi, t) + \pot_0, \qquad \pot_0 \sim \gkeps \fpot,
\end{equation}
and we have the following relationship between $\angvel(\psi,t)$ and $\fpot(\psi, t)$:
\begin{equation}
\angvel(\psi,t) = c \frac{d\fpot}{d\psi}.
\label{omdef}
\end{equation}

With these results for the fields, our orderings imply that the distribution function $f_s$ of species $s$ takes the following form:
\begin{eqnarray}
\label{fSoln}
f_s &=& F_s + \delta f_s,\\
\label{FSoln}
F_s &=& F_{0s} \left( \psi(\bm{R}_s), \energy, t \right) + F_{1s}\left( \bm{R}_s,\energy,\magmom,\sigma, t\right) + \Or\left(\gkeps^2 f\right),\\
		 \label{hDef}
\delta f_s &=& - \frac{Z_s e}{T_s} \gkupot(\bm{r},t) F_{0s} + h_s\left(\bm{R}_s,\energy,\magmom,\sigma,t\right) + \Or\left(\gkeps^2 f\right).
\end{eqnarray}
Let us explain this notation. We have followed \cite{flowtome1} and changed phase-space variables 
from $(\bm{r},\bm{v})$ to $(\bm{R}_s,\energy,\magmom,\gyr,\sigma)$:
\begin{eqnarray}
\label{Rdef}
\bm{R}_s &= \bm{r} - \frac{1}{\Omega_s} \Meanb\times\bm{w},\\
			 \label{energydef}
\energy &= \frac{1}{2} m_s v^2 + Z_s e \left( \fpot(\psi,t) + \pot_0 \right) - Z_s e\fpot(\psistar,t),\\
			  \label{mudef}
\magmom &= \frac{m_s w_\perp^2}{2\MeanMagB},
\end{eqnarray}
where $\Meanb = \MeanB / \MeanMagB$ is the unit vector along the mean magnetic field, the peculiar velocity $\bm{w}$ is
\begin{equation}
\label{wDef}
\bm{w} = \bm{v} - \bm{u}= w_\parallel \Meanb + w_\perp \left(\cos \gyr\,\bm{e}_2 - \sin\gyr\,\bm{e}_1 \right),
\end{equation}
where $\bm{e}_1$ and $\bm{e}_2$ are two arbitrary orthogonal unit vectors perpendicular to the magnetic field,
the direction of the parallel motion is $\sigma = w_\parallel / |w_\parallel|$, and we have defined
\begin{equation}
\label{psistardef}
\psistar(\bm{r},\bm{v},t) = \psi(\bm{r},t) + \frac{m_s c}{Z_s e} R^2\bm{v}\dg\tor.
\end{equation}

The quantity $\psistar(\bm{r},\bm{v},t)$ is equal to $c/(Z_s e)$ times the canonical toroidal angular momentum and is thus conserved in an exactly axisymmetric system -- i.e., in the absence of fluctuations.  In these variables, \eref{FSoln} splits the mean distribution function into the near Maxwellian equilibrium
\begin{eqnarray}
\label{F0def}
F_{0s} &=& \NotN(\psi(\bm{R}_s,t),t) \left[\frac{m_s}{2\pi T_s(\psi(\bm{R}_s,t),t)}\right]^{3/2} e^{- \energy / T_s(\psi(\bm{R}_s,t),t)},
\end{eqnarray}
and the neoclassical distribution function $F_{1s}$, which describes large-scale $\Or(\gkeps)$ deviations from a Maxwellian.
The function $\NotN$ is related to the mean density $n_s$ by
\begin{equation}
n_s= \NotN(\psi(\bm{r},t),t) \exp\left[- \frac{ Z_s e \pot_0(\bm{r},t)}{T_s(\psi(\bm{r},t),t)} + \frac{m_s\angvel^2(\psi(\bm{r},t),t) R^2}{2T_s(\psi(\bm{r},t),t)}\right].
\end{equation}
We now drop the explicit slow time dependence of all quantities -- this is for notational clarity only and is not an assumption that the mean quantities are constant.

The fluctuating distribution function, given by \eref{hDef}, is composed of the Boltzmann response (the first term of \eref{hDef}), where the fluctuating potentials $\delta\pot$ and $\delA$ have been combined to form the electrostatic potential in the frame rotating with the plasma 
\begin{equation}
\gkupot = \delta\pot - \frac{1}{c} \bm{u}\cdot\delA,
	\label{gkupotDef}
\end{equation}
and the gyrokinetic distribution function $h_s$.
Assuming that all mean quantities are known, then the gyrokinetic distribution function $h_s$ is given by the gyrokinetic equation~\cite{flowtome1,sugama1998neg}:
\begin{equation}
\fl
\begin{eqalign}{
\left[\pd{}{t}+\bm{u}(\bm{R}_s)\dgR{}\right]h_s + \left( w_\parallel \Meanb + \vdrift + \vchiR \right)\cdot\ddR{h_s} - \gyroR{\collop[h_s]} \\
\quad=\frac{Z_s e F_{0s}}{T_s}\left[\pd{}{t} + \bm{u}(\bm{R}_s)\dgR{}\right]\gyroR{ \gkpot}\\\qquad -
\left\{ \pd{F_{0s}}{\psi} + \frac{m_s F_{0s}}{T_s} \left[ \frac{I(\psi) w_\parallel}{\MeanMagB} + \angvel(\psi)R^2 \right]\frac{d\angvel}{d\psi}\right\} \vchiR\dg\psi,
}\end{eqalign}
\label{gke}
\end{equation}
where the guiding-centre drift velocity is
\begin{equation}
\label{vdrift}
\begin{eqalign}{
\vdrift = \frac{\Meanb}{\Omega_s} \times &\left[ w_\parallel^2 \Meanb\dg\Meanb + \frac{1}{2}w_\perp^2\nabla \ln \MeanMagB\right.\\
	 &\left.- \angvel^2(\psi,t) R\nabla R - 2 w_\parallel \angvel(\psi,t) \Meanb\times\nabla z +\frac{Z_s e}{m_s} \nabla\pot_0\right],
}\end{eqalign}
\end{equation}
and where we have defined the gyrokinetic potential
\begin{equation}
\gkpot = \delta \pot - \frac{1}{c}\bm{v}\cdot\delA = \gkupot - \frac{1}{c} \bm{w}\cdot\delA,
\label{chidef}
\end{equation}
and the fluctuating velocity due to $\gkpot$\footnote{
This can be shown to consist, physically, of the fluctuating $\bm{E}\times\bm{B}$ drift in the rotating frame, the motion of guiding centres along perturbed field lines and the fluctuating $\nabla B$ drift.}
\begin{equation}
\vchi = \frac{c}{\MeanMagB}\Meanb\times\nabla\gkpot,\qquad \vchiR = \frac{c}{\MeanMagB} \Meanb\times\ddR{\gyroR{\gkpot}} + \Or(\gkeps^2\vth).
\label{vchi}
\end{equation}
We have also introduced the gyroaverage at constant $\bm{R}_s$, $\energy$ and $\magmom$:
\begin{equation}
\gyroR{\gkpot(\bm{r},\bm{w},t)}  = \frac{1}{2\pi} \oint d\gyr \chi\left(\bm{r}(\bm{R}_s,\energy,\magmom,\gyr),\bm{w}(\bm{R}_s,\energy,\magmom,\gyr,\sigma),t\right).
\end{equation}

The equation for $h_s$ is closed through Maxwell's equations for the fluctuating fields.
The fluctuating potential $\gkupot$ obeys the quasineutrality condition:
\begin{equation}
\label{fluct-qn}
 \sum_s \frac{Z_s^2 e^2 n_s\gkupot}{T_s}   = \sum_s Z_s e \wint \gyror{h_s},
\end{equation}
where the integral over velocities is performed at constant $\bm{r}$ and the gyroaverage at constant $\bm{r}$, $w_\parallel$, and $w_\perp$ is defined by
\begin{equation}
\fl
\gyror{h_s} = \frac{1}{2\pi} \oint d\gyr h_s(\bm{R}_s(\bm{r},w_\parallel,w_\perp,\gyr),\energy(\bm{r},w_\parallel,w_\perp,\gyr),\magmom(\bm{r},w_\perp),\sigma,t).
\end{equation}
The fluctuating magnetic field is determined from $\delAp = \Meanb\cdot\delA$ and $\delBp = \Meanb\cdot\delB$, which obey the 
parallel component of Amp\`ere's law
\begin{equation}
\label{fluct-apar}
-\nabla_\perp^2 \delAp = \frac{4\pi}{c} \sum_s Z_s e \wint w_\parallel \gyror{h_s},
\end{equation}
and the perpendicular part of Amp\`ere's law (see the discussion in \Sref{P1-Sfmag} of \cite{flowtome1})
\begin{equation}
\label{fluct-bpar}
\nabla_\perp^2 \frac{\delBp \MeanMagB}{4\pi} + \bm{\nabla}_\perp\bm{\nabla}_\perp\bm{:} \sum_s \wint \gyror{m_s\bm{w}_\perp\bm{w}_\perp h_s} = 0,
\end{equation}
 respectively.
The evolution equation \eref{gke} and constituent relations \eref{fluct-qn}, \eref{fluct-apar} and \eref{fluct-bpar} form a closed system of equations for determining $h_s$, $\gkupot$, $\delAp$, $\delBp$ on the turbulent timescale. 
Typically, the system \eref{gke}, \eref{fluct-qn}, \eref{fluct-apar}, and \eref{fluct-bpar} is solved until statistical equilibrium is reached (a few nonlinear turnover times). The slowly-varying equilibrium is treated as constant during this evolution.
The slow time dependence of $F_{0s}$ is determined by transport equations for $\angvel(\psi,t)$, $\NotN(\psi,t)$, $T_s(\psi,t)$ and $I(\psi,t)$ and the equilibrium condition determining $\psi(\bm{r},t)$. Such equations are given in \cite{flowtome1}. 
\subsection{The Mass-Ratio Expansion}
\label{Sexpansion}
In order to express the scale separation of the electron and ion scales,
we introduce the secondary small parameter $\delta  = \sqrt{\infrac{m_e}{m_i}} \ll 1$. We will use a subscript $i$ to denote any ion species, of which there may be many.
We consider the expansion in $\massratio \ll 1$ to be a subsidiary expansion of the gyrokinetic system (equations~\eref{gke}, \eref{fluct-qn}, \eref{fluct-apar}, and \eref{fluct-bpar})\footnote{This is in contrast to the approach taken by \cite{hinton:168}, see \Sref{ap-comp} for details.}. Thus, we assume that $\gkeps = \rho_i / a$, i.e., that the fundamental gyrokinetic expansion parameter is that of the ions, and that
\begin{equation}
\gkeps \ll \massratio \ll 1.
\end{equation}
All other dimensionless parameters are treated as finite -- i.e., independent of $\delta$. Therefore, we are assuming that $T_i \sim  T_e$ and $\beta = 8\pi p / \MeanMagB^2 \sim 1$, where $p$ is the plasma pressure.  In typical fusion plasmas, $\beta$ is often small and one might worry about this assumption. A more detailed analysis shows that our expansion in $\delta$ requires that~\cite{Tome}
	\begin{equation}
	\beta \gg \frac{m_e}{m_i} = \delta^2,
	\end{equation}
which is equivalent to requiring that the Alfv\'en speed is smaller than the electron thermal speed.\footnote{For cases where $\beta$ is so low that this does not hold, an expansion of slab gyrokinetics for very low $\beta$ plasmas has been carried out in \cite{zoccoelectrons}.}  In plasmas of interest, this is indeed satisfied and we are justified in treating $\beta$ as finite.

We assume that fluctuating quantities vary on ion rather than electron scales.
Thus, we order the timescales of fluctuating quantities in the rotating frame and the size of the fluctuations by
\begin{eqnarray}
\hskip -0.6 truein \pd{ }{t} + \bm{u}\cdot\nabla \sim {\vth[i]}\Meanb \cdot\ddR{ }  \sim \frac{c}{B} |\left( \nabla \gkupot \times \Meanb \right) \cdot\nabla_\perp| \sim \nonumber\\  \frac{\vth[i]}{B} |\left( \nabla \delAp\times \Meanb \right) \cdot\nabla_\perp|\sim \frac{cT}{e\MeanMagB^2} |\left( \nabla \delBp \times \Meanb \right) \cdot\nabla_\perp| \sim \frac{\vth[i]}{a}
\end{eqnarray}
so that the typical fluctuation timescale, the ion parallel streaming time, and the ion nonlinear timescale due to the fluctuating fields are all comparable. We assume that the toroidal rotation velocity $\bm{u}$ is comparable to the ion thermal speed $\bm{u} \sim \vth[i]$,
which implies that the electric field
is ordered as
\begin{equation}
\MeanE \sim \frac{\vth[i]}{c} \MeanB.
\label{deltaEord}
\end{equation}
The perpendicular length scales are of course ordered as
\begin{equation}
|\nabla_\perp| \sim k_\perp  \sim \rho_i^{-1}.
\end{equation}
Clearly, the turbulent dynamics of the ions are not simplified by the subsidiary expansion in $\delta$ and hence $h_i$  is determined by the gyrokinetic equation~\eref{gke}.

The electron dynamics, however, are simplified because
\begin{equation}
k_\parallel{\vth[e]} \sim \massratio^{-1} \frac{\vth[i]}{a} \sim \frac{\vth[e]}{B} |\left( \nabla \delAp \times \Meanb \right) \cdot\nabla_\perp| \gg \pd{ }{t} + {\bm{u}}\cdot\nabla,
\end{equation}
so the parallel electron motion is much faster than the evolution of the fluctuations~\footnote{Examining the assumption that $\omega / k_\parallel \vth[e] \sim \delta$, we might worry that if $k_\parallel$ is small then this assumption would be violated. Thankfully, studies of strongly nonlinear turbulence (both numerically-simulated ITG and experimentally measured turbulence) suggest that it is, in fact, critically balanced~\cite{barnes2011critically,youngchul2012} -- all timescales are comparable and the nonlinear processes fix $k_\parallel$ such that $\omega \sim k_\parallel \vth[i]$. Thus, we do not expect our assumption to break down in practice.}
	and also
\begin{equation}
k_\perp \rho_e \sim \delta \ll 1,
\label{flre}
\end{equation}
so we can neglect finite-electron-gyroradius effects.  We choose the electron collision rate $\nu_e$ to be comparable to the fluctuation frequency, i.e.,
\begin{equation}
\nu_e \sim \frac{\vth[i]}{a}\sim \frac{1}{\delta}\nu_i.
\label{nue}
\end{equation}
Note that although electron collisions are comparable to the fluctuation frequency, and, therefore, important in the electron fluctuation dynamics, this would conventionally be considered a low collisionality plasma since 
$\nu_{\mathrm{eff}} = \infrac{\nu_i a}{\vth[i]}\sim \infrac{\nu_e a}{\vth[e]}\sim \delta\ll 1$.  Under these assumptions, we expand $h_e$, $\gkupot$, $\delAp$ and $\delBp$ as
\begin{eqnarray}
h_e = h_e^{(0)} + h_e^{(1)} + \cdots, \;\;\;\;\;\; & \gkupot = \gkupot^{(0)} + \gkupot^{(1)}+ \cdots, \;\; \nonumber \\
\delAp = \delAp^{(0)} + \delAp^{(1)}+ \cdots,     \;\; {\mathrm{and}} \;\; & \delBp = \delBp^{(0)} +   \delBp^{(1)}+ \cdots,
 \label{hExpand}
\end{eqnarray}
where $h_e^{(1)} \sim \massratio \, h_e^{(0)}$ \textit{etc.}\footnote{In fact, we will only require $\gkupot$ or $\delBp$ to lowest order in $\massratio$. Thus, we will drop the superscripts on these fields.}
In Sections~\ref{Smagfield} and \ref{Sreduced}, we systematically expand the gyrokinetic system of equations, \eref{gke}, \eref{fluct-qn}, \eref{fluct-apar} and \eref{fluct-bpar} in powers of $\delta$ to obtain $h_e^{(0)}$ and close the system at first order in $\delta$.

\section{Zeroth Order: Particular Solutions, Flux Conservation and Field-Aligned Coordinates}
\setcounter{footnote}{0}
\label{Smagfield}
In this section, we apply the orderings of \Sref{Sexpansion} to the gyrokinetic equation for electrons and examine the consequences to lowest order.

In order to expand the gyrokinetic equation, in \ref{apGke}, we express $\gyroR{\gkpot}$ explicitly in terms of the fields and find, \eref{gkpot}:
\begin{equation}
\gyroR{\gkpot} = \underbrace{\gkupot(\bm{R}_e)}_{\Or\left(\gkeps\frac{T}{e}\right)} -  \underbrace{\frac{w_\parallel}{c} \delAp(\bm{R}_e)}_{\Or\left(\frac{\gkeps}{\delta}\frac{T}{e}\right)} - \underbrace{\frac{m_e w_\perp^2}{2e }\frac{\delBp(\bm{R}_e)}{\MeanMagB}}_{\Or\left(\gkeps\frac{T}{e}\right)} + \Or\left(\gkeps\delta\frac{T_e}{e}\right).
\label{gkpotExp}
\end{equation}
Substituting this result and the expansion \eref{hExpand} into the gyrokinetic equation \eref{gke}, we find, to lowest order in $\massratio$:
\begin{equation}
\label{gk0thTmp}
\fl\begin{eqalign}{
w_\parallel\Meanb \dgR[e]{h_e^{(0)}}& - \frac{w_\parallel}{\MeanMagB} \left( \Meanb\times\ddR[e]{{\delAp^{(0)}}} \right) \cdot \ddR[e]{h_e^{(0)}} = \\
	&\frac{e w_\parallel  F_{0e}}{c T_e} \left[ \pd{ }{t} + \bm{u}(\bm{R}_e) \dgR[e]{} \right] {\delAp^{(0)}} + \frac{w_\parallel}{\MeanMagB}  \left( \Meanb\times\ddR[e]{{\delAp^{(0)}}} \right)\dgR[e]{F_{0e}},
}\end{eqalign}
\end{equation}
where $\delAp$ is evaluated at $\bm{R}_e$ and we have used $\bm{u}(\bm{R}_e) \cdot \binfrac{\partial w_\parallel}{\partial \bm{R}_e} = 0$ (axisymmetry).
These equations are written in terms of the variables $\bm{R}_e$, $\energy[e]$, $\magmom[e]$ and $\gyr$. It will be convenient to instead 
convert back to using $\bm{r}$ as our spatial variable. Since \eref{gk0thTmp} is only accurate up to corrections of order $\Or(\massratio\gkeps^2 \Omega_i F_{0e})$ and 
all functions in \eref{gk0thTmp} are evaluated at $\bm{R}_e$, we can use the fact that $k_\perp \rho_e \sim \massratio$ to replace $\bm{R}_e$ by $\bm{r}$ everywhere. The function $\energy[e]$ is now given by the simplified expression:
\begin{equation}
\energy[e] = \frac{1}{2} {m_e w_\parallel^2} + \magmom[e] \MeanMagB - e \pot_0 + \Or\left( \massratio^2 T_e  \right).
\end{equation}

Defining $\bhat = \infrac{\Bfield}{|\Bfield|}$ to be the unit vector along the exact magnetic field, we have
\begin{equation}
\bhat = \Meanb + \frac{\delB_\perp}{\MeanMagB} + \Or(\gkeps^2) = \Meanb - \frac{1}{\MeanMagB}  \left( \Meanb\times\nabla{{\delAp^{(0)}}} \right) + \Or(\delta\gkeps) .
	\label{bhatDef}
\end{equation}
Thus, after a little algebra (using \eref{bhatDef}), we can rewrite all but one term of \eref{gk0thTmp} as derivatives along the exact field line:
\begin{equation}
\fl\label{gketmp}
\begin{eqalign}{
w_\parallel \bhat\dg \left( \frac{\delta f_e}{F_{0e}} - \frac{e\gkupot}{T_e} \right) = &\frac{e w_\parallel}{cT_e} \left(\pd{}{t} + \bm{u}\dg\right)\delAp^{(0)} \\ 
	&\quad-w_\parallel \bhat\dg \ln \NotN[e] -  w_\parallel \left( \frac{\energy[e]}{T_e} - \frac{3}{2}\right) \bhat\dg \ln T_e,
}\end{eqalign}
\end{equation}
where we have used 
\begin{equation}
\delta f_e = \frac{e \gkupot}{T_e} F_{0e} + h_e^{(0)} + \Or\left(\gkeps \massratio F_{0e} \right),
	\label{dfe0}
\end{equation}
the Maxwellian form \eref{F0def} of $F_{0e}$, and divided through by $F_{0e}$.  \Eref{gketmp} is an inhomogeneous  ordinary differential equation along the exact field lines for $\delta f_e$.

In the remainder of this section, we will find the general solution of \eref{gketmp}.
The particular solution to the inhomogeneous problem is constructed in \Sref{Smax}.
In order to solve the homogenous part of \eref{gketmp}, it is necessary to discover the structure of the exact magnetic field.  Thus, in \Sref{Sflux}, we prove that magnetic flux is conserved to lowest order in $\massratio$ and hence that the topology of the magnetic field is fixed. This result then allows us in \Sref{Scoordinates} to introduce field-aligned coordinates in which it is simple to solve the homogenous part of \eref{gketmp}.
This general solution will provide us with an equation for $\delAp$ and place constraints upon the form of $h_e^{(0)}$.
\subsection{Particular Solutions of \eref{gketmp}}
\label{Smax}

We can now find particular solutions of \eref{gketmp}. To do this, we write $\delta f_e$ in the following form:
\begin{equation}
\delta f_e = \lambda(\bm{r},\energy[e],\magmom[e],\sigma)F_{0e} + g_e(\bm{r},\energy[e],\magmom[e],\sigma) + \Or\left(\gkeps\massratio F_{0e} \right),
	\label{ansatz}
\end{equation}
where $g_e$ satisfies the homogenous equation
\begin{equation}
\label{homog}
w_\parallel \bhat\dg g_e = 0.
\end{equation}
This can clearly be done for any $\delta f_e$ by suitable choice of $\lambda$.
Substituting \eref{ansatz} into \eref{gketmp} we obtain
\begin{equation}
\fl
\bhat\dg \left( \lambda  - \frac{e\gkupot}{T_e} \right) = \frac{e}{cT_e} \left(\pd{ }{t} + \bm{u}\dg\right) \delAp^{(0)} - \bhat\dg \ln \NotN[e] - \left(\frac{\energy[e]}{T_e} - \frac{3}{2}\right)\bhat\dg \ln T_e.
\label{tmp1stOrdDonkey}
	\end{equation}
To find the form of $\lambda$, we take the derivative of this equation with respect to $\magmom[e]$ and the second derivative with respect to $\energy[e]$ to obtain
\begin{equation}
\bhat\dg \pd{\lambda}{\magmom[e]} = 0 \mbox{ and } \bhat\dg \pd{^2\lambda}{\energy[e]^2} = 0,
\end{equation}
respectively.
Clearly, any solution to these equations is either independent of $\magmom[e]$ and a linear function of $\energy[e]$ or satisfies $\bhat\dg \lambda = 0$. Any solution of the second kind can be absorbed into the function $g_e$, leaving us with the general solution for $\lambda$ given by
\begin{equation}
\lambda = \frac{\delNe(\bm{r})}{n_e} + \left(\frac{\energy[e]}{T_e} - \frac{3}{2}\right)\frac{\delTe(\bm{r})}{T_e},
\label{lambdaSpec}
\end{equation}
where we have written the linear function of $\energy[e]$ in the natural way as a perturbed Maxwellian.  However, it should be noted that $\delNe(\bm{r})$ is not the perturbed electron density and $\delTe(\bm{r})$ is not the perturbed electron temperature because $g_e$ can have both density and energy moments.
Inserting \eref{lambdaSpec} into \eref{tmp1stOrdDonkey} and taking a derivative with respect to $\energy[e]$, we find the following equation for $\delTe$:
\begin{equation}
\label{deltaT}
\bhat\dg\left(\ln T_{e} + \frac{\delTe}{T_{e}}\right) = 0.
\end{equation}
Finally, substituting both \eref{lambdaSpec} and \eref{deltaT} into  \eref{tmp1stOrdDonkey}, we obtain:
\begin{equation}
\label{deltan}
\bhat\dg \left(\frac{\delNe}{n_e} - \frac{e \gkupot}{T_e} + \ln \NotN[e]\right) = \frac{e}{cT_e} \left(\pd{ }{t} + \bm{u}\dg\right) \delAp.
\end{equation}

Familiar physics is contained in these two equations. 
It will be shown in \Sref{SZerothOrder} that the contribution to $g_e$ from passing electrons is a function only of $\tpsi$ and not of $\talpha$ or $\tl$, and, therefore, so are its density and temperature. Thus, for the purposes of interpreting \eref{deltan} and \eref{deltaT}, we can add the density and temperature of the passing part of $g_e$ to $\delNe$ and $\delTe$ respectively.
Hence, \eref{deltaT} describes how (part of) the temperature of the electrons is equalised along field lines, due to rapid thermal conduction. The only part of the temperature that does not equalise along the field line, as
we shall see when we solve \eref{homog}, is the part of the temperature of the trapped particles contained in $g_e$. This is to be expected; the trapped particles do not traverse the entire field line and so cannot equalise their temperature along it.

\Eref{deltan} describes how the pressure gradient of the electrons (again, not including the trapped particle pressure) along the field line is balanced by a parallel electric field. As we shall demonstrate, it is useful to think of \eref{deltan} as an evolution equation for $\delAp$ where $\gkupot$ and $\delNe$ are given.

\subsection{Flux Conservation}
\label{Sflux}
To solve \eref{homog}, \eref{deltaT} and \eref{deltan} we will need to know the spatial structure of the total magnetic field.
As \eref{deltan} determines $\delAp$, it determines $\delB_\perp$ and thence $\bhat$.
In this section, we will prove that \eref{deltan} implies that magnetic flux is conserved and consequently that the evolution of the magnetic perturbation cannot change the structure of the magnetic field lines.

Magnetic flux is said to be conserved if there exists an effective velocity field $\veff$ such that closed curves moving with this velocity always enclose the same amount of magnetic flux. It is proved by Newcomb in \cite{newcombLinesOfForce} that if such a $\veff$ exists, it also preserves magnetic field lines and thus the magnetic topology is fixed. 
Now, if the parallel electric field satisfies
\begin{equation}
\Efield\cdot\bhat = -\bhat\dg\xi,
	\label{simpleXi}
\end{equation}
where $\xi$ is a single-valued scalar function, then the velocity field
\begin{equation}
\veff = \frac{c}{\MagBfield^2} \left(\Efield + \nabla\xi \right)\times\Bfield + U(\bm{r}) \bhat,
\label{veffDef}
\end{equation}
(where $\MagBfield = |\Bfield|$ and $U(\bm{r})$ is an arbitrary single-valued function) satisfies $\Efield + \infrac{\veff\times\Bfield}{c} = - \nabla\xi$.  Thus Faraday's law becomes $\infrac{\partial \Bfield }{\partial t} = \nabla\times (\veff\times\Bfield)$.  The familiar proof of the frozen-in theorem (that is reproduced in almost all MHD textbooks) clearly applies if we recognise $\veff $ as the flux- and field-line-preserving velocity. The condition \eref{simpleXi} is a special case of the general conditions for the existence of frozen flux that were investigated in~\cite{newcombLinesOfForce}.

In \ref{deriv}, we prove that we can write $\Efield\cdot\bhat$ in terms of potentials as
\begin{equation}
\label{simpleepar}
\fl
\begin{eqalign}{
\Efield\cdot\bhat = - \bhat\dg \pot_0 - \bhat\dg \gkupot - \frac{1}{c} \pd{\MeanA}{t}\cdot\Meanb - \frac{1}{c} \left(\pd{}{t}+ \bm{u}\dg\right)\delAp
	+ \Or\left(\gkeps^3\frac{\vth[i]}{c} \MeanMagB\right).
}\end{eqalign}
\end{equation}
In \ref{resistive}, we prove that the mean magnetic flux is conserved irrespective of the fluctuations, and so (see \eref{meanConsFlux})
\begin{equation}
 \pd{\MeanA}{t}\cdot\Meanb = \Meanb \dg \frac{T_{e} n_{1e}}{e {n}_{e}},
 \label{neoclass}
\end{equation}
where $n_{1e}$ is the $\Or(\gkeps)$ correction to the mean electron density (see discussion before \eref{meanConsFlux}).
Finally, using  \eref{deltan} and \eref{neoclass} in \eref{simpleepar}, we obtain
\begin{equation}
\Efield\cdot\bhat = -\bhat\dg\left[ \frac{T_e\left( \delNe  + n_{1e} \right)}{e n_{e}} + \frac{T_e}{e} \ln \NotN[e] + \pot_0\right],
\label{epar}
\end{equation}
which is indeed in the form \eref{simpleXi}.\footnote{
Strictly speaking, we have proven that $\Efield\cdot\bhat = -\bhat\dg\xi + \tilde{E}_\parallel \bhat$ where $\tilde{E}_\parallel \sim \gkeps^3 \vth[i] \MeanMagB / c$. 
Thus the electric field is precisely $\Efield = -\veff \times \Bfield / c - \nabla \xi + \tilde{E}_\parallel \bhat$. Substituting this into Faraday's law, we find that 
\begin{equation*}
\pd{\Bfield}{t} =  \curl \left( \veff \times\Bfield\right) - c\curl\left(\tilde{E}_\parallel \bhat\right).
\end{equation*}
Estimating the size of the left hand side and the second term on the right hand side, we see that the correction $\tilde{E}_\parallel$ will affect neither the lowest-order mean field $\MeanB$ nor the lowest-order fluctuation $\delB$. Thus, we can safely ignore it.}
	Thus, we can define
\begin{equation}
\label{xidef}
\xi = \frac{T_e}{e} \left(\ln \NotN[e] + \frac{\delNe + n_{1e}}{n_{e}}\right) +\pot_0 + \tilde{\xi},
\end{equation}
where $\tilde{\xi}$ is an arbitrary single-valued function that satisfies
\begin{equation}
\bhat\dg \tilde{\xi} = 0,
\label{tildexiDef}
\end{equation}
and with this definition, \eref{veffDef} defines the field-line-preserving velocity $\veff$.
In \eref{veffDef}, we are free to choose $U(\bm{r})$ and $ \tilde{\xi}$ subject to the constraint $\bhat\dg \tilde{\xi} = 0$. We postpone making choices for these parameters until \Sref{SZerothOrder}.  Since the field is frozen to $\veff$ the ion-scale turbulence cannot alter the magnetic field structure on the turbulent timescale.

\subsection{Structure of the Magnetic Field}
\label{SClebsch}
In the previous section we have demonstrated that ion-scale turbulence preserves field lines.  The topology of the field is thus fixed.  But what is this topology?
Electron-scale fluctuations \textit{can} alter the topology and so can generate a stochastic field\footnote{A stochastic field is one in which points on neighbouring field lines become exponentially separated as they travel along field lines \cite{rechester1978eht}.} from an initially regular one.  Such fluctuations may arise ``locally'' from electron-scale instabilities ({e.g.} Electron Temperature Gradient instabilities or Microtearing), but electron-scale fluctuations may also be driven ``non-locally'' by a cascade from ion-scale fluctuations.  In both cases, 
	some tangling of the magnetic field is possible, indeed probable.  Thus, at some intuitive level, the magnetic field is expected to be stochastic.  A stochastic field will be advected by $\veff$ and remain stochastic.

If the magnetic field were stochastic and stationary then the heat diffusivity of the electrons would be given by the well-known Rechester-Rosenbluth formula~\cite{rechester1978eht}: 
\begin{equation}
\chi_{T_e} = D_{st} \vth[e] \sim \vth[e] l_c\left(\frac{\delta B_{\mathrm{stoch}}}{B}\right)^2,
	\label{RechesterRosenbluth}
\end{equation}
where $D_{st}$ is a species-independent diffusion coefficient characterising the stochasticity of the field\footnote{\label{fieldlineDiffusion}The field-line diffusivity $D_{st}$ is defined as~\cite{rechester1978eht}
	\[
		D_{st} = \lim_{l\rightarrow\infty} \infrac{{\overline{\left(\delta r(l)-\delta r(0)\right)^2}}}{2l},
	\] where $\delta r(l)$ is the radial displacement of a point after following a field line for a distance $l$ and the overbar denotes statistical averaging over field lines (the factor of $2$ in the denominator is chosen to agree with the definition in \cite{rechester1978eht}).} and $l_c$ is the parallel correlation length of the stochastic component of the field $\delta B_{\mathrm{stoch}}$.  Let us assume the stochastic component of the field is comparable to the ion-scale field perturbations (i.e., $\infrac{\delta B_{\mathrm{stoch}}}{B}\sim \gkeps$) and  the correlation length is the system size (i.e., $l_c\sim a $). Then, \eref{RechesterRosenbluth} predicts that the electron heat diffusivity in a tokamak should be $\massratio^{-1}$ larger than the ion heat diffusivity (both the electrostatic and electromagnetic parts of the ion heat diffusivity).  This is not backed up by experimental observations; ion and electron heat diffusivities are observed to be of the same order of magnitude. Thus, we take this as experimental justification for assuming that the amount of magnetic field stochasticity is limited.
\footnote{
We can formulate this condition precisely as a condition on the field-line diffusivity (defined in the footnote on page\ \pageref{fieldlineDiffusion}) . The limit on stochasticity can be expressed as 
\[
D_{st} \ll \gkeps \rho_e.
\]
	where we obtain this estimate by insisting that the change in a function $g$ following the field line is dominated by explicit variation along a field line ($ k_\parallel g$) rather than the field line connecting neighbouring points in the perpendicular plane ($ D_{st} k_\perp^2 g$). The existence of a stochastic field giving rise to a field-line diffusivity of size $D_{st} \lesssim \gkeps\rho_e$ would not invalidate the existence of $\tpsi$ and $\talpha$, as (for the purposes of studying ion-scale turbulence) we could define $\tpsi$ surfaces by the mean radial location of a field line averaged along a magnetic correlation length $l_c$. This introduces a displacement $\lesssim \rho_e$ of the $\tpsi$ surface from the exact position of the field line, but for turbulence where $k_\perp\rho_e \ll 1$ this correction is negligible.
}

	We therefore assume good flux surfaces in solving Eqs.~\eref{homog}, \eref{deltaT} and \eref{deltan} -- these surfaces will be approximations to the exact field that we will refer to as the ``ion-scale field''.  The ion-scale turbulence will distort the ion-scale field but the surfaces will remain intact.   A smaller stochastic field component with $l_c \sim a$ and $\infrac{\delta B_{stoch}}{B}\sim\gkeps\delta^{1/2}$ would cause observable levels of transport and we cannot rule out a component of this size.  Indeed, such a field would change the final electron equations in this paper.  However, we will assume for simplicity that the stochastic component of the field is zero to order $\Or (\gkeps\delta)$ (i.e., $\infrac{\delta B_{\mathrm{stoch}}}{B} \ll \gkeps\delta$) in this paper.

\subsection{Field-Aligned Coordinates}
\label{Scoordinates}

If we write the ion-scale magnetic field in the Clebsch form
\begin{equation}
\Bfield = \nabla\tpsi \times\nabla\talpha,
\label{Clebsch}
\end{equation}
then $\tpsi$ is a single-valued function that labels the flux surfaces and $\talpha$ is a function that labels different field lines within a given flux surface. To ensure that $\talpha$ is single-valued, we make a cut in the poloidal domain on the inboard side of the tokamak.  This cut makes a ribbon forming a toroidal loop with one edge of the ribbon being the magnetic axis and the other the plasma boundary or separatrix. We allow the cut to move with the velocity $\veff$. This is legitimate as $\veff$ is continuous across the cut. Since the cut is perturbed by the turbulence, it will not now be axisymmetric. 
\begin{figure}[ht]
\centering
\includegraphics[width=\textwidth]{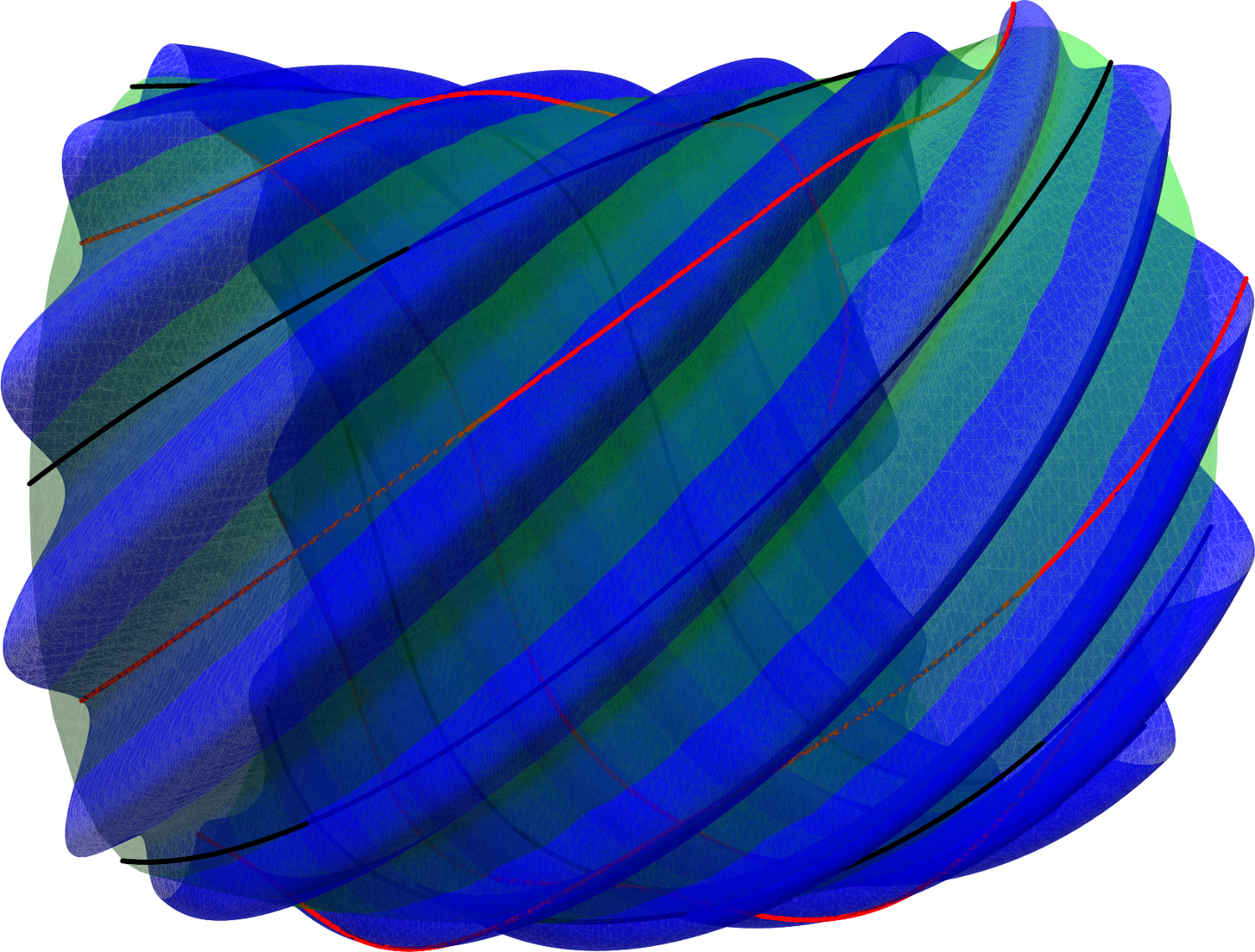}
\caption{The exact field $\Bfield$ and its $\tpsi=\mathrm{const.}$ surfaces (red lines, blue surface) plotted atop the mean field $\MeanB$ and a $\psi=\mathrm{const.}$ surface (black lines, green surface).}
\label{figSurfaces}
\end{figure}
The expression \eref{Clebsch} for the magnetic field allows us (see, e.g., \cite{beer1995field}) to introduce a system of field-aligned coordinates ($\tpsi$, $\talpha$, $\tl$) where $\tl$ measures distance along a given field line.  By definition, $\tl$ runs from one side of the cut in the poloidal plane to the other side of the cut.  As $\delB \ll \MeanB$, the variables aligned to the exact field only deviate slightly from those aligned to the background field:
\begin{equation}
|\tpsi - \psi | \sim \gkeps \psi, \qquad |\talpha - \alpha | \sim \gkeps \alpha, \qquad |\tl - l| \sim \gkeps l.
\end{equation}
Importantly, this means that all axisymmetric mean quantities are, to lowest order in $\gkeps$, functions of $\tpsi$ and $\tl$ but not of $\talpha$.

We need expressions for the spatial derivative operators $\bhat\dg$ and the Poisson bracket $\pb{a}{b} = \Meanb\cdot\left( \nabla a\times\nabla b \right)$ in our new coordinates.  By using the fact that $\bhat\dg\tpsi = \bhat\dg\talpha = 0$, we discover that
\begin{equation}
\bhat \dg a = \bhat\dg \tl \left.\pd{a}{\tl}\right|_{\tpsi,\talpha} = \left.\pd{a}{\tl}\right|_{\tpsi,\talpha}
\label{ddlDef}
\end{equation}
for any function $a(\bm{r})$. 
Similarly, by using the fact that $\bhat\cdot( \nabla\talpha\times\nabla\tpsi ) = \MeanMagB$ to lowest order, we see that the Poisson bracket takes the form
\begin{equation}
\fl
\pb{a}{b} = \bhat\cdot\left( \nabla a\times\nabla b \right) +\Or(\gkeps|\nabla a||\nabla b|) =  \MeanMagB\left( \pd{a}{\talpha} \pd{b}{\tpsi} - \pd{a}{\tpsi} \pd{b}{\talpha} \right) + \Or( \gkeps |\nabla a||\nabla b|).
\label{poissonSimple}
\end{equation}

Since $\veff$ is a field-line-preserving velocity, we have that
\begin{equation}
\left( \pd{}{t} + \veff\dg \right) \tpsi = 0,
	\label{psiEq}
\end{equation}
and
\begin{equation}
\left( \pd{}{t} + \veff\dg \right)\talpha = 0.
\label{alphaEq}
\end{equation}
As $\tl \sim l$ and $\dv \veff \sim \gkeps \omega$, $\veff$ also approximately preserves the length of field lines and we have, to lowest order in $\gkeps$,
\begin{equation}
\left( \pd{}{t} + \veff\dg \right) \tl = 0.
\label{lconst}
\end{equation}
It is of course convenient to choose the cut in the definition of $\talpha$ such that it is convected with $\veff$.  

Finally, the time derivative at constant $\tpsi$, $\talpha$ and $\tl$ is related to the time derivative at constant $\bm{r}$ by
\begin{equation}
\ddttwiddles = \pd{}{t} + \veff\dg.
\label{ddtTwiddles}
\end{equation}

\subsection{Solution of the Lowest-Order Equations}
\label{SZerothOrder}
We now use the magnetic coordinates introduced in \Sref{Scoordinates} to solve \eref{homog} and \eref{deltaT}.
We split velocity space ($\energy[e]$ and $\magmom[e]$)  into two regions: the passing region where, for a given $\tpsi$, the equation
\begin{equation}
  w_\parallel(\tl,\tpsi,\energy[e],\magmom[e]) =  \sqrt{{\frac{2}{m_e}}\left[ \energy[e] - \magmom[e] \MeanMagB + e\pot_0  \right]} = 0
\label{wparDef}
\end{equation}
has no solutions for any $\tl$ (i.e., $ \energy[e] > \magmom[e] \MeanMagB - e\pot_0$ for all $\tl$) and the trapped region in which there are two values of $\tl$ at which $w_\parallel = 0$. The points where $w_\parallel$ vanishes are called the bounce points of the electron's orbit.
In the passing region, \eref{homog} implies that
\begin{equation}
\pd{g_e}{\tl} = 0,
\end{equation}
so $g_e$ is constant along field lines. As most magnetic surfaces are irrational, one field line spans the entire surface and so, for passing electrons, $g_e$ cannot depend on $\talpha$:
\begin{equation}
g_e = g_{pe} ( \tpsi, \energy[e], \magmom[e], \sigma, t).
\end{equation}
In the trapped region of velocity space, \eref{homog} implies that
\begin{equation}
\mbox{either} \quad w_\parallel(\tl,\tpsi,\energy[e],\magmom[e]) = 0, \quad\mbox{or}\quad \pd{g_e}{\tl} = 0.
\end{equation}
Thus, $g_e$ must be constant along magnetic field lines, but only between the bounce points. So $g_e$ can depend on the location of the bounce points, and, therefore, on $\talpha$:
\begin{equation}
g_e = g_{te} (\tpsi,\talpha,\energy[e],\magmom[e],t),
\end{equation}
where the dependence on $\sigma$ has been dropped as $g_e$ must be continuous as $w_\parallel$ passes through zero and changes sign.
For convenience, we define $g_{pe}$ to be zero in the trapped region and $g_{te}$ to be zero in the passing region, so the complete solution is just the sum of the two functions:
\begin{equation}
g_e = g_{pe}(\tpsi,\energy[e],\magmom[e],\sigma,t) + g_{te}(\tpsi,\talpha,\energy[e],\magmom[e],t).
\end{equation}
The solution of \eref{deltaT} in our new coordinates is
\begin{equation}
T_e(\psi) + \delTe = \tilde{T}_e (\tpsi),
\end{equation}
for some function $\tilde{T}_e$. As we can absorb any part of $\delTe$ that is only a function of $\tpsi$ into $g_e$, we can pick $\tilde{T}_e(\tpsi) = T_e(\tpsi)$ and so
\begin{equation}
\delTe = \left(\tpsi - \psi\right) \frac{d T_e}{d\psi}.
\end{equation}

We can also now pick the arbitrary function $\tilde{\xi}$ in the definition \eref{xidef} of $\xi$ to be any function of $\tpsi$. In \ref{parveloc}, we find convenient forms for the arbitrary functions $\tilde{\xi}(\tpsi)$ and $U(\bm{r})$ to obtain (see \eref{veffE7})
\begin{eqnarray}
\veff = \bm{u} + \frac{c}{\MeanMagB}\bhat \times \nabla\left( \gkupot - \zeta \right),
	\label{uxi}
\end{eqnarray}
where we have defined
\begin{equation}
\zeta = \frac{T_e}{e}\left[ \frac{\delNe}{n_e} - \left(\tpsi - \psi\right) \frac{d\ln \NotN[e]}{d\psi}\right].
\label{zetaDef}
\end{equation}
With this definition, we can express the general solution of \eref{gketmp} as
\begin{equation}
\fl
\begin{eqalign}{
\delta f_e = &\frac{e\zeta}{T_e} F_{0e} + \left( \tpsi - \psi \right) \pd{F_{0e}}{\psi} + g_{pe}(\tpsi,\energy[e],\magmom[e],\sigma) + g_{te} (\tpsi,\talpha,\energy[e],\magmom[e]) \\
				 &\qquad+ \Or\left(\gkeps\massratio F_{0e} \right),
}\end{eqalign}
	\label{deltafSoln}
\end{equation}
and the evolution equation \eref{deltan} for $\delAp$ becomes
\begin{equation}
\left( \pd{ }{t} + \bm{u}\dg \right)\delAp = c \bhat\dg\left(\zeta - \gkupot\right) + \Or\left(\gkeps \delta \Omega_i \delAp \right).
\label{deltaApar}
\end{equation}
This equation allows us to solve for $\delAp$ correctly to lowest order in $\massratio$.  Equations for the evolution of $\zeta$, $g_{pe}$ and $g_{te}$ at the next order in our expansion are obtained in \Sref{Sreduced}.

\subsection{Solution for \texorpdfstring{$\delB_\perp$}{the fluctuating magnetic field}}
At this juncture, we have enough information to determine $\delB_\perp$ in two ways. 
Firstly, we have the evolution equations \eref{psiEq} and \eref{alphaEq} for $\tpsi$ and $\talpha$, with $\veff$ given by \eref{uxi}. 
This enables us to calculate $\tpsi$ and $\talpha$ correctly to first order in $\gkeps$.
 In \ref{repDeltaB}, we demonstrate that this gives us enough information to construct $\delAp$ from the equations
\begin{equation}
\pd{\delAp}{\tpsi} = -\pd{\talpha}{l} \quad\mbox{ and }\quad \pd{\delAp}{\talpha} = \pd{\tpsi}{l}.
\label{dAparFromT}
\end{equation}
Secondly, we could solve for $\delAp$ from the evolution equation \eref{deltaApar} and then use \eref{dAparFromT} to determine $\tpsi$ and $\talpha$.
Either of these methods is valid, and provides a solution for $\delAp$ correct to the requisite order. For the purposes of simulating our equations, a third option presents itself: determine $\delAp$ from \eref{deltaApar}, $\tpsi$ and $\talpha$ from \eref{psiEq} and \eref{alphaEq}, and then use \eref{dAparFromT} as consistency checks upon the solutions thus obtained.
We choose the first of these options, i.e., to eliminate $\delAp$ in favour of the fluctuating fields $\tpsi-\psi$ and $\talpha-\alpha$.

One point remains to be resolved. It would appear that knowing $\tpsi$ and $\talpha$ determines $\delB$ completely (via \eref{Clebsch}).
However, to determine $\delBp$ from $\tpsi$ and $\talpha$, one must know $\tpsi$ and $\talpha$ to $\Or(\gkeps^2)$ or, equivalently, the compressible part of $\veff$ to higher order ($\Or (\gkeps^2\vth[i])$; see \ref{repDeltaB}), but we do not.  Therefore, we determine $\delBp$ from \eref{fluct-bpar} which involves only lowest-order quantities -- quantities that we do calculate.

\subsection{Parallel Electron Currents}
As explained in the previous section, we have obtained a solution for $\delAp$. However, $\delAp$ is usually determined from the parallel component of Amp\`ere's Law, \eref{fluct-apar}.
Substituting our expansion for $h_e$ \eref{hExpand} into the parallel component of Amp\`ere's law \eref{fluct-apar}, we find
\begin{equation}
\nabla^2 \delAp = \frac{4\pi e}{c} \left( \wint w_\parallel \gyror{h_e^{(0)} + h_e^{(1)}}  \right) - \frac{4\pi}{c} \sum_{s=i} Z_s e \wint w_\parallel \gyror{h_s},
	\label{apartmp}
\end{equation}
where a sum over $s=i$ denotes a sum over all ion species. In this equation, the term due to $h_e^{(0)}$ is larger than the rest by $\massratio^{-1}$. Thus, to lowest order,
\begin{equation}
\wint w_\parallel h_e^{(0)} = \wint w_\parallel g_{pe} = 0.
\end{equation} 
This is a \textit{constraint} on $g_{pe}$ in the lowest-order solution given by \eref{deltafSoln}. Note that the trapped particles do not contribute to the current constraint. The parallel electron flow (and current) is comparable to that of the ions and contained in $h_e^{(1)}$. Defining the parallel electron flow 
\begin{equation}
\upar = \wint w_\parallel h_e^{(1)}
\label{uparDef}
\end{equation}
reduces \eref{apartmp} to an equation for $\upar$:
\begin{equation}
\label{uparEq}
\upar = \frac{c}{4\pi e} \nabla^2 \delAp + \sum_{s=i} Z_s \wint w_\parallel \gyror{h_s}.
\end{equation}
Thus, instead of determining $\delAp$ from the electron current, Amp\`ere's law determines the electron flow from $\delAp$~\cite{Tome}.  
As we are no longer using $\delAp$ to describe $\delB_\perp$ in our system of equations, we consider $\nabla_\perp^2 \delAp$ to be a shorthand for a complicated expression involving derivatives of $\tpsi-\psi$ and $\talpha-\alpha$ via the relations \eref{NablaApar} or \eref{NablaApar2}.
We require $\upar$ to complete the solution in the next order, but we do not need any information about $h_e^{(1)}$ other than $\upar$.

\section{First Order: Dynamical Equations for Electrons}
\label{Sreduced}
In this section, we return to the gyrokinetic equation \eref{gke} and proceed to the next order in the mass-ratio expansion in order to find evolution equations for $g_e$ and $\zeta$.

Subtracting \eref{gk0thTmp} from \eref{gke} and using the solutions for $h_e^{(0)}$ and $\delAp$ from \Sref{SZerothOrder}, we obtain the first-order part of the gyrokinetic equation:
\begin{equation}
\fl\begin{eqalign}{
\left( \pd{ }{t} + \bm{u}\dg\right) h_e^{(0)} + w_\parallel\bhat\dg h_e^{(1)} - \frac{w_\parallel}{\MeanMagB}\pb{\delApN{1}}{h_e^{(0)}}+ \vdrift[e]\dg h_e^{(0)} \\
	\quad+ \frac{c}{\MeanMagB}\pb{\gkupot - \frac{m_e w_\perp^2}{2 e} \frac{\delBp}{\MeanMagB}}{h_e^{(0)}}
- \gyroR{\collop[h_e^{(0)}]}\\
	= - \frac{e F_{0e}}{T_e} \left( \pd{ }{t} + \bm{u}\dg\right) \left( \gkupot - \frac{m_e w_\perp^2}{2 e} \frac{\delBp}{\MeanMagB}\right) 
	- \frac{c}{\MeanMagB} \pb{ \gkupot- \frac{m_ew_\perp^2}{2 e} \frac{\delBp}{\MeanMagB} }{F_{0e}} \\
	\quad+\, \frac{I(\psi) m_e w_\parallel^2 F_{0e}}{\MeanMagB^2 T_e} \frac{d\angvel}{d\psi} \left( \Meanb \times \nabla\delAp \right) \dg\psi\\
	\quad+\, {w_\parallel F_{0e}} \left[ \frac{e}{cT_e} \left( \pd{}{t}+\bm{u}\dg \right)\delApN{1} + \frac{1}{\MeanMagB}\pb{ \delApN{1}}{\ln F_{0e}}\right],
}\end{eqalign}
\label{gk1stTmp}
\end{equation}
where we have used \eref{bhatDef}, rewritten terms involving $\vchiR$ as Poisson brackets using \eref{vchi} and \eref{poissonSimple}, and, as in \Sref{Smagfield}, replaced $\bm{R}_e$ with $\bm{r}$ as our spatial variable. 
Thus, the collision operator in \eref{gk1stTmp} is evaluated at $\bm{R}_e = \bm{r}$.
The drift velocity in \eref{gk1stTmp} now contains only the leading-order part:
\begin{equation}
\label{vdrift2}
\fl
\begin{eqalign}{
\vdrift[e] = -\frac{c}{e\MeanMagB} \Meanb\times &\left(m_e w_\parallel^2 \Meanb\dg\Meanb + \frac{1}{2}m_e w_\perp^2\nabla \ln \MeanMagB -  e\nabla\pot_0\right) + \Or(\gkeps\delta\vth[i]).
}\end{eqalign}
\end{equation}

We now rewrite \eref{gketmp} in terms of $\delta f_e$ and $\zeta$:
\begin{equation}
\label{1stOrder}
\fl
\begin{eqalign}{
	\left( \pd{}{t} + \veff\dg \right) \delta f_e + w_\parallel \bhat\dg h_e^{(1)}- \frac{w_\parallel}{\MeanMagB}\pb{\delApN{1}}{h_e^{(0)}} + \vdrift[e] \dg \left(\delta f_e - \frac{e\gkupot}{T_e}F_{0e}\right)\\
		\qquad\qquad+ \frac{c}{\MeanMagB}\pb{\zeta - \frac{m_e w_\perp^2}{2e}\frac{\delBp}{\MeanMagB}}{\delta f_e} - {\gyroR{\collop[\delta f_e]}} \\
			\, = \left( \pd{}{t} + \veff\dg \right) \frac{w_\perp^2}{\vth[e]^2} \frac{\delBp}{\MeanMagB}F_{0e} + \pb{\zeta}{\frac{w_\perp^2}{\vth[e]^2}\frac{\delBp}{\MeanMagB}F_{0e}}
		- \frac{c}{\MeanMagB}\pb{\gkupot - \frac{m_e w_\perp^2}{2e}\frac{\delBp}{\MeanMagB} }{F_{0e}}
		\\
			\quad\,- \frac{I m_e w_\parallel^2 F_{0e}}{\MeanMagB^2 T_e} \frac{d\angvel}{d\psi} \left( \Meanb \times \nabla\delAp \right) \dg\psi\\
	\quad+\, {w_\parallel F_{0e}} \left[ \frac{e}{cT_e} \left( \pd{}{t}+\bm{u}\dg \right)\delApN{1} + \frac{1}{\MeanMagB}\pb{ \delApN{1}}{\ln F_{0e}}\right],
}\end{eqalign}
\end{equation}
where we have used \eref{dfe0}, \eref{zetaDef} and the fact that
\begin{equation}
\veff \dg  = \bm{u}\dg + \frac{c}{\MeanMagB}\pb{\gkupot-\zeta}{\cdot},
\end{equation}
which follows directly from \eref{uxi} and \eref{poissonSimple}.

Finally, we substitute the form of $\delta f_e$ from \eref{deltafSoln}, to obtain
\begin{equation}
\label{zugzug2}
\fl
\begin{eqalign}{
	&\left.\pd{g_e}{t}\right|_{\tpsi, \talpha,\tl} + w_\parallel \bhat\dg h_e^{(1)} + \vdrift[e] \dg g_e 
		+ \frac{c}{e\MeanMagB} \pb{e\zeta - \magmom[e]{\delBp}}{g_e}
		- \gyroR{\collop[g_e]}\\
			&\,= -\frac{F_{0e}}{T_e}\left.\pd{}{t}\right|_{\tpsi,\talpha,\tl} \left(e\zeta -\magmom[e]\delBp\right)- \vdrift[e] \dg (\tpsi-\psi) \pd{F_{0e}}{\psi} + \vdrift[e]\dg \left(\gkupot-\zeta\right) \frac{e F_{0e}}{T_e}
	\\
		&\quad+ \frac{c}{e}\pd{}{\talpha} \left(e\zeta - \magmom[e]\delBp\right)\pd{F_{0e}}{\tpsi}
			+ \frac{I(\psi) m_e w_\parallel^2 F_{0e}}{\MeanMagB^2 T_e} \frac{d\angvel}{d\psi} \left( \Meanb \times \nabla\delAp \right) \dg\psi \\
		&\quad+{w_\parallel F_{0e}} \left[ \frac{e}{cT_e} \left( \pd{}{t}+\bm{u}\dg \right)\delApN{1} + \frac{1}{\MeanMagB}\pb{ \delApN{1}}{\ln F_{0e}}\right]+ \frac{w_\parallel}{\MeanMagB}\pb{\delApN{1}}{h_e^{(0)}},
}\end{eqalign}
\end{equation}
where we have gathered all terms involving $g_e$ together on the left-hand side and all terms involving $\delApN{1}$ on the right.
Notationally, \eref{zugzug2} can be confusing as we have made some abbreviations for the sake of conciseness. Firstly, all spatial derivatives are taken at constant $t$, $\energy[e]$ and $\magmom[e]$. Thus, for equations that are solved in the moving coordinate system
\begin{equation}
\nabla = \nabla\tpsi \left.\pd{ }{\tpsi}\right|_{\talpha,\tl,t,\energy[e],\magmom[e]} + \left.\nabla\talpha \pd{ }{\talpha}\right|_{\tpsi,\tl,t,\energy[e],\magmom[e]} + \left.\nabla\tl\pd{ }{\tl}\right|_{\tpsi,\talpha,t,\energy[e],\magmom[e]}.
\end{equation}
As this notation is cumbersome, we will continue to use $\nabla$ where appropriate (i.e., where it would expand to more than one derivative). Secondly, all the Poisson brackets in \eref{zugzug2} are shorthand for the expression \eref{poissonSimple} in terms of $\tpsi$ and $\talpha$ derivatives.
Finally, this equation is, intrinsically, an equation {\textit{in the frame moving with the magnetic field}}, and all functions should be interpreted as functions of $t$, $\tpsi$, $\talpha$, and $\tl$ rather than $t$ and $\bm{r}$.

\Eref{zugzug2} describes how $g_e$, and thus $\delta f_e$, evolves dynamically in response to given perturbed fields. As $\veff$ is also considered known, \eref{psiEq}--\eref{lconst} can be solved to give the complete transformation from the fixed to the moving coordinate system.
Thus, we consider all the quantities in \eref{zugzug2} to be known in both the fixed coordinates (which will be needed when we come to the field equations) ($\psi, \alpha, l$),  and the moving coordinates, ($\tpsi, \talpha, \tl$). 
However, in its current form, \eref{zugzug2} is not closed, as we know neither $h_e^{(1)}$ nor $\delApN{1}$.  In the following sections we derive evolution equations for $\zeta$, $g_{te}$ and $g_{pe}$ as solubility constraints of \eref{zugzug2}.

\subsection{Fluctuating Continuity Equation}
\label{fluctcont}
The one piece of information that is known about $h_e^{(1)}$ is its parallel velocity moment $\upar$, determined by \eref{uparEq}.
Thus, we integrate \eref{zugzug2} over all velocities and find
\begin{equation}
\fl
\label{zetaeq}
\begin{eqalign}{
&\ddttwiddles  \left[\frac{e\zeta}{T_e} n_{e} + \wint g_e - \frac{\delBp}{\MeanMagB} n_e\right] + \MeanMagB \pd{ }{\tl} \frac{\upar}{\MeanMagB}
+ \wint \vdrift[e]\dg g_e	  \\
&\,+ \frac{c}{\MeanMagB}\pb{\zeta}{\wint g_e}- \frac{c}{e}\pb{\frac{\delBp}{\MeanMagB}}{\wint \magmom[e] g_e}
  \\
&\quad = - \wint {\bm{V}}_D\dg (\tpsi - \psi) \pd{F_{0e}}{\psi} + \left( \widehat{\bm{V}}_D + \bm{V}_\pot \right) \dg\left( \gkupot-\zeta\right) \frac{e n_{e}}{T_e} \\
&\qquad\quad+cn_e \pd{}{\talpha} \left(\zeta - \delBp \frac{T_e}{e\MeanMagB} \right) \left(\frac{d\ln\NotN[e]}{d\psi} - e\pot_0 \frac{d\ln T_e}{d\psi}\right) 
	- c \pd{\delBp}{\talpha} \frac{n_e T_e}{e\MeanMagB} \frac{d\ln T_e}{d\psi} \\
&\qquad\quad-\frac{In_e}{\MeanMagB}\frac{d\angvel}{d\psi}\pd{}{\tl}(\tpsi - \psi),
}\end{eqalign}
\end{equation}
where we have used $\MeanMagB \bhat = \Bfield + \Or(\gkeps \MeanMagB)$, the fact that $h_e^{(0)}$ carries no current, and
\begin{equation}
\wint  = \sum_\sigma \int \frac{B d\energy[e]d\magmom[e] d\gyr}{m_e^2|w_\parallel|}.
\end{equation}
We have also defined
\begin{equation}
\widehat{\bm{V}}_D = -\frac{cT_e}{e B} \Meanb \times \left( \Meanb\dg\Meanb + \nabla\ln \MeanMagB \right)
\end{equation}
and
\begin{equation}
\bm{V}_\pot = \frac{c}{B} \Meanb\times\nabla\pot_0
\end{equation}
so that
\begin{equation}
\wint \vdrift[e] F_{0e} = \left( \widehat{\bm{V}}_D + \bm{V}_\pot \right)n_e.
\end{equation}
We have also used $\dv\Bfield = 0$ and \eref{NablaApar2} to eliminate $\bhat$ and $\delAp$ from our equation.
It is important to note that this equation is written in the moving coordinate system. It relates $\zeta(\tpsi,\talpha,\tl,t)$ to $g_e$ and the other fluctuating fields, also given as functions of $\tpsi$, $\talpha$ and $\tl$. 
In this context, $\tpsi - \psi$ should be viewed as just another fluctuating field. Therefore, all gradients should be thought of in terms of derivatives with respect to $\tpsi$ and $\talpha$, for the Poisson bracket this is done via \eref{poissonSimple}.

That \eref{zetaeq} is essentially the fluctuating continuity equation is not immediatly obvious. Let us attempt to make this more transparent. Clearly, the first two terms under the time derivative in \eref{zetaeq} are the perturbed electron density due to the fluctuations; the time derivative of $\delBp$ adds to this the fluctuation in the electron density due to the compression of the flux surfaces.\footnote{In the derivation of the transport equations we show that
\[
	\ddttwiddles \frac{\delBp}{\MeanMagB} = \left(\idmat - \bhat\bhat\right)\bm{:}\nabla\veff,
	\]
		so $\partial \delBp / \partial t$ is equal to the rate of {\it perpendicular} compression of the flux surfaces.}
	This total density is changed by various flows -- the compression in the parallel flow $\upar$, the flow of electrons past the flux surface given by $(c/B)\bhat\times\nabla\zeta$, the grad-$B$ drift due to $\delBp$, and the magnetic drifts. Indeed, even the final term in \eref{zetaeq} can be interpreted as a compression -- it is the compression of the electron fluid due to the perturbed flux surfaces ``sticking out'' into regions with different toroidal angular velocities.

\Eref{zetaeq} provides us with an evolution equation for $\zeta$ if we know $g_e$ and the perturbed fields. The evolution equations for $g_e$ given $\zeta$ are found in the next two sections.

\subsection{Trapped Electrons and the Bounce Average}
\label{Sbounce}
In order to derive an equation for $g_e$ in the trapped region of velocity space, we need to define an averaging operator that eliminates the fast electron bounce motion (i.e., annihilates $w_\parallel \bhat\dg$ and thereby eliminates $h_e^{(1)}$).
The appropriate average is the bounce average along the perturbed field {\it i.e.,  at fixed $\tpsi$ and $\talpha$}. For an arbitrary function $a(\tpsi,\talpha,\tl,\energy[e],\magmom[e],\sigma)$ we define the bounce average to be
\begin{equation}
\bav{a} = \oint \frac{d\tl}{w_\parallel} a \left/ \oint \frac{d\tl}{w_\parallel}\right.,
\label{bavDef}
\end{equation}
where the integration is carried out from one bounce point where $w_\parallel = 0$, along the magnetic field, to the second bounce point where $w_\parallel$ passes through zero and changes sign, and then back to the first bounce point. This integral is carried out at constant $\tpsi$, $\talpha$, $\energy[e]$, $\magmom[e]$ and $\gyr$. Thus, equivalently, the bounce average is
\begin{equation}
\bav{a} = \left. \int \limits_{l_1}^{l_2} \frac{d\tl}{|w_\parallel|} \left[ a(\sigma = 1) + a(\sigma = -1) \right] \right/ 2\int\limits_{l_1}^{l_2} \frac{d\tl}{|w_\parallel|},
\end{equation}
where $l_1$ and $l_2$ are the bounce points of the particle orbit.

Clearly, the bounce average commutes with the time derivative at fixed moving coordinates $\tpsi$, $\talpha$ and $\tl$.  In \ref{apBounce}, we prove that 
\begin{equation}
\bav{w_\parallel \bhat\dg h_e^{(1)}} = 0
\end{equation}
and that, to lowest order in $\gkeps$,
\begin{equation}
\bav{\vdrift[e]\dg\tpsi} = 0.
\end{equation}
Using these results, we take \eref{zugzug2} in the trapped region of velocity space and perform the bounce average to find
\begin{equation}
\label{trap0}
\fl
\begin{eqalign}{
	&\ddttwiddles{g_{te}} + \bav{\vdrift[e]\dg g_{te}} + \frac{c}{e\MeanMagB} \pb{e\bav{\zeta}-\magmom[e]\bav{\delBp}}{g_{te}}  - \bav{\gyroR{\collop[g_{te}]}}\\
	&\,= - \ddttwiddles\left(\bav{\zeta} - \magmom[e] \bav{\delBp}\right) \frac{F_{0e}}{T_e} - \bav{\vdrift[e]\dg\left(\tpsi - \psi\right)} \pd{F_{0e}}{\psi} \\
	&\qquad+\bav{\vdrift[e]\dg\left(\gkupot-\zeta\right)}\frac{e F_{0e}}{T_e} 
			 + \frac{c}{e}\, \pd{}{\talpha} \left( e\bav{\zeta} - \magmom[e]\bav{\delBp}  \right) \pd{F_{0e}}{\psi}\\
	&\qquad+ \bav{ \frac{m_e w_\parallel^2 }{\MeanMagB} \pd{ }{\tl}(\tpsi-\psi) } \frac{I(\psi)}{T_e}\frac{d\angvel}{d\psi} F_{0e},
}\end{eqalign}
\end{equation}
where the last line in \eref{zugzug2} has vanished under the bounce average because it is antisymmetric in $\sigma$ (the sign of the parallel velocity).
This equation is a closed equation for $g_{te}$, once the fields and $\zeta$ are given. Note that all quantities must be expressed in the moving coordinates, in order to perform the bounce average.
Again, as with \eref{zetaeq}, all gradients should be interpreted in terms of derivatives with respect to the moving coordinates as should the Poisson bracket (via \eref{poissonSimple}).
\subsection{Passing Electrons and the Flux Surface Average}
\label{Sfluxav}
In the passing region of velocity space, the fast streaming of electrons along perturbed field lines causes a passing electron to sample an entire perturbed flux surface ($\tpsi=\mbox{const.}$ surface).
Thus, the averaging procedure we will need is the average over a perturbed flux surface. For any function $a$, this is defined by
\begin{equation}
\fluctfav{a(\bm{r},\energy[e],\magmom[e],\gyr,\sigma)} = \lim_{\Delta \psi \rightarrow 0} \left\{ \int\limits_\Delta d^3{\bm{r}} a(\bm{r},\energy[e],\magmom[e],\gyr,\sigma) \right/ \left. 
\int\limits_\Delta d^3{\bm{r}}\right\},
\label{favDef}
\end{equation}
where the region of integration $\Delta$ is the volume between the surface labelled by $\tpsi$ and that labelled by $\tpsi + \Delta \psi$, and the integral is taken at constant $\energy[e]$, $\magmom[e]$ and $\gyr$.
In \ref{apFlux}, we prove that
\begin{equation}
\fluctfav{\Bfield\dg h_e^{(1)}} = 0,
\end{equation}
that the flux surface average commutes with the time derivative following the magnetic field, that
\begin{equation}
\fluctfav{\frac{\MeanMagB}{|w_\parallel|}\vdrift[e]\dg\psi} = 0,
\end{equation}
and that, for any function $a(\bm{r})$,
\begin{equation}
\fluctfav{\frac{\MeanMagB}{|w_\parallel|}\left( \bhat\times\nabla a \right) \cdot \nabla\tpsi} = 0.
\label{favpb0}
\end{equation}
We also prove that, for any fluctuating function $\lambda$, 
\begin{equation}
\fluctfav{\frac{\MeanMagB}{|w_\parallel|} \vdrift[e]\dg\lambda} = 0.
\label{favVDlambd}
\end{equation}

Taking \eref{zugzug2} in the passing region of velocity space, multiplying by $\MeanMagB / |w_\parallel|$, and flux-surface averaging (to eliminate $w_\parallel \bhat\dg h_e^{(1)}$), we obtain:
\begin{equation}
\fl
\label{passtmp}
\begin{eqalign}{
	\fluctfav{\frac{\MeanMagB}{|w_\parallel|}}\ddttwiddles[g_{pe}] - \fluctfav{\frac{\MeanMagB}{|w_\parallel|} \gyroR{\collop[g_{pe}]}}\\
		\quad= -\frac{F_{0e}}{T_e}\ddttwiddles \fluctfav{\frac{\MeanMagB}{|w_\parallel|}\left(e\zeta-\magmom[e]\delBp\right)} 
- \fluctfav{ {m_e |w_\parallel| } \pd{ }{\tl}(\tpsi - \psi)}  \frac{I(\psi)}{T_e}\frac{d\angvel}{d\psi} F_{0e}\\
\qquad+ \sigma \fluctfav{\MeanMagB F_{0e}\left( \frac{e}{c{T_e}} \pd{\delApN{1}}{t} + \frac{1}{\MeanMagB}\pb{ \delApN{1}}{\ln F_{0s}}\right)},
}\end{eqalign}
\end{equation}
where the final term in \eref{zugzug2} vanishes because in the passing region $g_e$ is only a function of $\tpsi$ and we can use \eref{favpb0}. We still have to eliminate the final line of \eref{passtmp}. We note that we only require the part of $g_{pe}$ that is even in $\sigma$ -- we require its density in the quasineutrality condition \eref{fluct-qn} and its pressure in perpendicular Amp\`ere's law \eref{fluct-bpar}. Thus, we average \eref{passtmp} over the two signs of $\sigma$ and henceforth $g_{pe}$ denotes only the even part. Hence we obtain
\begin{equation}
\label{pass0}
\fl
\begin{eqalign}{
	\fluctfav{\frac{\MeanMagB}{|w_\parallel|}}\ddttwiddles[g_{pe}] - \fluctfav{\frac{\MeanMagB}{|w_\parallel|} \gyroR{\collop[g_{pe}]}}\\
		\quad= -\frac{F_{0e}}{T_e}\ddttwiddles \fluctfav{\frac{\MeanMagB}{|w_\parallel|}\left(e\zeta-\magmom[e]\delBp\right)} 
- \fluctfav{ {m_e |w_\parallel| } \pd{ }{\tl}(\tpsi - \psi)}  \frac{I(\psi)}{T_e}\frac{d\angvel}{d\psi} F_{0e}.
}\end{eqalign}
\end{equation}
This completes our solution for $g_e$, as \eref{pass0} is a closed equation for the part of $g_{pe}$ that is even in $\sigma$.  It is useful to note that only the zonal (i.e., $\talpha$-independent) parts of $\zeta$ and $\delBp$ will give a significant drive for $g_{pe}$ -- the $\talpha$-dependent parts will tend to cancel under the flux-surface average.
\subsection{Field Equations}
\label{Selectronequations}
In this section, we finally close our system of equations by writing the field equations in terms of our solution for $\delta f_e$ and ion quantities.

We find $\gkupot$ by using the solution \eref{deltafSoln} for $\delta f_e$ in the quasineutrality condition \eref{fluct-qn}:
\begin{equation}
\fl
\begin{eqalign}{
\frac{1}{n_e} \sum_{s=i} \frac{Z_s^2 e n_s \gkupot}{T_s} = \frac{e\zeta}{T_e} + \left(\tpsi-\psi\right)\frac{d\ln \NotN[e]}{d\psi} + \frac{1}{n_e}\wint g_e
	+\sum_{s=i} \frac{Z_s}{n_e} \wint \gyror{h_s}.
}\end{eqalign}
\label{final-qn}
\end{equation}
Similarly, we find $\delBp$ by using \eref{deltafSoln} in \eref{fluct-bpar} to find
\begin{equation}
\fl\begin{eqalign}{
-\nabla^2_\perp \frac{\delBp\MeanMagB}{4\pi} &= \sum_{s=i}\nabla_\perp\nabla_\perp : \int m_s\gyror{ \bm{w}_\perp \bm{w}_\perp h_s } + n_e\nabla_\perp^2 \left(\zeta-\gkupot\right) \\
	&\quad+ n_e \nabla_\perp^2\left(\tpsi-\psi\right) \left[T_e \frac{d\ln\NotN[e]}{d\psi} + \left( T_e - e \pot_0 \right) \frac{d\ln T_e}{d\psi}\right]\\
	&\quad+ \nabla^2_\perp \wint \frac{m_e w_\perp^2}{2} g_e.
}\end{eqalign}
\label{final-bpar}
\end{equation}
These two equations determine $\gkupot$ and $\delBp$.

It is instructive to note that \eref{final-qn} and \eref{final-bpar} are {\textit{not}} naturally solved in the moving coordinate system. It is because of this coupling to the fixed coordinates (and the ion physics) that we require the mapping between the coordinate systems (from the solution of \eref{psiEq}--\eref{lconst}).

\section{Summary of the Reduced Electron Model}
\label{Turb_evol}

The set of electron equations we have derived is complicated: the electron time-scale has been removed at the expense 
of introducing more dependent variables and more equations, some integro-differential.   In these equations, we have the nine unknowns $\tpsi$, $\talpha$, $\upar$, $\zeta$, $g_{te}$, $g_{pe}$, $h_i$, $\gkupot$, and $\delBp$. We have provided a complete set of nine equations to determine these quantities.  These are:
\begin{itemize}
\item \eref{psiEq} and \eref{alphaEq} to evolve $\tpsi$ and $\talpha$ which define the moving coordinate system that moves with velocity $\veff$ given by \eref{uxi},
\item \eref{uparEq} to determine $\upar$,
\item and the coupled set \eref{zetaeq}, \eref{trap0}, \eref{pass0}, and \eref{gke} to evolve $\zeta$, $g_{te}$, $g_{pe}$, and $h_i$ with the constituent relations \eref{final-qn} and \eref{final-bpar} determining $\gkupot$ and $\delBp$.
\end{itemize}
Together, these equations comprise a rigorously-derived model for ion-scale turbulence, accurate up to corrections of order $\delta$. The only assumption we have had to make, beyond our orderings, is that 
any stochastic component of the magnetic field (if present) satisfies $\delta B_{\mathrm{stoch}}/B \ll \gkeps \delta$.

\section{The Collisional Limit}
\label{Scollisional}
The equations for $\delta f_e$ in \Sref{SZerothOrder} and \Sref{Selectronequations} are derived under the assumption that
$\nu_{ee} \sim \omega \sim k_\parallel \vth[i]$. In many cases of interest, the collision frequency is in fact much larger than this: $\nu_{ee} \gg \omega$.
We can take this limit as a subsidiary expansion, subsidiary to our expansion in the mass ratio.\footnote{Strictly speaking, to treat this as a subsidiary expansion we also require
that $\nu_{ee} \ll k_\parallel \vth[e]$, so that it does not change the form of \eref{gk0thTmp} and \eref{gketmp}. However, if $\nu_{ee}$ is large enough that $\nu_{ee}\sim k_\parallel \vth[e]$, then
the homogenous equation for $g_e$ is $w_\parallel\bhat\dg g_e = \collop[g_e]$. The only solution of this equation is a perturbed Maxwellian with perturbed densities and temperatures that are functions of $\tpsi$ only. Thus, the results of this section still hold.}

Formally, we introduce the parameter $\nustar = \nu_{ee} q R / \vth[e]$, where $q(\psi)$ is the safety factor (see \Sref{P1-magevolve} of \cite{flowtome1}) and $R$ the cylindrical radial coordinate (the major radius of the tokamak). As our initial ordering was $\nu_{ee} \sim \omega$, our ordering for $\nustar$ is $\nustar \sim \massratio$; for our collisional ordering we let $\nustar \gg \massratio$ and expand in
\begin{equation}
\frac{\massratio}{\nustar} \ll 1.
\end{equation}
The results of this expansion will be presented in the following sections.

\subsection{Maxwellian Electrons}
In our collisional expansion, $\collop[g_e]$ is the dominant term in \eref{zugzug2}. 
Thus, up to corrections that are small in $\omega/\nu_{ee} \ll 1$, $g_e$ is given by
	\begin{equation}
	\collop[g_e] = 0.
	\end{equation}
Hence, $g_e$ is a perturbed Maxwellian to leading order:
\begin{equation}
g_e = \left\{\frac{\delta n_H}{\NotN[e]} + \left( \frac{\energy[e]}{T_e} - \frac{3}{2} \right) \frac{\delta T_e}{T_e} \right\}F_{0e} +\Or\left(\frac{\massratio}{\nustar}\gkeps f_e\right).
\label{gIsMax}
\end{equation}
where $\delta n_H$ and $\delta T_e$ are arbitrary functions of $\bm{r}$ -- $\delta n_H$ will shortly be shown to be the homogenous (i.e., a function only of $\tpsi$) part of the electron density perturbation and $\delta T_e$ the perturbed electron temperature.

Now, we know from \Sref{Smax} that $g_e$ is a solution of \eref{homog}.
Substituting this into \eref{homog}, we see that $\delta n_H / \NotN[e]$ and $\delta T_e / T_e$ are functions of $\tpsi$ only (the derivatives in \eref{homog} being taken at constant $\energy[e]$). 
Thus, in the this limit, collisions trap and de-trap electrons rapidly enough that there is no longer a distinction between passing and trapped electrons ($\talpha$-dependence being the signature of kinetic trapped electrons)
-- all electrons sample the entire flux surface on the fluctuation timescale.

Let us now further simplify our solution for $g_e$. Substituting \eref{gIsMax} into \eref{deltafSoln}, we obtain
\begin{equation}
\fl
\delta f_e = \left(\frac{e \zeta}{T_e} + \frac{\delta n_H}{\NotN[e]}\right) F_{0e} + \left( \tpsi - \psi \right) \pd{ F_{0e}}{\psi} + \left( \frac{\energy[e]}{T_e} - \frac{3}{2} \right) \frac{\delta T_e(\tpsi)}{T_e} F_{0e} + \Or\left( \gkeps \massratio F_{0e} \right).
\label{tmptmpBlah}
\end{equation}
From this expression, we observe that we were justified in our notation for $\delta T_e$ -- it really is the perturbed electron temperature relative to the moving surfaces.
We also note that $\zeta$ and $\delta n_H$ both contribute to the perturbed density. As we have the freedom to add an arbitrary function of $\tpsi$ to the definition of $\zeta$ -- it does not change the velocity of the flux surface (see the end of \Sref{Sflux}), we redefine $\zeta$ to be $\zeta - T_e \delta n_H / e \NotN[e]$ and so eliminate $\delta n_H$ entirely. Thus, our solution for $\delta f_e$ becomes
\begin{equation}
\fl
\delta f_e = \frac{e \zeta}{T_e} F_{0e} + \left( \tpsi - \psi \right) \pd{ F_{0e}}{\psi} + \left( \frac{\energy[e]}{T_e} - \frac{3}{2} \right) \frac{\delta T_e(\tpsi)}{T_e} F_{0e} + \Or\left( \frac{\massratio}{\nustar} \gkeps F_{0e} \right),
\label{tmpBlah}
\end{equation}
where the entire density perturbation is contained in the Boltzmann-like first term.

\subsection{Fluctuating Continuity Equation}
We now derive an equation for $\zeta$ in the collisional limit. 

Using the solution \eref{gIsMax} for $g_e$ in \eref{zetaeq}, we obtain, to lowest order in $\massratio / \nustar$,
\begin{equation}
\label{bibble}
\fl
\begin{eqalign}{
&\ddttwiddles \left( \frac{e\zeta}{T_e} - \frac{e\pot_0}{T_e}\frac{\delta T_e}{T_e} -\frac{\delBp}{\MeanMagB} \right) n_{e} + \MeanMagB\pd{}{\tl}\frac{ \upar}{\MeanMagB}
- \frac{cT_e}{e\MeanMagB}\pb{\frac{\delBp}{\MeanMagB}}{\frac{\delta T_e}{T_e}}n_e\left(1-\frac{e\pot_0}{T_e}\right)\\
&\quad=-\MaxVDrift\dg \left(\tpsi-\psi\right) \left(\frac{d\ln \NotN[e]}{d\psi} + \frac{d\ln T_e}{d\psi}\right)n_{e} - \bm{V}_\pot\dg\left(\tpsi-\psi\right)\frac{d\ln\NotN[e]}{d\psi}n_e  \\
&\qquad+ \frac{ec\pot_0}{\MeanMagB T_e}\pb{\zeta}{\frac{\delta T_e}{T_e}}- \MaxVDrift\dg\frac{\delta T_e}{T_e} n_e + \left(\widehat{\bm{V}}_D + \bm{V}_\pot\right) \dg\left(\gkupot-\zeta\right) \frac{e n_{e}}{T_e} \\
&\qquad+c \pd{}{\talpha} \left(\zeta - \delBp \frac{T_e}{e\MeanMagB} \right) \pd{\ln\NotN[e]}{\psi} n_e
- c \pd{\delBp}{\talpha} \frac{n_e T_e}{e\MeanMagB} \pd{\ln T_e}{\psi} -\frac{In_e}{\MeanMagB}\frac{d\angvel}{d\psi} \pd{ }{\tl} (\tpsi - \psi),
}\end{eqalign}
\end{equation}
where we have used the fact that, as we have absorbed the density perturbation into $\zeta$, $\wint g_e = \Or( (\massratio/\nustar)\gkeps n_e )$. 
As in \Sref{fluctcont}, this is just the fluctuating continuity equation for electrons.

\subsection{Fluctuating Energy Equation}
To close our system of collisional equations, we now need the fluctuating energy equation from which we will determine $\delta T_e$.
Thus, we multiply \eref{zugzug2} by $m_e w^2 / (2 T_e) - 3 / 2$ and integrate over all velocities to find
\begin{equation}
\fl
\label{bibble2}
\begin{eqalign}{
&\ddttwiddles\left(\frac{3}{2}\frac{\delta T_e}{T_e} - \frac{\delBp}{B}\right) n_e+ \dv\left(\bhat q_{\parallel e}\right)
	+\frac{c}{\MeanMagB}\pb{\zeta - \frac{T_e}{e}\frac{\delBp}{\MeanMagB}}{\frac{\delta T_e}{T_e}}n_e \\
&\quad = - \frac{1}{2}\left(7\widehat{\bm{V}}_D+3\bm{V}_\pot\right)\dg \left( \tpsi-\psi \right) \pd{}{\tpsi}\left( \frac{\delta T_e}{T_e} + \ln T_e \right) n_e \\
	&\qquad- \widehat{\bm{V}}_D\dg\left( \tpsi -\psi \right)\pd{\ln\NotN[e]}{\tpsi} n_e  + \widehat{\bm{V}}_D \dg\left(\gkupot-\zeta\right) \frac{e n_{e}}{T_e}\\
&\qquad- \pd{\delBp}{\talpha}  \frac{cT_e n_e}{e\MeanMagB} \left( \pd{\ln\NotN[e]}{\psi} +\pd{\ln T_e}{\psi}\right) -\frac{In_e}{\MeanMagB}\frac{d\angvel}{d\psi}\pd{ }{\tl} (\tpsi-\psi),
}\end{eqalign}
\end{equation}
where $q_{\parallel e} = \wint \left( \infrac{m_e w^2}{2T_e} - \infrac{3}{2}\right) h_e^{(1)}$. 
As we do not know $q_{\parallel e}$, we wish to eliminate it from this equation.
Therefore, we average \eref{bibble2} over the perturbed flux surfaces and obtain
\begin{equation}
\fl
\label{dTH}
\begin{eqalign}{
 &\pd{}{t} \left(\frac{\delta T_e}{T_e}\fluctfav{n_e} + \fluctfav{\frac{2\delBp}{3B}n_e}\right) + \fluctfav{\bm{V}_\pot\dg\left(\frac{\delta T_e}{T_e} n_e\right)}= 
 \\
&\quad-\fluctfav{\left(\frac{7}{3}\widehat{\bm{V}}_D+\bm{V}_\pot\right)\dg (\tpsi-\psi) \pd{\ln T_e}{\tpsi} n_e}
-\frac{2}{3}\fluctfav{\widehat{\bm{V}}_D\dg(\tpsi-\psi)\pd{\ln\NotN[e]}{\tpsi} n_e} \\
 &\quad- \frac{2}{3}\fluctfav{ \frac{n_e}{\MeanMagB}\pd{ }{\tl}(\tpsi-\psi)} I(\psi) \frac{d\angvel}{d\psi},
}\end{eqalign}
\end{equation}
where we have used \eref{pbFav} to eliminate some terms.

Let us examine the terms in \eref{dTH}. Other than the last term on the right-hand side, which is due to the viscous heating of electrons, these terms seem fairly opaque. However, they all vanish if we apply the low-Mach-number ordering from \Sref{P1-SLowMachLimit} of \cite{flowtome1} -- we can move $n_e$ outside the flux-surface averages and then use the properties of the flux-surface average and the drift velocities to eliminate the terms in question. Thus, we conclude that these terms, which disappear if the Mach number is small, are to do with the exchange of potential energy of the electrons with thermal energy.

\Eref{dTH} completes our solution for $\delta f_e$. In the next section, we close the system by working out the electron contributions to the field equations in the collisional limit.

\subsection{Field Equations}
We now use our solution for $\delta f_e$ to simplify the field equations. Substituting the solution \eref{tmpBlah} into \eref{final-qn} gives the following equation for $\gkupot$:
\begin{equation}
\fl
\begin{eqalign}{
\frac{1}{n_e} \sum_{s=i} \frac{Z_s^2 e n_s \gkupot}{T_s} =& \frac{e\zeta}{T_e} + \left(\tpsi-\psi\right)\frac{d\ln \NotN[e]}{d\psi}- \frac{e\pot_0}{T_e}\frac{\delta T_e}{T_e}  +\sum_{s=i} \frac{Z_s}{n_e} \wint \gyror{h_s}.
}\end{eqalign}
\label{coll-qn}
\end{equation}
Similarly, by substituting \eref{tmpBlah} into \eref{final-bpar}, we obtain an equation for $\delBp$ in the collisional limit:
\begin{equation}
\fl
\begin{eqalign}{
-\nabla^2_\perp \frac{\delBp\MeanMagB}{4\pi} &= \sum_{s=i}\nabla_\perp\nabla_\perp : \int m_s\gyror{ \bm{w}_\perp \bm{w}_\perp h_s } + n_e\nabla_\perp^2 \left(\zeta-\gkupot\right)\\
	&\quad+ n_e \nabla_\perp^2\left(\tpsi-\psi\right) \left[T_e \frac{d\ln\NotN[e]}{d\psi} + \left( T_e - e \pot_0 \right) \frac{d\ln T_e}{d\psi}\right]\\
	&\quad+ \left(1-\frac{e\pot_0}{T_e}\right)n_e \nabla^2_\perp \delta T_e.
}\end{eqalign}
\label{coll-bpar}
\end{equation}

The field equations \eref{coll-qn} and \eref{coll-bpar} close the collisional electron model, which we now summarise.

\subsection{The Collisional Electron Model}
The collisional electron model is a simplified version of the model given in \Sref{Turb_evol}. We have replaced the distribution functions $g_{pe}$ and $g_{te}$ by a single perturbed electron density and temperature. 
The complete set of equations describing this model is:
\begin{itemize}
\item \eref{psiEq} and \eref{alphaEq} to evolve $\tpsi$ and $\talpha$ which define the moving coordinate system,
\item \eref{uparEq} to determine $\upar$,
\item $\zeta$ is given by \eref{bibble},
\item $\delta T_e$ is given by \eref{bibble2},
\item \eref{coll-qn} and \eref{coll-bpar} determine the fields $\gkupot$ and $\delBp$ respectively,
\item and the ion dynamics ($h_i$) are given by \eref{gke}.
\end{itemize}
\section{The Transport Timescale}
\label{STransp}
The procedure of \Sref{Turb_evol} is sufficient to calculate the evolution of our plasma on the fast (turbulent) timescale. Our expansion in $\massratio$ also has consequences for the evolution of $n_e$, $T_e$ and $\angvel$ on the slow (transport) timescale. In \Sref{P1-transport} of \cite{flowtome1}, transport equations for $n_e$ and $T_e$ were derived by considering the transport of particles and energy across $\psi=\mathrm{const.}$ surfaces. It is clear from the results of \Sref{Sflux} and \Sref{SClebsch} that, for electron transport, it is more appropriate to consider $\tpsi=\mathrm{const.}$ surfaces. 

To this end, in \ref{magicalTransp}, we rewrite the electron kinetic equation relative to the moving flux surfaces. We then average this equation over the turbulence and over the fluctuating flux surfaces. Finally, we integrate over velocity space to find transport equations for particles and multiply by $\energy[e]$ and integrate over velocities to obtain the electron heat transport equation. This procedure is entirely analogous to that undertaken in \Sref{P1-transport} of \cite{flowtome1} save for the extra complexity introduced by the rapid fluctuation of the $\tpsi={\mathrm{const.}}$ surfaces. We summarise the results of these calculations in the following sections.
\subsection{Particle Transport}
In \ref{EPTransP}, we derive the electron particle transport equation, correct to lowest order in $\massratio$:
\begin{equation}
\frac{1}{V'}\ddtpsitwiddles V'\fluctfav{n_e} + \frac{1}{V'} \pd{  }{\tpsi} V'\fluctfav{\Gamma_e}  = \fluctfav{\Psource[e]},
\label{dndt}
\end{equation}
where $\Psource[e]$ is defined by \eref{P1-psource} and the particle flux is given by
\begin{equation}
\fluctfav{\Gamma_e} = \fluctfav{\ensav{ c \wint g_{te} \pd{ }{\talpha} \left( \zeta + \frac{w_\perp^2}{2\cycfreq[e]} \frac{\delta B_\parallel}{B}\right)}}.
\end{equation}
Note that whilst we use the same notation for the particle flux $\Gamma_e$ as in \cite{flowtome1}, it is not the same quantity.
In this paper the transport equations are derived for transport across a $\tpsi=\mathrm{const.}$ surface not a $\psi=\mathrm{const.}$ surface.

Several important features of this flux immediatly present themselves. Firstly, only the trapped electrons contribute -- the passing particles play no part in the transport. Secondly, 
there is no ``flutter'' transport (transport due to correlations between $\delta \bm{u}$ and $\delB_\perp$ despite the fact that the contribution to \eref{P1-partflux} of \cite{flowtome1} from $\delAp$ is clearly not zero.
This is, in fact, the primary advantage of writing our transport equations with respect to the moving surfaces, we observe this cancellation analytically rather than having to recover it via a very careful evaluation of \eref{P1-dndt} and \eref{P1-partflux} of \cite{flowtome1}.
Finally, we see that drifts that do cause transport are the drift due to $\zeta$ and the $\nabla B$ drift due to $\delBp$. This supports our interpretation of the drift due to $\zeta$ being the drift of electrons across the moving surfaces.

\subsubsection{Electron Transport in the Collisional Model}
As we proved in \Sref{gIsMax}, in the collisional limit the electrons are trapped and de-trapped fast enough to remove the distinction between trapped and passing particles -- in effect all electrons are passing. This immediatly implies that
if the electrons are collisional, there is {\textit{no}} transport of electrons.
Thus, the electron transport time is $\Or(\massratio^{-1}\left(\infrac{\omega}{\nu_{ee}}\right))$ longer than the ion transport time -- collisions are actually beneficial. Naturally, with sufficiently large collisions, neoclassical and classical electron transport reappear in the fluxes and the confinement worsens again.

\subsection{Electron Heat Transport}
Similarly, in \Sref{eHeat}, we derive the electron heat transport equation, to lowest order in $\massratio$:
\begin{eqnarray}
		\label{dpdt}
\frac{1}{V'}\frac{3}{2}&\ddtpsitwiddles V' \fluctfav{n_e}T_e + \frac{1}{V'} \pd{ }{\tpsi} V' \fluctfav{q_e} = \\\nonumber
		&\qquad\Pturb + \CompHeat + \PotEng + \CollEnergy + \Esource,
\end{eqnarray}
where the collisional energy exchange $\CollEnergy$ and the explicit heat source $\Esource$ are defined by \eref{P1-CollEnergyDef} and \eref{P1-esource}, of \cite{flowtome1}, respectively, the heat flux is given by
\begin{equation}
\fl
		\begin{eqalign}{
\fluctfav{q_e} =& \fluctfav{\ensav{{c}\wint \left( \frac{1}{2} m_e w^2 - e \pot_0 \right) g_{te}\pd{ }{\talpha}\left( \zeta + \frac{w_\perp^2}{2\Omega_e} \frac{\delBp}{B}\right)}},
}\end{eqalign}
\end{equation}
and the heat sources are given by
\begin{equation}
\fl
\label{CompheatMain}
\CompHeat = -T_e \fluctfav{n_e \dv\vpsi},
\end{equation}
\begin{equation}
\fl
\begin{eqalign}{
\PotEng = &\fluctfav{ e\pot_0 \left(\ddttwiddles n_e -\Psource[e]\right) } + \fluctfav{ e\pot_0 n_e \dv\vpsi} \\
			 &\quad+ \fluctfav{ e\pot_0 \ensav{ \left( \frac{e\zeta}{T_e} n_e + \wint g_e  \right) \dv\veff }},
}\end{eqalign}
\label{PotHeatMain}
\end{equation}
and
\begin{equation}
\fl
\begin{eqalign}{
\Pturb = -\fluctfav{\wint \ensav{g_e\ddttwiddles  \left(e\zeta - \magmom[e]\delBp\right)}} + e\fluctfav{n_e\ensav{\zeta\dv\veff}}.
}\end{eqalign}
\label{turbHeat}
\end{equation}
In these expressions for the heating, the divergence of $\veff$ is given by
\begin{equation}
\fl
\dv\veff =\ddttwiddles\frac{\delBp}{\MagBfield} +\left( \bhat\dg\psi \right)  \frac{I(\psi)}{\MeanMagB} \frac{d\omega}{d\psi} - \frac{c}{B}\left( \Meanb\dg\Meanb \right)\cdot \Meanb \times \nabla\left( \zeta-\gkupot \right).
\label{DvVeffMain}
\end{equation}
This expression is correct and closed to the required order in $\gkeps$ and $\massratio$.
Again, it is crucial to note that whilst we use the same notation for $q_e$, $\CompHeat$, $\PotEng$, and $\Pturb$ as in \cite{flowtome1}, these are not the same quantities. 

Let us now interpret the heat flux and the heating terms.
The heat flux parallels the particle flux: there is no flutter transport, only trapped particles contribute to the flux, and they contribute via the $\bm{E}\times\bm{B}$-drift across flux surfaces and the $\nabla B$-drift due to $\delBp$.

Of the heating terms, the first, $\CompHeat$ is just the compressional heating due to the mean motion of the flux surfaces. The second, $\PotEng$, is the change in the thermal energy of the electrons due to the change in their electrostatic potential energy.\footnote{For further discussion of the nature of this potential-energy-exchange term, see \Sref{P1-potengsec} of \cite{flowtome1}.}
Finally, we have the turbulent heat source $\Pturb$. We interpret the terms in the expression \eref{turbHeat} for $\Pturb$ as follows: the first term is heating due to work done against the parallel electric field (through $\zeta$); the term involving the time-derivative of $\delBp$ is in fact composed of the electron viscous heating, compressional heating due to the fluctuating motion of the $\tpsi=\mathrm{const.}$ surfaces and work done bending the magnetic field. This can be seen by substituting for $\delBp$ from \eref{DvVeffMain} -- the term on the left-hand side of \eref{DvVeffMain} is the compressional heating, the second term on the right is part of the electron viscous heating.
The final term is the potential energy contribution to the compressional heating due to the fluctuating motion of the exact flux surfaces.

\subsubsection{Electron Heating in the Collisional Model}
Again, in the collisional limit, the flux in \eref{dpdt} vanishes. We are just left with local heating terms, of which only the turbulent heating simplifies:
\begin{equation}
\begin{eqalign}{
\Pturb &= \fluctfav{\ensav{n_e\delta T_e\ddttwiddles \frac{\delBp}{\MeanMagB}}} + e\fluctfav{n_e\ensav{\zeta\dv\veff}},
}\end{eqalign}
\end{equation}
where $\dv\veff$ is given by \eref{DvVeffMain} as before. 
Thus, despite the lack of any transport, the electron response to ion-scale turbulence can still dissipate free energy.
All other terms in \eref{dpdt} remain unchanged.

\subsection{Momentum Transport}
As the momentum transport involves all species, not merely electrons, we do not write a new equation for it in the moving frame.
The transport equation for toroidal angular momentum is (Equation \eref{P1-domegadt} of \cite{flowtome1}) 
\begin{equation}
\begin{eqalign}{
\frac{1}{V'}\ddtpsi&{{V'\inertia \angvel(\psi)}} + \frac{1}{V'}\pd{}{\psi} V'\fav{\TotMomFlux}
= \fav{\Msource},
}\end{eqalign}
\label{domegadt}
\end{equation}
where the moment of inertia of the plasma is now given by
\begin{equation}
\inertia = \sum_{s=i} m_s \fav{n_s R^2},
\end{equation}
and the total momentum flux is 
\begin{equation}
\begin{eqalign}{
\fav{\TotMomFlux} &= \fav{\MomentumFlux[e]} + \sum_{s=i} \left( \fav{\MomentumFlux[s]} + m_s\angvel(\psi)\fav{R^2\ParticleFlux[s]}\right) \\
	&\quad- \sum_{s=i}\frac{Z_s e}{c}\fav{\wint \ensav{\gyror{h_s \bm{w}\dg\psi}\delA}\cdot R^2\nabla\tor} \\
	&\quad- \frac{1}{4\pi}\fav{\nabla\psi \cdot \ensav{\delB\delB}\cdot R^2\nabla\tor},
}\end{eqalign}
\label{totMflux}
\end{equation}
where the sums over $s=i$ are taken over all ion species and fluxes $\MomentumFlux[s]$ and $\ParticleFlux[s]$ for the ions are given by \eref{P1-momflux} and \eref{P1-partflux} of \cite{flowtome1}, respectively. The electron momentum flux is now (see \ref{mfluxap})
\begin{equation}
\fav{\MomentumFlux[e]} = \fav{ I(\psi)\ensav{\wint \left( m_e w_\parallel^2 - \frac{1}{2} m_e w_\perp^2\right) g_{e} \pd{ }{\tl}(\tpsi-\psi) }}.
\label{eMomFlux}
\end{equation}

The three evolution equations \eref{dndt}, \eref{dpdt} and \eref{domegadt} close our system on the transport timescale. The fluxes are written entirely in terms of fluctuating quantities that our reduced electron model provides (note that nowhere do we require the part of $g_{pe}$ that is odd in $\sigma$). 
\subsubsection{The Collisional Limit}
In the collisional limit, $g_e$ is Maxwellian, and so isotropic in velocity space -- see \eref{gIsMax}.
Therefore, the velocity integral in \eref{eMomFlux} vanishes and so there is no angular momentum transport due to the electrons.

\section{Summary} 
\label{Sconclusions}
In this section we summarise our work, first by comparing it in detail to previous models of the electron response, then by briefly collecting our results and outlining future applications of this work.
\subsection{Comparison to Previous Models}
\label{ap-comp}

Many previous works have introduced physically-motivated model responses for the electrons. Some are based on the linear electron response~\cite{linelectrons01} and others use fluid moments of the electron distribution function to try and capture the passing particle response~\cite{chenparker2001,snyder:3199,cohen:251}. Our model improves upon this by being a rigorous consequence of the electron gyrokinetic equation, clearly demonstrating how the electron response fits into the full turbulent transport framework. Also, by retaining the bounce- and flux-surface-averaged kinetics of the electrons, we retain the distinction between the passing and trapped populations and retain drift resonances that may be lost in a fluid picture.
We recover some of the results of the fluid models in the collisional limit, where there are no trapped electrons. In particular, the fact that $\delta T_H$ is only a function of $\tpsi$ means that the temperature has equilibrated along perturbed field lines as expected. Even in this limit, the previous models did not allow for a rapid toroidal rotation (comparable to the ion thermal speed).  Also all these models neglect $\delBp$ and only retain $\delAp$ in the perturbed magnetic field.

One of the original and still widely-used models for the electron response is the adiabatic electron response corrected for zonal perturbations~\cite{hammett1993developments}. It can be shown that this is a rigorous limit of our model in the limit of vanishing $\beta$ (the electrostatic limit) and collisional electrons.
For collisionless electrostatic turbulence, similar equations have been derived rigorously in \cite{gang1990nonlinear}. Our model extends this to include fully electromagnetic turbulence and rapid toroidal rotation. The electrostatic model of \cite{gang1990nonlinear} also neither handles general magnetic geometry nor discusses the behaviour of passing electrons.
This limit is useful as an easily-implemented stepping stone between the adiabatic electron model (strictly only valid for quite collisional plasmas) and simulating the full electron gyrokinetic equation (which is either computationally prohibitive or formally invalid depending on whether one chooses to resolve the electron scales or not).

The most complete previous model is that of \cite{hinton:168}. However, this model is derived under two restrictive assumptions. Firstly, the mass ratio is assumed to be much smaller than we assume here ($\massratio\sim \gkeps$ rather than $1 \gg \massratio\gg \gkeps$ which we assume) and secondly $\delBp$ is assumed to be identically zero. 
We note that \eref{deltafSoln} corresponds directly to equation (54) of \cite{hinton:168}, but that the further assumption made in \cite{hinton:168} of (in our notation) $g_{pe} = 0$ is unjustified ($g_{pe} =0$ is not a solution of \eref{pass0}) and hence neglects the electron response that is zonal with respect to the perturbed flux surfaces. In addition, the model in \cite{hinton:168} does not include the effects of sonic toroidal rotation.

Finally, we must discuss how our model compares to the common practice of simulating the full electron gyrokinetic equation at ion scales (so-called ``kinetic electrons''). Firstly, the electron gyrokinetic equation clearly contains electron-gyroscale and electron-bounce-frequency physics so simulating only the ion scales is formally invalid -- deliberately leaving the simulation unresolved in some sense. Secondly, how an unresolved simulation of the electron gyrokinetic equation behaves will be sensitive to the numerical algorithm used -- it is possible to design a numerical algorithm that numerically bounce-averages the kinetic equation in the long-timestep limit, but of the commonly-used gyrokinetic codes only \verb#GS2#~\cite{gs2ref} does this. Indeed, the only way to confirm that simulating ``kinetic electrons'' gives the correct physics is to either compare with full two-scale simulations that resolve the electron dynamics as well as the ion dynamics or to compare with simulations of the electron model presented in this paper. As well-resolved two-scale simulations are generally beyond current computational resources, simulations of our model are the only way to confirm the validity of this practice.

\subsection{Conclusions}
\label{ASConcFinal}
In this paper we have derived equations describing the electron response to ion-scale turbulence. The fast electron timescales have been eliminated from this system.
The scale separation between electron and ion scales both motivates and enables the derivation of such a model. Starting from electron gyrokinetics~\cite{flowtome1}, we systematically expand our equations in the square root of the electron-to-ion mass ratio to obtain our model.
As discussed in the introduction, as studies of turbulence at non-zero $\beta$ become more and more important -- both for conventional and spherical tokamaks -- having an electron response model that handles this rigorously will be necessary. We have derived full ($\beta \sim 1$) collisional and arbitrary-collisionality models -- summarised in \Sref{Scollisional} and \Sref{Turb_evol} respectively.

The procedure by which we constructed this model, a rigorous expansions in $\sqrt{m_e/m_i}\ll 1$, allowed us to clearly delineate which effects are due to electron-scale physics and which to ion-scale. Most importantly, 
the elimination of the electron scales has a profound impact upon the structure of the fluctuating magnetic field. We prove in \Sref{Sflux}, that purely ion-scale turbulence conserves the total magnetic flux, and, moreover that
a velocity field can be found into which the magnetic field is frozen. This generalisation of the usual MHD frozen-in theorem demonstrates that it is {\it only} electron-scale effects which can change the magnetic topology. The inclusion of $\beta \sim 1$, trapped particles, sonic rotation, and arbitrary axisymmetric geometry do {\it not} change this fundamental fact. 

Similarly, whilst deriving transport equations across the exact flux surfaces, we have obtained another general property of ion-scale turbulence: passing electrons do not contribute to transport fluxes (see \Sref{STransp}). This result could prove useful in designing magnetic geometries that have beneficial trapped-particle properties, safe in knowledge that the passing particles never contribute to the transport.

In addition to these general results, two other ancillary results have been proved in the course of this work.
Firstly, in \ref{resistive}, we proved that the topology of the mean magnetic field (i.e.,  the $q(\psi)$ profile) evolves on the resistive rather than the transport timescale. Importantly, this result holds independently of the nature of the turbulence. 
Secondly, in \ref{electrostaticLimit}, we prove that the commonly-used adiabatic electron model is, in fact, the correct model for collisional ($\nu \gg \omega$) and electrostatic ($\beta \ll 1$) electrons.

Throughout our derivation, we have simply assumed that the magnetic stochasticity produced by electron-scale fluctuations is sufficiently small that we can neglect it. This assumption must be investigated in detail and will be explored in future work.
Similarly, the potential effects of small resonant regions around rational surfaces (which would give a finite width to each $\tpsi$-constant surface) should be examined closely. Indeed, a natural extension of this work would be to include the reaction of a small amount
of purely electron-scale turbulence upon the ion scales.
As always, the final arbiter of the usefulness of these models will be their numerical implementation and application to simulating moderate- and high-$\beta$ ion-scale turbulence, in e.g. nonlinear studies of finite-$\beta$ ITG turbulence or kinetic ballooning mode turbulence.

\ack
Part of this work was carried out at meetings supported by the Wolfgang Pauli Institute and the Leverhulme Trust Academic Network for Magnetized Plasma Turbulence. During this work, IGA was supported by a CASE EPSRC studentship jointly with the EURATOM/CCFE Fusion Association and by a Junior Research Fellowship at Merton College, Oxford.
This work was carried out within the framework of the European Fusion Development Agreement, the views and opinions expressed herein do not necessarily reflect those of the European Commission.

\appendix

\section{Derivation of \eref{gkpotExp}}
\label{apGke}
The gyrokinetic potential $\gkpot$ is
\begin{equation}
\gkpot(\bm{r}) = \gkupot(\bm{r}) - \frac{w_\parallel}{c} \delta {A}_\parallel(\bm{r}) - \frac{1}{c} \bm{w}_\perp \cdot \delA_\perp(\bm{r}).
\end{equation}
Using \eref{Rdef}, we perform a Taylor expansion about $\bm{r} = \bm{R}_e$ to find
\begin{equation}
\fl\begin{eqalign}{
\gkpot(\bm{r}) = &\gkupot(\bm{R}_e) - \frac{w_\parallel}{c} \delAp(\bm{R}_e) - \frac{w_\parallel}{c} \frac{\Meanb\times\bm{w}}{\cycfreq[e]} \dgR[e]{} \delAp(\bm{R}_e)\\
		&- \frac{1}{c} \bm{w}_\perp \cdot\delA_\perp (\bm{R}_e) - \frac{1}{c} \left[ \frac{\Meanb\times\bm{w}}{\cycfreq[e]} \dgR[e]{}\delA_\perp(\bm{R}_e)  \right]\cdot\bm{w}_\perp + \Or\left(\massratio^2\gkpot \right),
}\end{eqalign}
\end{equation}
where we have used the fact that $\bm{w}\cdot\delA / c \gg \gkupot$. Gyroaveraging this expression, the third and fourth terms vanish to leave
\begin{equation}
\gyroR{\gkpot} = \gkupot(\bm{R}_e) - \frac{w_\parallel}{c} \delAp(\bm{R}_e) - \frac{m_e w_\perp^2}{2e }\frac{\delBp(\bm{R}_e)}{\MeanMagB} + \Or\left(\massratio^2\gkpot \right),
\label{gkpot}
\end{equation}
where we have used $\gyroR{\bm{w}_\perp\bm{w}_\perp}= w_\perp^2 / 2(\idmat - \Meanb\,\Meanb)$, with $\idmat$ the unit dyadic, and $\delBp \Meanb = \curl \delA_\perp$.
\section{Derivation of \eref{simpleepar}}
\label{deriv}
Expressing $\Efield$ in terms of potentials, we arrive at the following expression for $\Efield\cdot\bhat$:
\begin{equation}
\fl\Efield\cdot\bhat = -\bhat\dg\fpot - \bhat\dg\pot -\bhat \dg\delta\pot
	- \frac{1}{c} \pd{\MeanA}{t}\cdot\Meanb - \frac{1}{c} \pd{\delA}{t}\cdot\bhat + \Or\left(\gkeps^3\frac{\vth[i]}{c} \MeanMagB\right).
\label{tmpEpar}
\end{equation}
Manipulating the first term of this expression, we find that
\begin{equation}
\fl
\bhat \dg \fpot = \frac{1}{c}\angvel(\psi,t)\bhat \dg \psi = -\frac{1}{c} \bhat \cdot\left( \MeanB\times\bm{u} \right),
\end{equation}
where we have used \eref{magfield}, \eref{torrot}, and \eref{omdef}.
Using $\MeanB = \Bfield - \delB$, we obtain
\begin{equation}
\fl
\begin{eqalign}{
\bhat\dg\fpot &= -\frac{1}{c} \bhat \cdot \left[ \left( \Bfield - \delB\right) \times\bm{u}\right] = \frac{1}{c} \bhat \cdot \left[\left( \curl \delA \right)\times\bm{u} \right] \\
		&= \frac{1}{c} \left[ \bm{u} \cdot \left( \nabla\delA \right) \cdot\bhat - \bhat\cdot\left( \nabla\delA \right)\cdot\bm{u} \right] \\
		&= \frac{1}{c} \left[ \bm{u} \dg\left( \delA \cdot\bhat\right) - \bhat \dg \left( \delA \cdot\bm{u}\right) - \left( \bm{u}\dg\bhat - \bhat \dg\bm{u} \right)\cdot\delA \right] \\
		&= \frac{1}{c} \left[ \bm{u}\dg \delAp - \bhat\dg\left(\delA\cdot\bm{u}\right) + \frac{1}{\MeanMagB} \bm{u}\cdot\left( \nabla\delA \right)\cdot\delB_\perp\right] + \Or\left(\gkeps^3\frac{\vth[i]}{c} \MeanMagB\right),
}\end{eqalign}
\label{ick}
\end{equation}
where in the last line we have used \eref{bhatDef}, $\bm{u}\dg\Meanb = \Meanb\dg\bm{u}$ (\eref{P1-udgb} of \cite{flowtome1}), and neglected $\delB\cdot\left(\nabla\bm{u}\right)\cdot\delA$ as it is $\Or(\gkeps^3\binfrac{\vth[i]}{c} \MeanMagB)$.
Substituting \eref{ick} back into \eref{tmpEpar}, we obtain
\begin{equation}
\fl
\begin{eqalign}{
\Efield\cdot\bhat &= - \bhat \dg\pot -\bhat \dg\left( \delta\pot - \frac{1}{c} \delA\cdot\bm{u} \right) - \frac{1}{c} \pd{\MeanA}{t}\cdot\Meanb - \frac{1}{c} \left(\pd{}{t}+ \bm{u}\dg\right)\delAp\\
	&-\frac{1}{cB} \left[\left(\pd{}{t} + \bm{u}\dg\right)\delA\right]\cdot\delB_\perp +  \Or\left(\gkeps^3\frac{\vth[i]}{c} \MeanMagB^2\right).
}\end{eqalign}
\end{equation}
The final term in this expression is $\Or( \gkeps^3 \binfrac{\vth[i]}{c} \MeanMagB)$ and can be neglected. We can then use $\gkupot = \delta\pot - \delA\cdot\bm{u}/c$ 
to finally write the parallel electric field as
\begin{equation}
\fl
\Efield\cdot\bhat = - \bhat\dg \pot - \bhat\dg \gkupot - \frac{1}{c} \pd{\MeanA}{t}\cdot\Meanb - \frac{1}{c} \left(\pd{}{t}+ \bm{u}\dg\right)\delAp
	+ \Or\left(\gkeps^3\frac{\vth[i]}{c} \MeanMagB\right).
	\label{apSimpleEpar}
\end{equation}
\section{Derivation of \eref{neoclass} and Conservation of Mean Magnetic Flux}
\label{resistive}
In this Appendix, we first derive \eref{neoclass}, as required for the derivation in \Sref{Sflux}. Then we prove that there is a field-line-preserving velocity for the mean magnetic field, irrespective of the fluctuations. Finally, we demonstrate that this implies that there are two timescales for the evolution of the mean magnetic field: the transport timescale $\tau_E$ on which the flux $\psi(R,z,t)$ evolves and the resistive timescale $\tau_\eta \sim \delta^{-1} \tau_E$ on which the safety factor $q(\psi,t)$ evolves.

We now proceed to derive \eref{neoclass} in a way that exactly parallels the derivation of \eref{deltan} from \eref{gke}. 
The neoclassical drift-kinetic equation that determines $F_{1e}$ is (\eref{P1-neoclass} of \cite{flowtome1})
\begin{equation}
\label{e-ncdke}
\fl
w_\parallel \Meanb\cdot\ddR[e]{F_{1e}} - \gyroR{\collop[F_{1e}]} = - \vdrift[e]\cdot\ddR[e]{F_{0e}} +\frac{e}{T_e c} w_\parallel F_{0e} \pd{\MeanA}{t} \cdot \Meanb + \gyroR{\collop[F_{0e}]}.
\end{equation}
The first term on the left-hand side and the second term on the right-hand side are $\Or(\gkeps^2 \massratio^{-1} \Omega_i F_{0e})$, all other terms are $\Or(\gkeps^2\Omega_i F_{0e})$.
Thus, to lowest order in $\massratio$, we obtain
\begin{eqnarray}
	w_\parallel \Meanb \dg F_{1e} = \frac{e}{T_e c} w_\parallel F_{0e}\pd{\MeanA}{t}\cdot\Meanb,
	\label{low-encdke}
\end{eqnarray}
where we have dropped the distinction between $\bm{R}_e$ and $\bm{r}$ as it is small in $\gkeps$.
This equation is completely similar to the first line of \eref{gketmp} and so we seek solutions in an identical fashion. Let (c.f. \eref{ansatz})
\begin{equation}
\label{F1esol}
F_{1e} = \lambda(\bm{r},\energy[e],\magmom[e],\sigma) F_{0e} + G_e(\bm{r},\energy[e],\magmom[e]) +  \Or\left(\gkeps \delta F_{0e} \right),
\end{equation}
where $G_e$ satisfies
\begin{equation}
w_\parallel \Meanb\dg G_e = 0.
\label{GeEq}
\end{equation}
As $G_e$ is axisymmetric and so cannot depend on $\alpha$, we immediatly solve \eref{GeEq} to find $G_e = G_e(\psi,\energy[e],\magmom[e])$.
Substituting \eref{F1esol} back into \eref{low-encdke} and dividing by $F_{0e}$, we obtain
\begin{equation}
	w_\parallel \Meanb \dg\lambda  = \frac{e}{T_e c} w_\parallel \pd{\MeanA}{t}\cdot\Meanb.
	\label{foo}
\end{equation}
By the same logic as in \Sref{Smax}, $\lambda$ is a function of $\bm{r}$ only (this time there is no gradient term to drive a temperature perturbation). Moreover, by absorbing any density moment of $G_e$ into $\lambda$, we can assume that $\wint G_e = 0$ and so
\begin{equation}
\lambda = \frac{n_{1e}}{n_{e}}
\label{lambdaApC}
\end{equation}
where $n_{1e}$ is the first-order correction to the mean electron density: $n_{1e} = \wint F_{1e}$. 
Using this result in \eref{foo}, we finally obtain
\begin{equation}
\Meanb\dg \frac{n_{1e}}{n_e} = \frac{e}{T_e c}\pd{\MeanA}{t} \cdot\Meanb.
\label{meanConsFlux}
\end{equation}
This is precisely \eref{neoclass} as required in \Sref{Sflux}.

We now go to next order in $\massratio$ to obtain an equation for $G_e$. To this order, we have
\begin{equation}
w_\parallel \nabla_\parallel F_{1e}^{(1)} - \collop[G_e] = - \vdrift[e]\dg\psi \pd{F_{0e}}{\psi},
\end{equation}
where $F_{1e}^{(1)}$ is the next correction to $F_{1e}$. Using \eref{P1-VDpsi} of \cite{flowtome1}, we have
\begin{equation}
w_\parallel \MeanB\dg \left( F_{1e}^{(1)} + \frac{Iw_\parallel}{\cycfreq[e]}\pd{F_{0e}}{\psi}\right) = \collop[G_e].
\end{equation}
Multiplying this equation by $G_e/F_{0e}$, integrating over all velocities, and flux-surface averaging, we obtain
\begin{equation}
\fav{\wint \frac{G_e}{F_{0e}}\collop[G_e]} = 0.
\end{equation}
The only solution to this equation is for $G_e$ to be a Maxwellian with a density and temperature that are functions of $\psi$ only. We can absorb any such solution 
into $F_{0e}$. Thus, $G_e$ vanishes and the solution for $F_{1e}$ is merely
\begin{equation}
F_{1e} = \frac{n_{1e}}{n_e} F_{0e} + \Or(\gkeps\massratio F_e)
\label{F1SOL}
\end{equation}

Substituting \eref{meanConsFlux} into the expression for $\MeanE\cdot\MeanB$ in terms of the potentials $\pot$ and $\MeanA$, we have
\begin{equation}
\fl
\label{EdotB}
\MeanE\cdot\MeanB = -\MeanB\cdot\left(\nabla\pot + \frac{1}{c}\pd{\MeanA}{t}\right) = -\MeanB\dg\left( \pot + \frac{T_e n_{1e}}{e n_e}\right) + \Or\left(\gkeps^2\massratio\frac{\vth[i]}{c} B^2\right).
\end{equation}
This form of the mean electric field is entirely analogous to \eref{simpleXi}. Thus, by the arguments of the second paragraph of \Sref{Sflux}, the mean magnetic flux is conserved and there is a field-line-preserving velocity for the mean magnetic field. 

Finally, we look at the implications of this for the evolution of the mean magnetic field given by \eref{magfield}.
This field is determined by the two functions $I(\psi,t)$ and $\psi(R,z,t)$, which {\it a priori} we expect to evolve on the transport timescale $\tau_E$.
In \cite{flowtome1}, it is shown that instead of $I$, we can use the safety factor $q(\psi)$ given by
\begin{equation}
q(\psi) = \frac{1}{4\pi^2} V' I(\psi) \fav{R^{-2}},
	\label{C9}
\end{equation}
to characterise the magnetic field. It is this quantity that will evolve slowly in the limit of $\massratio \ll 1$.
The evolution of $q(\psi)$ is given by~\cite{flowtome1}:
\begin{equation}
  \ddtpsi q = \frac{c}{4\pi^2} \pd{}{\psi} V'\fav{\MeanE\cdot\MeanB}.
  \label{dqdt-C}
\end{equation}
Using the expression \eref{EdotB} for $\MeanE\cdot\MeanB$, we see that
\begin{equation}
\fav{\MeanE\cdot\MeanB} = \Or\left(\gkeps^2\massratio\frac{\vth[i]}{c} B^2\right),
\end{equation}
and so  the right hand side of \eref{dqdt-C}  is $\Or(\gkeps^3 \massratio \Omega_i) = \Or(\tau_\eta^{-1})$, and thus $q$ evolves on the resistive timescale, not the transport timescale.
Since $q(\psi)$ is the number of times a field line winds around the vertical symmetry axis for one poloidal transit around the magnetic axis, a topological quantity, it changes only when the magnetic topology changes. However, the shape of the flux surfaces changes on the transport timescale in response to changes in the pressure profile (via the Grad-Shafranov equation, see \Sref{P1-magevolve}). It is therefore clear from \eref{C9} that $I(\psi)$ also changes on the transport timescale.
\section{Representations of \texorpdfstring{$\delB$}{the perturbed magnetic field}}
\label{repDeltaB}
Once we introduce the Clebsch form of $\Bfield$ we have two equivalent expressions for $\delB$: firstly, the usual expression in terms of $\delAp$ and $\delBp$,
\begin{equation}
\delB = \Meanb \times \nabla_\perp \delAp + \delBp \Meanb
\label{deltaBnormal}
\end{equation}
and secondly the expression in terms of $\tpsi$ and $\talpha$,
\begin{equation}
\delB = \nabla\tpsi\times\nabla\talpha - \nabla\psi\times\nabla\alpha.
\label{deltaBtwiddle1}
\end{equation}
Let us investigate this second expression for $\delB$.
Writing the gradients of $\tpsi$ and $\talpha$ in terms of the ($\psi$,$\alpha$,$l$) coordinates we have
\begin{equation}
\delB = \left( \pd{\tpsi}{\psi} \pd{\talpha}{\alpha} - \pd{\tpsi}{\alpha}\pd{\talpha}{\psi} -1\right) \MeanB + \pd{\tpsi}{l} \nabla l \times \nabla \talpha + \pd{\talpha}{l} \nabla \tpsi\times\nabla l.
\label{deltaBtwiddle}
\end{equation}
Examining the size of the various terms in this equation we see that
\begin{equation}
\pd{\tpsi}{\psi} - 1 \sim \Or(1) \; \mathrm{ and }\; \pd{\talpha}{\alpha} -1 \sim \Or(1),
\end{equation}
because $\tpsi$ and $\talpha$ can vary on the short perpendicular spatial scale ($\nabla_\perp \sim \rho_i^{-1}$). Thus, the first term in \eref{deltaBtwiddle} is comparable to $\MeanB$. Similarly, we have that
\begin{equation}
\pd{\tpsi}{l} = \pd{ }{l} \left(\tpsi - \psi\right) = \Or\left( \gkeps\frac{\psi}{a} \right),
\end{equation}
as parallel variation is much weaker than perpendicular. Thus, the second and third terms in \eref{deltaBtwiddle} are smaller than the first by one power of $\gkeps$. In order that $\delB$ is indeed smaller than $\MeanB$ we require that
\begin{equation}
\pd{\tpsi}{\psi} \pd{\talpha}{\alpha} - \pd{\tpsi}{\alpha}\pd{\talpha}{\psi} - 1 = \Or(\gkeps).
\label{incompressible}
\end{equation}
Thus, to lowest order, $\tpsi$ and $\talpha$ are not ``independent''. This represents the fact that, to lowest order, $\veff$ is an incompressible flow and thus cannot generate any $\delBp$. Taking the scalar product of \eref{deltaBtwiddle1} with $\nabla\tpsi$ we obtain
\begin{equation}
\delB\dg\tpsi = - \MeanMagB \pd{\tpsi}{l}.
\end{equation}
Substituting for $\delB$ via \eref{deltaBnormal}, we are able to relate $\tpsi$ to $\delAp$:
\begin{equation}
\pd{\delAp}{\talpha} = \pd{\tpsi}{l}.
\label{AparTpsi}
\end{equation}
Similarly, we can find an expression for $\delAp$ in terms of $\talpha$:
\begin{equation}
\pd{\delAp}{\tpsi} = -\pd{\talpha}{l}.
\label{AparTalpha}
\end{equation}
Hence, given any one of $\delAp$, $\tpsi$ or $\talpha$ we can determine the other two from the appropriate pair of equations from \eref{incompressible}, \eref{AparTpsi} and \eref{AparTalpha}.
We do not use \eref{deltaBtwiddle} to determine $\delBp$ since for this we must know $\tpsi$ and $\talpha$ to $\Or(\gkeps^2)$ or equivalently the $\Or(\gkeps^2\vth[i])$ compressible corrections to $\veff$ -- and we do not.  Instead we use \eref{fluct-bpar} to determine $\delBp$ since it involves only lowest order quantities.

From the above results we can also form two expressions for $\nabla\delAp$ in terms of $\talpha$ and $\tpsi$:
\begin{equation}
\nabla_\perp\delAp = -\pd{ }{l}(\talpha - \alpha) \nabla\tpsi + \pd{ }{l} (\tpsi - \psi) \nabla\talpha,
\label{NablaApar}
\end{equation}
and
\begin{equation}
\nabla_\perp\delAp = -\pd{ }{\tl}(\talpha - \alpha) \nabla\psi + \pd{ }{\tl} (\tpsi - \psi) \nabla\alpha,
\label{NablaApar2}
\end{equation}
We will require these expressions when we come to eliminate $\delAp$ from our equations.
\section{Derivation of \eref{uxi}}
\label{parveloc}
In the definition \eref{veffDef} of $\veff$ we expand $\Efield$ in terms of potentials, keeping terms up to $\Or(\gkeps \vth[i])$, to obtain
\begin{equation}
\fl
\begin{eqalign}{
\veff =& \frac{c}{\MeanMagB} \left( 1 - \frac{\delBp}{\MeanMagB}\right) \bhat \times \nabla\left( \fpot + \pot_0 +\delta \pot - \xi \right) + \frac{1}{\MeanMagB}\left( \bm{u}\dg\delA \right) \times \bhat \\
		 &\qquad+ U(\bm{r}) \bhat + \Or(\gkeps^2 \vth[i]).
}\end{eqalign}
\label{tmpB2}
\end{equation}
where we have used the fact that $\left( \infrac{\partial}{\partial t} + \bm{u}\dg \right)\delA \sim\gkeps \Omega_i \delA$ to rewrite the time derivative of $\delA$ in the second term.
Using \eref{omdef} in the first term in \eref{tmpB2}, we find
\begin{equation}
\fl\begin{eqalign}{
\frac{c}{\MeanMagB} \left(1-\frac{\delBp}{\MeanMagB}\right) \bhat \times \nabla \fpot &= \left( 1- \frac{\delBp}{\MeanMagB} \right) \left[ \bm{u} - \frac{I\angvel(\psi,t)}{\MeanMagB} \Meanb \right] + \frac{\angvel(\psi,t)}{\MeanMagB} \delB_\perp  \times \nabla\psi \\
	&= \left( 1- \frac{\delBp}{\MeanMagB} \right) \left[ \bm{u} - \frac{I\angvel(\psi,t)}{\MeanMagB} \Meanb \right] - \frac{\delB_\perp}{\MeanMagB} \cdot\bm{u} \, \bhat  +\Or(\gkeps^2\vth[i]),
}\end{eqalign}
\label{1stSubB}
\end{equation}
where we have also used \eref{magfield} and \eref{bhatDef}.
The second tern in \eref{tmpB2} can be rewritten as follows:
\begin{equation}
\fl\begin{eqalign}{
\frac{1}{\MeanMagB} \left(\bm{u}\dg\delA\right)\times\bhat &= \frac{1}{\MeanMagB}\left[ \left( \nabla\delA \right)\cdot\bm{u} - \bm{u}\times\left( \curl\delA \right)\right]\times\bhat \\
	&= -\frac{1}{\MeanMagB}\bhat\times\nabla\left( \delA\cdot\bm{u} \right)  -\frac{1}{\MeanMagB} \left( \bm{u}\times\delB \right) \times\bhat +\Or(\gkeps^2\vth[i])\\
	&= -\frac{1}{\MeanMagB}\bhat\times\nabla\left( \delA\cdot\bm{u} \right) + \frac{\delBp}{\MeanMagB}\bm{u} - \frac{\Meanb\cdot\bm{u}}{\MeanMagB}\, \delB + \Or(\gkeps^2 \vth[i]).
}\end{eqalign}
\label{2ndSubB}
\end{equation}
Substituting \eref{1stSubB} and \eref{2ndSubB} back into \eref{tmpB2}, we obtain
\begin{equation}
\begin{eqalign}{
\veff =& \bm{u} +  \frac{c}{\MeanMagB} \bhat \times \nabla\left( \pot_0 +\gkupot - \xi \right) + \left[ U(\bm{r}) - \bhat\cdot\bm{u}\right] \bhat + \Or(\gkeps^2 \vth[i]),
}\end{eqalign}
\label{veffTmp}
\end{equation}
where we have used $\Meanb\cdot\bm{u}= I\angvel(\psi,t)/\MeanMagB$ and \eref{bhatDef}.
Letting $U(\bm{r}) = \bhat\cdot\bm{u}$ and substituting for $\xi$ from \eref{xidef}, we find
\begin{equation}
\veff = \bm{u} + \frac{c}{\MeanMagB} \bhat \times \nabla\left( \gkupot - \frac{T_e}{e} \ln \NotN[e] - \frac{T_e \delta n_e}{e n_e} - \tilde{\xi}\right) + \Or(\gkeps^2 \vth[i]),
\end{equation}
where we have used the fact that $\binfrac{c}{\MeanMagB} \bhat\times\nabla\left( \infrac{T_e n_{1e}}{e n_e}\right) \sim \gkeps^2\vth[i]$ and can thus be neglected.
Finally, picking $\tilde{\xi} = \ln \NotN[e] (\tpsi)$, we arrive at
\begin{equation}
\fl
\veff = \bm{u} + \frac{c}{\MeanMagB} \bhat \times \nabla\left[ \gkupot + \frac{T_e}{e}\left(\tpsi - \psi\right) \frac{d\ln \NotN[e]}{d\psi} - \frac{T_e \delta n_e}{e n_e}\right] + \Or(\gkeps^2 \vth[i]).
\label{veffE7}
\end{equation}
This is precisely \eref{uxi}, in which $\zeta$ is defined by \eref{zetaDef}.
\section{Properties of the Bounce Average}
\label{apBounce}
In this appendix we prove the properties of the bounce average that were stated in \Sref{Sbounce}.

First, as the second portion of the integration only differs from the first in the sign of $\sigma$ and the direction of integration, we have that, for any function $a$ that is independent of $\sigma$,
\begin{equation}
\bav{ w_\parallel a(\tl,\tpsi,\talpha,\energy[e],\magmom[e],t)} = 0.
\end{equation}
Secondly, combining \eref{ddlDef} and \eref{bavDef}, we discover that, for any single-valued function $a$,
\begin{equation}
\fl
\begin{eqalign}{
\bav{w_\parallel \bhat\dg a(\tl,\tpsi,\talpha,\energy[e],\magmom[e],\sigma,t)} 
	&= \bav{ w_\parallel \pd{a}{\tl}} \\
	  &= \int\limits_{l_1}^{l_2} \frac{d\tl}{|w_\parallel|} \left[\pd{a(\sigma=1)}{l} -\pd{a(\sigma=-1)}{l} \right]  = 0,
}\end{eqalign}
\label{wpardg}
\end{equation}
where the contributions from the endpoints has vanished as at the bounce points $a(\sigma = +1) = a(\sigma = -1)$ in order that $a$ is continuous through $w_\parallel = 0$.

We now prove that $\bav{\vdrift[e] \dg\tpsi} = 0$. Starting from the formula for $\vdrift[e]$, \eref{vdrift2},
we can show that, to leading order in $\massratio$,
\begin{equation}
\vdrift[e] = \frac{w_\parallel}{\cycfreq[e]}\left.\nabla\right|_{\energy[e],\magmom[e],\gyr}\times\left({w_\parallel\Meanb}\right),
\label{VDCurl}
\end{equation}
where the gradient is taken at constant energy $\energy[e]$ and magnetic moment $\magmom[e]$.
Using this form for $\vdrift[e]$, we have that
\begin{equation}
\vdrift[e]\dg\tpsi = \frac{w_\parallel}{\cycfreq[e]} \dv\left( {w_\parallel} \Meanb\times\nabla\tpsi\right).
\label{VDPsiDiv}
\end{equation}
By using the expression of a divergence of a vector field $\MeanA$:
\begin{equation}
\dv \MeanA = \MagBfield \pd{}{\tpsi} \left(\frac{\MeanA\dg\tpsi}{\MagBfield}\right)+ \MagBfield\pd{}{\talpha} \left(\frac{\MeanA\dg\talpha}{\MagBfield}\right)+\MagBfield\pd{}{\tl} \left(\frac{\MeanA\cdot\nabla \tl}{\MagBfield}\right),
\end{equation}
we can write \eref{VDPsiDiv} as
\begin{equation}
\fl
{\vdrift[e]\dg\tpsi} = \frac{w_\parallel\MagBfield}{\cycfreq[e]} \left\{\pd{}{\talpha} \left[ \frac{w_\parallel}{\MagBfield} \left( \Meanb\times\nabla\tpsi \right)\dg\talpha\right] + \pd{ }{\tl} \left[  \frac{w_\parallel}{\MagBfield} \left( \Meanb\times\nabla\tpsi \right)\dg\tl\right]\right\},
\end{equation}
where we have used \eref{wpardg}.
Now,
\begin{equation}
\frac{1}{\MagBfield} \left( \Meanb\times\nabla\tpsi\right)\dg\talpha = \bhat \cdot\Meanb = 1 + \Or(\gkeps^2).
\end{equation}
Thus,
\begin{equation}
\fl
{\vdrift[e]\dg\tpsi} = -\frac{m_e cw_\parallel}{e}\pd{ }{\tl} \left[  \frac{w_\parallel}{\MeanMagB} \left( \Meanb\times\nabla\tpsi \right)\dg\tl\right] = \frac{w_\parallel}{\cycfreq[e]} \Bfield\dg\left[ \frac{w_\parallel}{\MeanMagB} \left( \Meanb\times\nabla\tpsi \right)\dg\tl\right],
	\label{VDPsi}
\end{equation}
as $w_\parallel$ is axisymmetric and so independent of $\alpha$ and thus $\talpha$ (to lowest order).
We can now take the bounce average of this expression and use \eref{wpardg} to get the required result.

Finally, we prove that the bounce average commutes with the time derivative in the field-aligned coordinates to the required order in $\gkeps$.
The integration over $\tl$ is taken at constant $t$, $\tpsi$, $\talpha$, $\energy[e]$ and $\magmom[e]$ so clearly commutes with the time derivative in these variables. The fact that the denominator in \eref{bavDef} varies slowly in time, follows from the fact that it is in fact just the time taken for an electron with a given location, energy $\energy[e]$ and magnetic moment $\magmom[e]$ to travel along a field line from one bounce-point, to the other, and back again. As the $\delB_\perp \sim \gkeps \MeanB$ and $k_\parallel \sim l$, the contribution to the distance along the perturbed field line from the fluctuations is small. As $w_\parallel$ varies slowly in space, its value along the perturbed field line is almost the same as its value along the unperturbed field line. Thus, the bounce time along the perturbed field is a slowly varying function.

\section{Properties of the Flux-Surface Average}
\label{apFlux}
As the flux surface average is defined in exactly the same way as the average over equilibrium flux surfaces defined in \cite{flowtome1} (the fact that the integral is taken at constant $\energy[e]$ and $\magmom[e]$ rather than $\bm{v}$ does not affect the derivation), we can use the results of \Sref{P1-Sfluxav} of \cite{flowtome1} 
with a simple change of notation ($\psi$ becomes $\tpsi$). Thus, for any function $A(\bm{r},\bm{v})$ we have
\begin{equation}
\fluctfav{A} = \frac{1}{V'} \int \frac{dS}{|\nabla\tpsi|} A = \frac{1}{V'} \int \frac{d\tl d\talpha}{\MagBfield} A,
	\label{simpleFav}
\end{equation}
where the integration is taken over the surface labelled by $\tpsi$ and
\begin{equation}
\label{Vprimdef}
V' = \lim_{\Delta\tpsi \rightarrow 0} \frac{1}{\Delta\tpsi} \int_\Delta d^3\bm{r} = \int \frac{d\talpha d\tl}{\MagBfield}.
\end{equation}
Similarly, for any vector field $\MeanA(\bm{r},\bm{v})$ we have
\begin{equation}
\fluctfav{\dv \MeanA} = \frac{1}{V'} \pd{ }{\tpsi} \left( V' \fluctfav{\MeanA\dg\tpsi}\right),
	\label{favDiv}
\end{equation}
where the derivative are all taken at constant $\energy[e]$, $\magmom[e]$ and $\gyr$.

We now prove the results that are stated without proof in \Sref{Sfluxav}.
Using \eref{favDiv}, we can show that
\begin{equation}
\fluctfav{\Bfield \dg A} = \fluctfav{\dv \left( A\Bfield \right)} = \frac{1}{V'} \pd{ }{\tpsi} \left( V'\fluctfav{A \Bfield\dg\tpsi} \right) = 0.
\label{favAnnhil}
\end{equation}
Using \eref{simpleFav}, we see that for any single valued function $a(\bm{r})$
\begin{equation}
\fluctfav{\pd{a}{\talpha}} = \int \frac{d\tl}{\MeanMagB} d\talpha \pd{a}{\talpha} = 0,
	\label{ddAlpha}
\end{equation}
to lowest order in $\gkeps$, where we have used $\MagBfield = \MeanMagB+\Or(\gkeps \MeanMagB)$, axisymmetry, and periodicity in $\talpha$.
For any fluctuating function $a(\bm{r})$ and any fluctuating function $b(\tpsi)$ we have that, to lowest order in $\gkeps$,
\begin{equation}
\fl
\fluctfav{\frac{\MeanMagB}{|w_\parallel|} \pb{a}{b}} = \fluctfav{ \frac{\MeanMagB^2}{|w_\parallel|} \pd{a}{\talpha}} \pd{b}{\tpsi}  = \fluctfav{\pd{}{\talpha}\left( \frac{\MeanMagB^2 a}{|w_\parallel|} \right)} \pd{b}{\tpsi} = 0,
	\label{pbFav}
\end{equation}
where we have used \eref{ddAlpha} and \eref{poissonSimple}.

We now prove some results on flux-surface averages involving $\vdrift[e]$ and $\MaxVDrift$.
First, multiplying \eref{VDPsi} by $\MeanMagB / |w_\parallel|$ and using \eref{favAnnhil} gives
\begin{equation}
\fluctfav{\frac{\MeanMagB}{|w_\parallel|} \vdrift[e]\dg\tpsi } = 0.
\label{favVDdpsi}
\end{equation}
Secondly, we prove \eref{favVDlambd}. Working from \eref{VDCurl}, we have
\begin{equation}
\begin{eqalign}{
\fluctfav{\frac{\MeanMagB}{|w_\parallel|} \vdrift[e] \dg\lambda} &= -\fluctfav{\frac{c}{e m_e} \dv\left( |w_\parallel|\Meanb\times\nabla\lambda \right)} \\
	&= -\frac{c}{em_e}\frac{1}{V'}\pd{ }{\tpsi}V'\fluctfav{|w_\parallel|\Meanb\times\nabla\lambda\dg\tpsi}.
}\end{eqalign}
\label{tmpG8}
\end{equation}
However, as, to lowest order in $\gkeps$, $\MeanMagB$ and $|w_\parallel|$ are independent of $\talpha$, we have
\begin{equation}
\fluctfav{|w_\parallel|\Meanb\times\nabla\lambda\dg\tpsi} = \fluctfav{\pd{ }{\talpha}\left( |w_\parallel|\MeanMagB \lambda \right)}, 
\end{equation}
where we have used the fact that $\lambda$ is a fluctuating quantity to neglect $\infrac{\partial\lambda}{\partial\tl}$ compared to $\infrac{\partial\lambda}{\partial\talpha}$. Inserting this into \eref{tmpG8} and using \eref{ddAlpha}, we obtain \eref{favVDlambd} as promised.
Finally, as we can write $\MaxVDrift$ in the form
\begin{equation}
\MaxVDrift = \frac{cT_e}{e} \curl \left(\frac{\Meanb}{\MeanMagB}\right),
\end{equation}
we can prove that
\begin{equation}
\fluctfav{\MaxVDrift\dg\tpsi} = 0,
	\label{tmpG11}
\end{equation}
and
\begin{equation}
\fluctfav{\MaxVDrift\dg\lambda} = 0,
	\label{tmpG12}
\end{equation}
in the same manner as \eref{favVDdpsi} and \eref{favVDlambd}, by using the fact that $T_e = T_e(\tpsi) + \Or(\gkeps T_e)$ which commutes with the flux-surface average.

As the magnetic perturbation is small, the fluctuations cannot change the area of a flux surface by more than an a fraction of order $\Or(\gkeps)$. Thus $V'$ can change by at most $\gkeps V'$ in one fluctuation turnover time, and so $\infrac{\partial V'}{\partial t} \ll \omega V'$. This means that the flux-surface average commutes with the time derivative at constant $\tpsi$, $\talpha$, $\tl$:
\begin{equation}
\fluctfav{\ddttwiddles[a]}  = \ddttwiddles[\fluctfav{a}].
\end{equation}
\section{Transport Equations for Electrons}
\label{magicalTransp}
In this Appendix we derive the transport equations presented without proof in \Sref{STransp}.
We proceed from the kinetic equation written in $\bm{r}$ and $\bm{v}$ variables
\begin{equation}
\fl
\left( \pd{}{t}+\bm{u}\dg \right)f_e + \bm{w}\dg f_e + \frac{e}{m_e} \left(\Efield+ \frac{1}{c}\bm{v}\times\Bfield\right) \cdot\pd{f_e}{\bm{w}} = \collop[f_e] + \source[e].
\label{transpke-pre}
\end{equation}
It will be convenient to introduce the peculiar velocity $\tw = \bm{v} -\veff$, at which point we can write \eref{transpke-pre} as
\begin{equation}
\fl
\begin{eqalign}{
\left.\pd{f_e}{t}\right|_{\tw}+\left.\nabla\right|_{\tw}\cdot\left( {\veff f_e} \right) &+ \tw\left.\dg\right|_{\tw} f_e +  \pd{ }{\tw} \cdot\left[ f_e\left(\frac{e}{m_e}\nabla\xi-\pd{\veff}{t} - \bm{v}\dg\veff\right)\right]\\
	& + \frac{e\MagBfield}{cm_e}\tw\times\bhat\cdot \pd{f_e}{\tw} = \collop[f_e] + \source[e].
}\end{eqalign}
\label{transpke}
\end{equation}

To explicitly evaluate fluxes, we will need to use \eref{transpke} to evaluate the gyrophase-dependent piece of $f_e$ correct to second order in $\gkeps$. Taking \eref{transpke}, dropping 
terms smaller than $\Or(\gkeps^2\cycfreq[i] f_e)$, and dividing by ${e\MagBfield}/{cm_e}$, we have
\begin{equation}
\begin{eqalign}{
\bhat\times\tw\cdot \pd{f_e}{\tw}= & \frac{cm_e}{e\MagBfield}\tw\dg f_e + \left( \frac{c}{\MagBfield}\nabla \xi - \frac{cm_e}{e\MagBfield}\bm{v}\dg\veff \right)\cdot \pd{f_e}{\tw} \\
	& + \frac{cm_e}{e \MagBfield}\left[\left(\pd{ }{t} +\veff\dg\right) f_e - \collop[f_e]\right].
}\end{eqalign}
\label{fluxke}
\end{equation}
In this equation, the second-order parts of $f_e$ appear only on the left-hand side and so we will use this to eliminate the second-order parts of $f_e$ from our calculation.

The structure of this appendix is as follows: first, in \ref{tavFluct} and \ref{favFluctT} we introduce some averages that will be needed later. In \ref{EPTransP}, we derive the particle transport equation for electrons across the moving surfaces. In \ref{EETransp}, we derive the transport equation for electron heat. Finally, we evaluate the flux of toroidal angular momentum due to the electrons.

Detailed derivations of the intermediate results used in this appendix are provided in \ref{derivTransp}.

\subsection{The Turbulence Average at constant \texorpdfstring{$\tpsi$}{psi-tilde}:}
\label{tavFluct}
In order to derive the transport equations, we will need some new averages.
We define a new average over the fluctuations where the perpendicular spatial integrals are taken between surfaces of constant $\tpsi$ and $\talpha$ rather
than $\psi$ and $\alpha$. Similarly, the time integral in this average is taken at constant $\tpsi$,$\talpha$ and $\tl$. We denote this new average by $\ensav{\cdot}'$:
\begin{equation}
\ensav{g}' = \left. \int d\tpsi \int d\talpha \int\limits_{T/2}^{T/2} dt g \right/ T\int d\tpsi \int d\talpha,
\label{primav}
\end{equation}
where the spatial integral is taken over a small patch that has dimension $\lambda$ (the intermediate length scale) in each direction -- for reference compare with \eref{P1-PerpAvDef} of \cite{flowtome1}.

This average commutes with time derivatives at constant $\tpsi$ and with flux-surface averages over $\tpsi$-constant surfaces.

As usual, the composition of this average and the average over $\tpsi$-constant surfaces is an average over time composed with an average over an annular region between $\tpsi = \tpsi_0$ and $\tpsi = \tpsi_0+\Delta$.
It is important to note that this average differs from the turbulence average in \cite{flowtome1} only by terms that are small in $\gkeps$. This will be crucial, as we wish to replace all instances of this average by their unprimed version in the final equations.

\subsection{Flux Surface Averaging on the Transport Timescale:}
\label{favFluctT}
We will want to average the transport equations over the fluctuating flux surfaces. For this we need two identies. The first is just a specific case of \eref{favDiv}. If $\bm{A}(\bm{r},t)$ is a function only of space and time then
\begin{equation}
\fluctfav{\dv\bm{A}} = \frac{1}{\tvp} \pd{ }{\tpsi} \fluctfav{\tvp \bm{A}\cdot\tpsi},
\label{favDivNoV}
\end{equation}
where the derivative on the right is taken at constant $t$ only.

Secondly we need to interchange flux-surface averages and time derivatives. We do this via
\begin{equation}
\fluctfav{\pd{g}{t}} = \frac{1}{\tvp} \ddtpsitwiddles \tvp\fluctfav{g} - \frac{1}{\tvp} \pd{ }{\tpsi} \fluctfav{\tvp g \veff\dg\tpsi},
\label{movingflux}
\end{equation}
which is proven in an identical fashion to \eref{P1-movingflux} of \cite{flowtome1} except that we retain the velocity explicitly rather than replacing it with ${\partial\tpsi}/{\partial t}$ via \eref{psiEq}.
\subsection{Particle Transport}
\label{EPTransP}
We now move to the meat of the problem. Deriving our first transport equation.
Integrating \eref{transpke} over all velocities 
\begin{equation}
\fl
\ddttwiddles {\twint f_e} + \dv \left(\twint f_e \tw\right) + \left( \dv\veff \right) \twint f_e  = {\Psource},
\label{pTransp}
\end{equation}
where we have used the definition of the time derivative at constant $\tpsi$, $\talpha$ and $\tl$.
Averaging this equation over the fluctuating flux surfaces we obtain
\begin{equation}
\frac{1}{\tvp}\ddtpsitwiddles \tvp\fluctfav{\twint f_e} + \frac{1}{\tvp} \pd{}{\tpsi}\tvp \fluctfav{\twint f_e \tw\dg\tpsi} = {\Psource}.
\end{equation}
where we have used \eref{favDivNoV} and \eref{movingflux}. We now multiply this equation by $\tvp$, apply the new average over the fluctuations, and divide by $\tvp$ to find
\begin{equation}
\frac{1}{\tvp}\ddtpsitwiddles \tvp\fluctfav{n_e} + \frac{1}{\tvp} \pd{ }{\tpsi} \tvp\fluctfav{ \Gamma_e} = \fluctfav{\Psource},
\end{equation}
where we have used the fact that $\ensav{\tvp}' = V'(\tpsi) = \tvp + \Or(\gkeps V')$ and the fact that $\Psource \sim n_e / \tau_E$. The particle flux $\Gamma_e$ is defined by
\begin{equation}
\fluctfav{\Gamma_e} = \fluctfav{ \twint \ensav{ f_e \tw\dg\tpsi }' }.
\label{gammaDef}
\end{equation}

In \ref{SAPFLUX}, we use the kinetic equation to find the following explicit expression for the particle flux:
\begin{equation}
\fluctfav{\Gamma_e} = \fluctfav{ \twint \ensav{ g_{te} \pd{ }{\talpha}\left( c\zeta + \frac{w_\perp^2}{2\cycfreq[e]} {\delBp} \right) } }.
\label{pflux-dev}
\end{equation}

\subsection{Energy Transport and Heating}
\label{EETransp}
Multiplying \eref{transpke} by $\infrac{m_s\Magtw^2}{2}$, integrating over all velocities and averaging over flux-surfaces and the turbulence we obtain
\begin{equation}
\fl
\begin{eqalign}{
\frac{1}{\tvp}\frac{3}{2}\ddtpsitwiddles &\tvp \fluctfav{n_e}T_e + \frac{1}{\tvp} \pd{ }{\tpsi} \tvp \fluctfav{Q_e} = \\
		&\ensav{\fluctfav{\twint \left(e\nabla\xi - \pd{\veff}{t} - \bm{v}\dg\veff\right)\cdot\tw f_s }}  + \CollEnergy + \Esource,
}\end{eqalign}
\label{HeatTrans}
\end{equation}
where the electron thermal energy flux is defined by
\begin{equation}
\fluctfav{Q_e} = \ensav{\fluctfav{\twint \frac{1}{2} m_e \Magtw^2 f_e\tw\dg\tpsi}}.
\end{equation}
We show in \ref{eHeat} that we can rewrite this equation as
\begin{equation}
\fl
\begin{eqalign}{
\frac{1}{\tvp}\frac{3}{2}\ddtpsitwiddles &\tvp \fluctfav{n_e}T_e + \frac{1}{\tvp} \pd{ }{\tpsi} \tvp \fluctfav{q_e} = \\
		&\Pturb + \CompHeat + \PotEng + \CollEnergy + \Esource,
}\end{eqalign}
\label{HeatTransFinal}
\end{equation}
The heating terms are, in order, the turbulent heating
\begin{equation}
\fl
\begin{eqalign}{
\Pturb = -\fluctfav{\wint \ensav{g_e\ddttwiddles  \left(e\zeta - \magmom[e]\delBp\right)}} + e\fluctfav{n_e\ensav{\zeta\dv\veff}},
}\end{eqalign}
\label{apTurbHeat}
\end{equation}
the mean compressional heating
\begin{equation}
\CompHeat = -T_e \fluctfav{n_e \dv\vpsi},
\end{equation}
where $\vpsi$ is the velocity of the mean flux surfaces defined in \eref{P1-VpsiDef} of \cite{flowtome1},
and the exchange between thermal energy and electrostatic potential energy:
\begin{equation}
\fl
\begin{eqalign}{
\PotEng = &\fluctfav{ e\pot_0 \left(\ddttwiddles n_e -\Psource[e]\right) } + \fluctfav{ e\pot_0 n_e \dv\vpsi} \\
			 &\quad+ \fluctfav{ e\pot_0 \ensav{ \left( \frac{e\zeta}{T_e} n_e + \wint g_e  \right) \dv\veff }}.
}\end{eqalign}
			 \label{apPotHeat}
\end{equation}
In \eref{apPotHeat} and \eref{apTurbHeat}, the divergence of $\veff$ is given by
\begin{equation}
\fl
\dv\veff  =\ddttwiddles\frac{\delBp}{\MagBfield} +\left( \bhat\dg\psi \right)  \frac{I}{\MeanMagB} \frac{d\omega}{d\psi} - \frac{c}{B}\left( \Meanb\dg\Meanb \right)\cdot \Meanb \times \nabla\left( \zeta-\gkupot \right).
\end{equation}
This expression is correct and closed to the required order in $\gkeps$ and $\massratio$.

In \ref{qflux-deriv}, we show that the heat flux can be written in terms of $g_e$ as
\begin{equation}
\fl
\fluctfav{q_e} = \fluctfav{ \twint \left(\frac{1}{2}m_e w^2 - e\pot_0 - e \zeta\right)\ensav{ g_{te} \pd{ }{\talpha}\left( c\zeta + \frac{w_\perp^2}{2\cycfreq[e]} {\delBp} \right) } },
	\label{qflux-dev}
\end{equation}

\subsection{The Electron Angular Momentum Flux}
\label{mfluxap}

We start from \eref{P1-torflux} of \cite{flowtome1} and drop all terms that are small in $\massratio$ to obtain
\begin{equation}
\fl
\left(\nabla\psi\right)\cdot\viscosity\cdot\left(R^2\nabla\tor\right) = \ensav{ m_e R^4 \wint \gyror{\left(\bm{w}\dg\tor\right)\left[\left(\bm{w}\times\delB\right) \dg\tor\right]\delta f_e}},
\end{equation}
where we have expanded $\daccel$ and only kept terms that are of order $\gkeps^2 R^2 B n_i T_i$, i.e.,  comparable to the ion momentum flux.
Substituting for $\delta f_e$ from \eref{deltafSoln} and performing the gyroaverage gives \eref{totMflux}.

\section{Derivations for \ref{magicalTransp}}
\label{derivTransp}
\subsection{The solution for \texorpdfstring{$f_e$}{the electron distribution function}}
\label{ApSolnF}
We now collate results from the rest of the paper to write down a convenient form for $f_e$. 
First, using the general solution \eref{fSoln} for $f_e$, we can write
\begin{equation}
\begin{eqalign}{
f_e = &F_{0e}(\tpsi(\bm{R}_e'),\energy[e])\left(1+\frac{e\gkupot}{T_e}\right) + h_e'(\tpsi(\bm{R}_e'),\talpha(\bm{R}_e'),\energy[e],\magmom[e],\gyr,t) \\
		&+ F_{1e}(\bm{R}_e',\energy[e],\magmom[e]) + f_{2e},
}\end{eqalign}
\label{fSolnE2}
\end{equation}
where $f_{2e}\sim \gkeps^2 f_e$ and we have defined $h_e' = h_e - [\tpsi(\bm{R}_e') - \psi(\bm{R}_e')] \infrac{\partial F_{0e}}{\partial\tpsi}$ in order to be able to evaluate $F_{0e}$ at $\tpsi(\bm{R}_e')$. In addition, we have introduced the guiding centre position defined with respect to the exact field
\begin{equation}
\bm{R}_e' = \bm{r} - cm_e \frac{\bhat\times\tw}{e\MagBfield},
\end{equation}
which differs from the guiding center with respect to $\MeanB$ -- $\bm{R}_e$ -- by $\Or(\gkeps^2 \bm{R}_e')$.

We now use the solutions for the distribution function obtained in this paper.
For $F_{1e}$, we have from \eref{F1SOL}, that
\begin{equation}
F_{1e} = \frac{n_{1e}}{n_e} F_{0e} + F_{1e}^{(H)}(\bm{r},\energy[e],\magmom[e]) + \Or(\massratio^2 \gkeps^2 f_e),
\end{equation}
where $F_{1e}^{(H)} \sim \massratio\gkeps f_e$ contains the higher-order (in $\massratio$) contributions to $F_{1e}$.
We also wish to write $F_{0e}$ in terms of the new velocity variable $\tw$. To this end, we introduce the following quantity
\begin{equation}
\fl
\begin{eqalign}{
F_e' &= \left( \frac{m_e}{2\pi T_e(\tpsi(\bm{R}_e'))} \right)^{3/2} \exp\left[ \frac{e\xi}{T_e(\tpsi(\bm{R}_e'))} + \frac{m_e \angvel^2(\tpsi(\bm{R}_e')) R^2}{2 T_e(\tpsi(\bm{R}_e'))} - \frac{m_e \Magtw^2}{2T_e(\psi(\bm{R}_e'))}\right] \\
	&\qquad+ \frac{m_e \left(\veff-\bm{u}\right)\cdot\tw}{T_e} F_{0e}.
}\end{eqalign}
	\label{F0cute}
\end{equation}
Using \eref{xidef} and \eref{zetaDef}, we obtain the following expression for $\xi$
\begin{equation}
\frac{e\xi}{T_e} = \ln \NotN[e](\tpsi) + \frac{e\pot_0}{T_e} + \frac{n_{1e}}{n_e} + \frac{e\zeta}{T_e},
	\label{transpxi}
\end{equation}
in which $T_e$ is evaluated at $\tpsi$ and we have let $\widetilde{\xi} = 0$.
Substituting this into \eref{F0cute}, and using the definition \eref{F0def} of $F_{0e}$, we can show that
\begin{equation}
F_{0e}\left(1 + \frac{n_{1e}}{n_{e}} + \frac{e \gkupot}{T_e}\right) + F_{1e}^{(H)}
 = F_e' - \frac{e\left(\zeta-\gkupot\right)}{T_e} F_{0e} + F_{1e}^{(H)'} + \Or(\gkeps^2 f_e)
\end{equation}
where $F_{1e}^{(H)'} \sim \massratio\gkeps f_e$ differs from $F_{1e}^{(H)}$ by some small gyrophase-independent pieces -- we will never need $F_{1e}^{(H)}$ explicitly so can absorb any gyrophase-independent function into it.
Because of the convenience of the form of $F_e'$, we choose to write $f_e$ as
\begin{equation}
f_e = F_e' - \frac{e\left(\zeta-\gkupot\right)}{T_e} F_{0e} + F_{1e}^{(H)'} + h_e' + f_{2e}.
\label{fSolnDE}
\end{equation}
Using our solution \eref{deltafSoln} for $\delta f_e$ and the definition of $h_e'$, we have that, to lowest order in $\massratio$,
\begin{equation}
h_e' = \frac{e\left(\zeta-\gkupot\right)}{T_e} F_{0e} + g_e.
\label{hTog}
\end{equation}
We will use this equation later to eliminate $h_e'$ from our final results.

\subsection{An equation for \texorpdfstring{$f_{2e}$}{the second order part of f\_e}}
\label{subApFhigh}
As mentioned above, we will need properties of $f_{2e}$ to calculate the fluxes in our transport equations.
We now derive such an equation by using the solution \eref{fSolnDE} in \eref{fluxke} -- we will need to retain corrections to $f_{2e}$ up to $\Or(\massratio\gkeps^2 f_e)$. We perform this substitution step-by-step. The preliminary step is to notice that all the terms in the second line of \eref{fluxke} are $\Or(\massratio^2\gkeps^2 f_e)$ or smaller and so do not contribute in this calculation.

Next, we compute the left-hand-side of \eref{fluxke} for $F_e'$:
\begin{equation}
\fl
\begin{eqalign}{
&\left(\tw\times\bhat\right)\cdot\pd{F_e'}{\tw} \\
	&\quad= \frac{cm_e}{e\MagBfield} \tw\dg \tpsi \left\{ \frac{d T_e}{d \tpsi} \left[ \frac{e\xi}{T_e} + \frac{m_e \left( \angvel^2(\psi) R^2 - \Magtw^2 \right)}{2T_e} - \frac{3}{2} \right] + \frac{m_e \angvel(\tpsi) R^2}{2T_e} \frac{d\angvel}{d\tpsi} \right\} F_e'\\
	&\qquad + \frac{m_e \left(\veff - \bm{u}\right)\cdot\bhat\times\tw}{T_e} F_{0e} + \Or(\massratio^2 \gkeps^2 f_e)
}\end{eqalign}
\end{equation}
where we have used the definition of $\bm{R}_e'$ and the fact that $\bhat\dg\tpsi = 0$. Using this result and the definition \eref{F0cute} of $F_e'$, we can show that
\begin{equation}
\fl
\begin{eqalign}{
\left(\tw\times\bhat\right)\cdot\pd{F_e'}{\tw} - &\frac{cm_e}{e\MagBfield} \tw\dg F_e' = -\frac{cm_e}{e\MagBfield T_e} \tw\cdot \left( e\nabla\xi + \frac{1}{2} m_e \angvel^2(\psi) \nabla R^2 \right) F_e' \\
	&+ \frac{c m_e}{e \MagBfield} \tw\dg\left(\zeta-\gkupot\right) F_{0e} + \frac{m_e}{T_e} \tw\tw\bm{:}\nabla\left( \veff-\bm{u} \right) F_{0e},
}\end{eqalign}
\label{ap-f2eDgyr1}
\end{equation}
where we have used \eref{uxi} to show that $(\veff-\bm{u})\cdot\bhat\times\tw = (c/\MagBfield) \tw\dg\left(\zeta-\gkupot\right)$.
However, we also have that 
\begin{equation}
\fl
\begin{eqalign}{
\left(\frac{c}{\MagBfield}\nabla\xi - \frac{cm_e}{e\MagBfield}\bm{v}\dg\veff\right)\cdot\pd{F_e'}{\tw} =& -\frac{cm_e}{e\MagBfield T_e} \tw\cdot \left( e\nabla\xi + \frac{1}{2} m_e \angvel^2(\psi) \nabla R^2 \right) F_e' \\
	&- \frac{cm_e}{e\MagBfield} \frac{m_e}{T_e} \tw\tw\bm{:}\nabla\veff F_e'
}\end{eqalign}
\label{tmpDhoc}
\end{equation}
where we have used the fact that $\bm{v} = \veff + \tw$ and the fact that, to lowest order in $\gkeps$, $\veff = \bm{u}$. Thus, the contributions from $F_e'$ to \eref{fluxke} cancel (up to corrections of order $\Or(\massratio^2\gkeps^2 f_e)$) save for the last term in \eref{tmpDhoc}.

Then, using these results and substituting \eref{fSolnDE} into \eref{fluxke}, we have
\begin{equation}
\fl
\begin{eqalign}{
\left(\bhat\times\tw\right)\cdot\pd{f_{2e}}{\tw} = &\frac{cm_e}{e\MagBfield} \left(\tw\dg g_e -\left.\tw\dg\right|_{\energy[e],\magmom[e]}g_e\right)  + \frac{c}{\MagBfield}\nabla\left(\pot_0 + \zeta\right)\cdot\pd{g_e}{\tw} \\
	&\qquad- \frac{cm_e}{e\MagBfield} \frac{m_e}{T_e}\tw\tw\bm{:}\nabla\bm{u} F_{0e}+ \Or(\gkeps^2\massratio^2 f_e),
}\end{eqalign}
\label{ap-f2eDgyr}
\end{equation}
where we have used \eref{hTog} to write the result in terms of $g_e$ and used the fact that, for any function $H(\bm{R}_e',\energy[e],\magmom[e])$,
\begin{equation}
\fl
\left(\bhat\times\tw\right)\cdot\pd{H}{\tw} = \frac{cm_e}{e\MagBfield} \tw\left.\dg\right|_{\energy[e],\magmom[e]} H.
\end{equation}

We now convert the derivative of $g_e$ at constant $\tw$ into a derivative at constant $\energy[e]$ and $\magmom[e]$.
\begin{equation}
\fl
\begin{eqalign}{
\frac{cm_e}{e\MagBfield} \tw\dg g_e &= \frac{cm_e}{e\MagBfield} \left.\tw\dg\right|_{\energy[e],\magmom[e]} g_e - \frac{cm_e}{e\MagBfield} \tw\cdot\left( \left.\nabla\right|_{\tw}\energy[e] \pd{g_e}{\energy[e]} + \left.\nabla\right|_{\tw}\magmom[e] \pd{g_e}{\magmom[e]}\right) \\
	&= \frac{cm_e}{e\MagBfield} \left.\tw\dg\right|_{\energy[e],\magmom[e]} g_e - \frac{cm_e}{e\MagBfield} \left(  - e\tw\dg\pot_0 \pd{g_e}{\energy} + \tw\dg\magmom[e] \pd{g_e}{\magmom[e]}\right),
}\end{eqalign}
\end{equation}
in which we have used the definition $\energy[e] = (m_e/2)\Magtw^2 - e \pot_0 + \Or(\massratio^2 T_e)$. 
Combining this equation with the term from \eref{ap-f2eDgyr} involving $\pot_0$, we have
\begin{equation}
\fl
\begin{eqalign}{
\frac{cm_e}{e\MagBfield} \tw\dg g_e  + \frac{c}{\MagBfield} \nabla\pot_0 \cdot\pd{g_e}{\tw} =\\
	\qquad\quad\frac{cm_e}{e\MagBfield} \left.\tw\dg\right|_{\energy[e],\magmom[e]} g_e - \frac{cm_e}{e\MagBfield} \left(\tw\left.\cdot \nabla\right|_{\tw}\magmom[e] - \frac{e}{\MeanMagB}\tw_\perp\dg\pot_0\right)\pd{g_e}{\magmom[e]}.
}\end{eqalign}
\label{muStuffs}
\end{equation}
By taking the derivative of the definition of $\magmom[e]$ we can show that
\begin{equation}
\fl
\tw\dg\magmom - e\tw_\perp\dg\pot_0 = -m_e\tw \cdot \left( \frac{w_\perp^2}{2\MeanMagB} \nabla \ln \MeanMagB + \frac{w_\parallel}{\MeanMagB} \tw\dg\Meanb\right) - e\tw_\perp\dg\pot_0
\end{equation}
We can rewrite this in terms of $\vdrift[e]$ as
\begin{equation}
\begin{eqalign}{
\tw\dg\magmom - e\tw_\perp\dg\pot_0 =\\
	\qquad m_e\tw_\perp \cdot \left( \bhat\times\vdrift[e]\right) - \frac{m_ew_\parallel}{\MeanMagB} \left( \tw_\perp\tw_\perp \bm{:} \nabla \Meanb + \frac{w_\perp^2}{2} \Meanb\dg\ln\MeanMagB \right)
}\end{eqalign}
\end{equation}
Substituting this back into \eref{muStuffs} and thence back into \eref{ap-f2eDgyr} we arrive at:
\begin{equation}
\fl
\begin{eqalign}{
\left(\bhat\times\tw\right)\cdot\pd{f_{2e}}{\tw} = \frac{c}{\MagBfield}\nabla\zeta\cdot\pd{g_e}{\tw} - \frac{cm_e}{e\MagBfield} \frac{m_e}{T_e}\tw\tw\bm{:}\left( \nabla\bm{u} \right) F_{0e} \\
	\qquad \frac{cm_e}{e\MagBfield}\left[ m_e\tw\cdot\bhat\times\vdrift[e] - \frac{m_ew_\parallel}{\MeanMagB} \left( \tw_\perp\tw_\perp \bm{:} \nabla \Meanb + \frac{w_\perp^2}{2} \Meanb\dg\ln\MeanMagB \right) \right]\pd{g_e}{\magmom[e]}\\
	\qquad+ \Or(\gkeps^2\massratio^2 f_e).
}\end{eqalign}
\label{hoc}
\end{equation}
Estimating $f_{2e}$ from this equation, we see that $f_{2e} \sim \massratio\gkeps^2 f_e$, rather than $f_{2e}\sim\gkeps^2 f_e$ which we would na\"ively expect -- this is because it is only the gyrophase-dependent piece of $f_{2e}$ we need and, to lowest order in $\massratio$, $f_{2e}$ is gyrophase-independent.

\subsection{Derivation of \eref{pflux-dev}}
\label{SAPFLUX}
Using the solution \eref{fSolnDE} for $f_e$ in the definition of the particle flux, we have
\begin{equation}
\fl
\begin{eqalign}{
\tvp\Gamma_e = \ensav{\tvp\fluctfav{ \twint  \left( F_{e}' - \frac{e\left(\zeta-\gkupot\right)}{T_e} F_{0e} + h_e' + f_{2e} \right) \tw\dg\tpsi }}.
}\end{eqalign}
\end{equation}
Substituting for $F_e'$ from \eref{F0cute} and performing the velocity integration, we have
\begin{equation}
\begin{eqalign}{
\tvp\Gamma_e = &\ensav{\tvp{\fluctfav{n_e \left(\veff - \bm{u}\right)\dg\tpsi}}} \\
	&\quad+\ensav{\tvp\fluctfav{ \twint  \left( h_e' + f_{2e} \right) \tw\dg\tpsi }}.
}\end{eqalign}
\end{equation}
Using \eref{uxi} for $\veff$ and using the definition of the flux-surface average in the first term, we have
\begin{equation}
\begin{eqalign}{
\tvp\Gamma_e = &\ensav{\int d\talpha d\tl {\frac{ cn_e}{\MagBfield} \pd{}{\talpha}\left(\zeta-\gkupot\right) }} \\
	&\quad+\ensav{\tvp\fluctfav{ \twint  \left( h_e' + f_{2e} \right) \tw\dg\tpsi }}.
}\end{eqalign}
\end{equation}
Integrating by parts with respect to $\talpha$ in the first term, we find that
\begin{equation}
\begin{eqalign}{
\tvp\Gamma_e = &\ensav{\int d\talpha d\tl { \frac{cn_e}{\MagBfield} {(\zeta-\gkupot)}\pd{ }{\talpha}\left(\frac{\delBp}{\MagBfield}\right) }} \\
	&\quad+\ensav{\tvp\fluctfav{ \twint  \left( h_e' + f_{2e} \right) \tw\dg\tpsi }}.
}\end{eqalign}
\label{gammaTmp}
\end{equation}

The calculation of the term involving $h_e'$ is particularly involved. In \ref{heprimDeriv}, we prove the following identity:
\begin{equation}
\fl
\begin{eqalign}{
\ensav{\tvp\fluctfav{h_e'\tw\dg\tpsi}} = &\ensav{\tvp\fluctfav{  g_{te}\pd{}{\talpha}\left( \frac{w_\perp^2}{2\cycfreq[e]}{\delBp} \right) }}\\
		&\qquad- \ensav{\int d\talpha d\tl \frac{c n_e}{\MagBfield} (\zeta-\gkupot) \pd{ }{\talpha} \left( \frac{\delBp}{\MagBfield} \right) }.
}\end{eqalign}
\label{heIdent}
\end{equation}
Using this in \eref{gammaTmp}, we find
\begin{equation}
\begin{eqalign}{
\Gamma_e = 
	& \ensav{\tvp\fluctfav{ \twint   g_{te}\pd{}{\talpha}\left( \frac{w_\perp^2}{2\cycfreq[e]}{\delBp} \right)  } }\\
	&+\ensav{\tvp\fluctfav{\twint \bhat\times\nabla\tpsi\cdot\tw \left(\tw\times\bhat\right)\cdot\pd{f_{2e}}{\tw}}},
}\end{eqalign}
\label{gammaTmp2}
\end{equation}
where we have integrated by parts in the term involving $f_{2e}$.

Now, the derivative of $f_{2e}$ in \eref{gammaTmp2} is precisely the one appearing in \eref{hoc}. Thus, substituting \eref{hoc} into \eref{gammaTmp2}, we have
\begin{equation}
\begin{eqalign}{
\Gamma_e = 
	& \ensav{\tvp\fluctfav{ \twint   g_{te}\pd{}{\talpha}\left( \frac{w_\perp^2}{2\cycfreq[e]}{\delBp} \right)  } }\\
	&+ \ensav{\tvp\fluctfav{\twint \frac{c}{\MagBfield}\left( \bhat\times\nabla\zeta\dg\tpsi \right) g_e}} \\
			  &\quad- \ensav{\tvp\fluctfav{\twint \frac{cm_e}{e\MagBfield}\frac{\Magtw_\perp^2}{2} \vdrift[e]\dg\tpsi \pd{g_e}{\magmom[e]}}}, \\
}\end{eqalign}
\label{closer}
\end{equation}
where we have integrated by parts with respect to $\tw$ in the second line and used the gyrophase-independence of $g_e$ in the third line. We have also used the isotropy of $F_{0e}$ and the gyrotropy of $g_e$ to eliminate terms in \eref{hoc} that contain an even power of $\tw_\perp$ (we are multiplying by one power of $\tw_\perp$ so this becomes an odd power and thus vanishes upon integration over $\gyr$).

We now prove that the last line of \eref{closer} is small and can be dropped.
Writing it in terms of $\magmom[e]$ and using \eref{VDPsi}, we obtain
\begin{equation}
\fl
\begin{eqalign}{
&\ensav{\tvp\fluctfav{\twint \frac{cm_e}{e\MagBfield}\frac{\Magtw_\perp^2}{2} \vdrift[e]\dg\tpsi \pd{g_e}{\magmom[e]}}} \\
	&\qquad= \frac{c}{e}\ensav{\tvp\fluctfav{\twint \magmom[e] \frac{w_\parallel}{\cycfreq[e]} \Bfield\left.\dg\right|_{\energy[e],\magmom[e]}\left[\frac{w_\parallel}{B} \left(\Meanb\times\nabla\tpsi\right)\dg\tl\right]\pd{g_e}{\magmom[e]}}}\\
   &\qquad= \frac{c^2m_e}{e^2}\ensav{\tvp\fluctfav{\twint  {\Magtw_\parallel} \bhat\left.\dg\right|_{\energy[e],\magmom[e]}\left[\magmom[e]\pd{g_e}{\magmom[e]}\frac{\Magtw_\parallel}{B} \left(\Meanb\times\nabla\tpsi\right)\dg\tl\right]}},\\
	&\qquad=0,
}\end{eqalign}
\label{I26}
\end{equation}
where we have been able to write \eref{VDPsi} in terms of $\tw$ because the difference between $\tw$ and $\bm{w}$ is $\Or(\gkeps\vth[i])$ and thus negligible.

Finally, integrating by parts with respect to $\talpha$ in the second line of \eref{closer} and dropping all the references to $\tw$, we have
\begin{equation}
\Gamma_e = \fluctfav{ \wint \ensav{ g_{te} \pd{ }{\talpha}\left( c\zeta + \frac{w_\perp^2}{2\cycfreq[e]} {\delBp} \right) } },
\label{success}
\end{equation}
where the passing electron response has vanished because $\infrac{\partial g_{pe}}{\partial\talpha} = 0$ (at constant $\energy[e]$ and $\magmom[e]$). This is \eref{pflux-dev} as required.
\subsection{Derivation of the Heat Transport Equation}
\label{eHeat}
As in \ref{P1-dpdtderiv} of \cite{flowtome1}, it is convenient to adjust the form of the pressure evolution equation before calculating the heat flux.
In particular, we subtract $\ensav{\fluctfav{\dv[e(\pot_0+\zeta) \twint \tw f_e]}}$ from both sides of \eref{HeatTrans}, and use 
\begin{equation}
\begin{eqalign}{
\dv&\ensav{e\pot_0\twint \tw f_e}\\
	&= \wint e\nabla\pot_0 \cdot \ensav{\tw f_e} + e\pot_0 \dv\ensav{ \twint \tw f_e } \\
									 &=\ensav{\wint e\nabla\pot_0 \cdot \tw f_e} \\
									 &\qquad- e\pot_0 \left( \ddttwiddles{n_e} + \ensav{\left( \dv\veff \right)\wint f_e} - \Psource  \right),
}\end{eqalign}
\end{equation}
where we have used \eref{pTransp}.
This results in the following heat transport equation
\begin{equation}
\fl
\begin{eqalign}{
\frac{1}{\tvp}\frac{3}{2}\ddtpsitwiddles &\tvp \fluctfav{n_e}T_e + \frac{1}{\tvp} \pd{ }{\tpsi} \tvp \fluctfav{Q_e - e\ensav{\left( \pot_0+\zeta \right) \twint f_e\tw\dg\tpsi}} = \\
		&\ensav{\fluctfav{\twint \left[e\nabla\left( \xi - \pot_0 \right) - \pd{\veff}{t} - \bm{v}\dg\veff\right]\cdot\tw f_s }} \\
		&-\ensav{\fluctfav{\dv\left(e\zeta\twint f_e\tw\right)}} + \CollEnergy + \Esource \\
&+ \fluctfav{e\pot_0\left(\ddttwiddles{n_e} +\ensav{\left( \dv\veff \right)\wint f_e}- \Psource\right)},
}\end{eqalign}
\label{HeatTmp1}
\end{equation}

\subsubsection{Electron Heating.}
We now evaluate the heat sources in the temperature evolution equation, from left to right.
The first heating term is
\begin{equation}
e\ensav{\fluctfav{\wint f_e \tw\dg \left(\xi-\pot_0\right)}} =  e\ensav{\fluctfav{\wint f_e \tw\dg \zeta}},
\end{equation}
where we have used \eref{transpxi} for $\xi$.
Rearranging the derivative to act on $f_e$, we have
\begin{equation}
\fl
\begin{eqalign}{
e\ensav{\fluctfav{\wint f_e \tw\dg \zeta}} = \\
		\qquad e\ensav{\fluctfav{ \dv\wint \zeta \tw f_e}} - e \ensav{\fluctfav{ \wint \zeta \tw\left.\dg\right|_{\tw} f_e}}.
}\end{eqalign}
\label{tmpHeat}
\end{equation}
Now, integrating \eref{fluxke} multiplied by $(e\MagBfield / cm_e)\zeta$ over $\tw$, we have that
\begin{equation}
0 = \zeta \wint \tw\left.\dg\right|_{\tw} f_e + \zeta n_e \dv\veff + \zeta \wint \ddttwiddles f_e,
\end{equation}
where we have integrated by parts with respect to $\tw$ as needed. Substituting this back into \eref{tmpHeat}, we have
\begin{equation}
\fl
\begin{eqalign}{
e&\ensav{\fluctfav{\wint f_e \tw\dg \zeta}} = e\ensav{\fluctfav{ \dv\wint \zeta \tw f_e}}\\
&\qquad\qquad+ e\ensav{\fluctfav{\zeta n_e \dv\veff}}-  e \ensav{\fluctfav{ g_e \ddttwiddles{\zeta}}},
}\end{eqalign}
\label{blah1}
\end{equation}
where we have integrated by parts with respect to time in the second term.
The first of these terms cancels with the corresponding term in \eref{HeatTmp1} and the second and third will become part of the turbulent heating.
We will need to process terms involving $\dv\veff$ several times in this section. In \ref{apDvVeff}, we show that we can write it in a closed (if unwieldly) form -- see \eref{DvVeff}. For conciseness we will not expand $\dv\veff$ in this section.

Now, the time-derivative term is small:
\begin{equation}
\ensav{\fluctfav{ \wint m_e \pd{\veff}{t} \cdot\tw f_e }} = \Or(\massratio \gkeps^3 \cycfreq[i] T_e),
\end{equation}
and can thus be neglected.

Substituting the solution \eref{fSolnDE} for $f_e$ into the remaining heating term gives
\begin{equation}
\fl
- m_e\ensav{\fluctfav{ \wint \bm{v}\dg\veff\cdot\tw \left( F_e' + h_e' + \frac{e\left(\gkupot-\zeta\right)}{T_e} F_{0e} + F_{1e}^{(H)'} + f_{2e}  \right)}}.
\end{equation}
Estimating the size of the contribution from the gyrophase-dependent piece of $f_{2e}~\sim~\massratio\gkeps^2 f_e$, we see that it is $\Or(\massratio\gkeps^3\cycfreq[i]n_eT_e)$ and so we can neglect it.
Now, as $\ensav{\dv\veff} \sim \ensav{\bhat\cdot\nabla\veff\cdot\bhat} \sim \Or(\gkeps^3 \cycfreq[i])$, we can also neglect the contributions from $F_{1e}^{(H)'}$ and any gyrophase-independent piece of $f_{2e}$.
Of the remaining terms, we handle the $F_e'$ term first. Using \eref{F0cute}, we have
\begin{equation}
\fl
\begin{eqalign}{
-m_e \ensav{\fluctfav{ \wint \bm{v}\dg\veff\cdot\tw F_e' }} \\
	\quad\qquad = - \fluctfav{\ensav{\dv\veff} n_e} T_e - m_e \fluctfav{\ensav{\left( \veff\dg\veff \right)\cdot\left( \veff-\bm{u} \right)} n_e},
}\end{eqalign}
\end{equation}
where we have used the fact that, to lowest order in $\gkeps$, the flux-surface average and the turbulence average commute. As $\veff-\bm{u} \sim \gkeps \vth[i]$ we can use \eref{uxi} to express $\veff$ in the second term. Doing this, we find that the entire second term
is $\Or(\massratio\gkeps^3 n_e T_e\cycfreq[i])$ or smaller and can thus be neglected.
In the first term, we cannot use \eref{uxi}, but instead, we interchange the divergence and the turbulence average to find that
\begin{equation}
-m_e \ensav{\fluctfav{ \wint \bm{v}\dg\veff\cdot\tw F_e' }} = - T_e\fluctfav{n_e{\dv\vpsi} } ,
	\label{blah2}
\end{equation}
where we have used $\ensav{\veff} = \vpsi$ -- the average of the velocity of the exact surfaces is just the velocity of the average surfaces.
We now work out the corresponding term involving $g_e$. Using the gyrophase-independence of $g_e$, we have
\begin{equation}
\begin{eqalign}{
-m_e\ensav{\fluctfav{ \wint \bm{v}\dg\veff\cdot\tw g_e}} \\
	\qquad= -\ensav{\fluctfav{\wint \MeanMagB\left(\idmat-\bhat\bhat\right)\bm{:}\left( \nabla\veff \right) \magmom[e] g_e }}.
}\end{eqalign}
\end{equation}
Substituting for $\dv\veff$ from \eref{DvVeff}, we obtain 
\begin{equation}
\fl
-m_e\ensav{\fluctfav{ \wint \bm{v}\dg\veff\cdot\tw g_e}} = -\ensav{\fluctfav{\wint  \magmom[e] g_e \ddttwiddles \delBp }}.
\label{blah3}
\end{equation}

Finally, we have to deal with the term involving $\pot_0$ and the divergence of $\veff$.
Using our solution for $f_e$, we have
\begin{equation}
\fl
\begin{eqalign}{
e\fluctfav{\pot_0 \ensav{\left(\dv\veff\right) \twint f_e }} = \\
	\qquad e \fluctfav{\pot_0 n_e \dv\vpsi} + e \fluctfav{\pot_0 \ensav{\left( \frac{e\zeta}{T_e} n_e + \wint g_e\right) \dv\veff}}.
}\end{eqalign}
\end{equation}
Using \eref{DvVeff} to substitute for $\dv\veff$ in the last term, we have
\begin{equation}
\fl
\begin{eqalign}{
e\fluctfav{\pot_0 \ensav{\left(\dv\veff\right) \twint f_e }} = e \fluctfav{\pot_0 n_e \dv\vpsi} \\
		\qquad+ e \fluctfav{\pot_0 \ensav{\left( \frac{e\zeta}{T_e} n_e + \wint g_e\right) \left( \ddttwiddles\frac{\delBp}{\MagBfield} + \bhat\bhat\bm{:}\nabla\veff\right)}},
}\end{eqalign}
\label{blah4}
\end{equation}
where the parallel compression is given by \eref{bbnablavef}.

Using \eref{blah1}, \eref{blah2}, \eref{blah3}, and \eref{blah4} in \eref{HeatTmp1}, we obtain
\begin{equation}
\fl
\begin{eqalign}{
&\frac{1}{\tvp}\frac{3}{2}\ddtpsitwiddles \tvp \fluctfav{n_e}T_e + \frac{1}{\tvp} \pd{ }{\tpsi} \tvp \fluctfav{Q_e - e\ensav{\left( \pot_0+\zeta \right) \twint f_e\tw\dg\tpsi}} = \\
	& -\ensav{\fluctfav{\wint   g_e \ddttwiddles \left( e\zeta + \magmom[e]\delBp \right) }} +e\ensav{\fluctfav{\zeta n_e\dv\veff}}\\
	&	+\fluctfav{e\pot_0\left[\ddttwiddles{n_e} + n_e\dv\vpsi + \ensav{\left(\frac{e\zeta}{T_e} n_e + \wint g_e\right) \dv\veff} - \Psource\right]} \\
	&- T_e\fluctfav{n_e{\dv\vpsi} }+\CollEnergy + \Esource,
}\end{eqalign}
\label{HeatTmp2}
\end{equation}
where $\dv\veff$ is a shorthand for the expression given in \eref{DvVeff}.
Gathering the terms in this equation together according to their physical interpretation, we obtain the heat transport equation \eref{HeatTransFinal} as required.
\subsubsection{The Divergence of \texorpdfstring{$\veff$}{v\_eff}:}
\label{apDvVeff}
We wish to calculate the divergence of $\veff$ correct to $\Or(\gkeps^2 \cycfreq[i])$. We start from Faraday's Law:
\begin{equation}
\pd{\Bfield}{t} = \curl\left(\veff\times\Bfield\right) + \Or(\gkeps^2\massratio \Bfield \cycfreq[i]).
\end{equation}
Taking the component of this equation along $\bhat$, we obtain
\begin{equation}
\fl
\left(\pd{}{t}+\veff\dg\right)\delBp = \MagBfield \left(\idmat - \bhat\bhat\right)\bm{:}\nabla\veff + \Or(\gkeps^2\massratio \Bfield \cycfreq[i]).
\end{equation}
Rearranging this into an equation for $\dv\veff$, we have
\begin{equation}
\dv\veff = \ddttwiddles \frac{\delBp}{\MagBfield} + \bhat\bhat\bm{:}\nabla\veff+ \Or(\gkeps^2\massratio \Bfield \cycfreq[i]).
\label{DvVeff}
\end{equation}
To the required order, we can use \eref{uxi} to write the parallel compression due to $\veff$ as
\begin{equation}
\fl
\bhat\bhat\bm{:}\nabla\veff = \left( \bhat\dg\psi \right)  \frac{I}{\MeanMagB} \frac{d\omega}{d\psi} - \frac{c}{B}\left( \Meanb\dg\Meanb \right)\cdot \Meanb \times \nabla\left( \zeta-\gkupot \right) + \Or(\gkeps^2\massratio \cycfreq[i]).
\label{bbnablavef}
\end{equation}
\subsection{Derivation of \eref{qflux-dev}}
\label{qflux-deriv}
The derivation of the heat flux proceeds in exactly the same way as the particle flux derivation.

Using the solution \eref{fSolnDE} for $f_e$ in the definition of the heat flux, we have
\begin{equation}
\fl
\begin{eqalign}{
\tvp q_e = \ensav{\tvp\fluctfav{ \twint \left(\frac{1}{2}m_e \Magtw^2 - e \pot_0 - e\zeta\right) \left( F_{e} + h_e' + f_{2e} \right) \tw\dg\tpsi }}.
}\end{eqalign}
\end{equation}
Substituting for $F_e'$ and integrating, we have
\begin{equation}
\fl
\begin{eqalign}{
\tvp q_e = &\ensav{\tvp{\fluctfav{ \tee \left(\veff - \bm{u}\right)\dg\tpsi}}} +\ensav{\tvp\fluctfav{ \twint \tee \left( h_e' + f_{2e} \right) \tw\dg\tpsi }}\\
			  &\quad-\ensav{\tvp{\fluctfav{ e \zeta f_e \tw\dg\tpsi}}},
}\end{eqalign}
\label{q-gammaTmp}
\end{equation}
where, for conciseness, we have defined $\tee = (1/2)m_e \Magtw^2 - e\pot_0$.

As with the calculation of the particle flux, we use the identity \eref{heIdent} to handle the term involving $h_e'$.
Using \eref{heIdent} in \eref{q-gammaTmp} and integrating by parts where required, we obtain
\begin{equation}
\begin{eqalign}{
 q_e = 
	& \tvp\fluctfav{ \twint \ensav{  \tee g_{te}\pd{}{\talpha}\left( \frac{w_\perp^2}{2\cycfreq[e]}{\delBp} \right)  } }\\
	&+\ensav{\tvp\fluctfav{\twint \tee\bhat\times\nabla\tpsi\cdot\tw \left(\tw\times\bhat\right)\cdot\pd{f_{2e}}{\tw}}}\\
			  &\quad-\ensav{\tvp{\fluctfav{ e \zeta f_e \tw\dg\tpsi}}},
}\end{eqalign}
\end{equation}
where we have used \eref{uxi} and the definition of the flux-surface average to cancel the first term in \eref{q-gammaTmp} with the contribution arising from the first term of \eref{heIdent}.

Substituting for $f_{2e}$ from \eref{hoc}, we have
\begin{equation}
\begin{eqalign}{
 q_e = 
	& \ensav{\tvp\fluctfav{ \twint   \tee g_{te}\pd{}{\talpha}\left( \frac{w_\perp^2}{2\cycfreq[e]}{\delBp} \right)  } }\\
	&+ \ensav{\tvp\fluctfav{\twint \tee\frac{c}{\MagBfield}\left( \bhat\times\nabla\zeta\dg\tpsi \right) g_e}} \\
	&+ \ensav{\tvp\fluctfav{\twint \frac{cm_e}{\MagBfield}\tw\dg\zeta\left( \bhat\times\tw\dg\tpsi \right) g_e}} \\
			  &\quad- \ensav{\tvp\fluctfav{\twint \frac{cm_e}{e\MagBfield}\frac{\Magtw_\perp^2}{2} \tee\vdrift[e]\dg\tpsi g_e}} \\
			  &\quad-\ensav{\tvp{\fluctfav{\twint e \zeta f_e \tw\dg\tpsi}}},
}\end{eqalign}
\label{q-closer}
\end{equation}
where we have integrated by parts with respect to $\tw$ and used the gyrophase-independence of $g_e$. We have also used the isotropy of $F_{0e}$ and the gyrotropy of $g_e$ to show that most terms in \eref{hoc} do not contribute to the heat flux.
The proof that the term involving $\vdrift[e]$ can be dropped is identical to the proof for the particle flux -- the manipulations of \eref{I26} go through in exactly the same way with an extra multiple of $\energy[e]$.

Next, we show that the third and fifth lines of \eref{q-closer} cancel. Taking the fifth line and expressing $f_e$ via \eref{fSolnDE}, we have
\begin{equation}
\fl
\ensav{\tvp{\fluctfav{\twint e \zeta f_e \tw\dg\tpsi}}} = \ensav{\tvp{\fluctfav{\twint e \zeta \left(F_e' + h_e'\right) \tw\dg\tpsi}}},
\end{equation}
where we have been able to drop terms associated with $F_{1e}^{(H)'}$ as it is both small and gyrophase-independent. Substituting for $F_{e}'$ from \eref{F0cute}, we have
\begin{equation}
\fl
\begin{eqalign}{
\ensav{\tvp{\fluctfav{\twint e \zeta f_e \tw\dg\tpsi}}} = \\
\qquad	\ensav{\tvp{\fluctfav{\twint e \zeta \left[ \frac{m_e \left(\veff-\bm{u}\right)\cdot\tw}{T_e} F_{0e} + \left(\frac{\bhat\times\tw}{\cycfreq[e]}\right)\dg h_e' \right] \tw\dg\tpsi}}},
}\end{eqalign}
\label{tmpQ2}
\end{equation}
where we have expanded $h_e'$ around $\bm{r} = \bm{R}_e'$ and dropped terms that are small in $\massratio$. Substituting for $h_e'$ from \eref{hTog} and using \eref{uxi}, we see that the first term in the brackets in \eref{tmpQ2} cancels with the contribution from the first term on the right-hand side of \eref{hTog} leaving us with
\begin{equation}
\begin{eqalign}{
\ensav{\tvp{\fluctfav{\twint e \zeta f_e \tw\dg\tpsi}}} \\
		\qquad= \ensav{\tvp{\fluctfav{\twint e \zeta \left( \frac{\bhat\times\tw}{\cycfreq[e]} \right)\dg g_e \tw\dg\tpsi}}}.
}\end{eqalign}
\end{equation}
Performing the gyroaverage explicitly we have, 
\begin{equation}
\fl
\begin{eqalign}{
\ensav{\tvp{\fluctfav{\twint e \zeta f_e \tw\dg\tpsi}}} 
	&= \ensav{\tvp{\fluctfav{\twint {cm_e} \zeta w_\perp^2 \pd{g_e}{\talpha}}}} \\
	&= -\ensav{\tvp{\fluctfav{\twint {cm_e} g_e w_\perp^2 \pd{\zeta}{\talpha}}}},
}\end{eqalign}
\end{equation}
where we have used the definition of the flux-surface average to integrate by parts with respect to $\talpha$.
Now, taking the third line of \eref{q-closer} and performing the gyroaverage, we obtain
\begin{equation}
\begin{eqalign}{
\ensav{\tvp\fluctfav{\twint \frac{cm_e}{\MagBfield}\tw\dg\zeta\left( \bhat\times\tw\dg\tpsi \right) g_e}} \\
		\qquad = -\ensav{\tvp{\fluctfav{\twint {cm_e} g_e w_\perp^2 \pd{\zeta}{\talpha}}}}.
}\end{eqalign}
\end{equation}
Thus, the third and fifth lines of \eref{q-closer} cancel.

Finally, using this result and integrating by parts with respect to $\talpha$ in the second line of \eref{q-closer} and dropping all the references to $\tw$, we have
\begin{equation}
 q_e = \fluctfav{ \wint \ensav{ g_{te} \left(\frac{1}{2} m_e w^2 - e \pot_0\right)\pd{ }{\talpha}\left( c\zeta + \frac{w_\perp^2}{2\cycfreq[e]} {\delBp} \right) } },
\label{q-success}
\end{equation}
where the passing electron response has vanished because $\pd{g_{pe}}{\talpha} = 0$ (at constant $\energy[e]$ and $\magmom[e]$). This is \eref{qflux-dev} as required.

\subsection{Derivation of \eref{heIdent}}
\label{heprimDeriv}
In order to derive this result, we will need a concise notation for the following operation:
\begin{equation}
\widetilde{\nabla}_\perp  = \nabla - \nabla\tl \pd{ }{\tl}  = \nabla\talpha \pd{ }{\talpha} + \nabla\tpsi \pd{ }{\tpsi}.
\end{equation}
This is an important operation as $\nabla\tl$ is not parallel to the exact field, but the strong perpendicular variation is contained in $\tpsi$ and $\talpha$.

Now, the quantity we have to evaluate is
\begin{equation}
\ensav{\tvp{\fluctfav{h_e' \tw\dg\tpsi}}} =  \int d\talpha d\tl \ensav{\gyror{\frac{h_e'}{\MagBfield} \tw}\dg\tpsi}.
\end{equation}
First, we evaluate $\MagBfield(\tpsi,\talpha,\tl)$ at $\tpsi(\bm{R}_e')$, $\talpha(\bm{R}_e')$, and $\tl$ to obtain
\begin{equation}
\fl
\begin{eqalign}{
\int d\talpha d\tl \ensav{ \gyror{\frac{h_e'}{\MagBfield} \tw_\perp}\dg\tpsi} = \\
		\quad \int d\talpha d\tl \ensav{ \gyror{\frac{h_e'}{\MagBfield(\bm{R}_e')}\left(1 - \frac{\bhat\times\tw}{\cycfreq[e]}\dgt\ln\MagBfield\right) \tw_\perp} \dg\tpsi }.
}\end{eqalign}
\end{equation}
In the term involving the gradient of $\ln \MagBfield$ we can perform the gyroaverage to find
\begin{equation}
\fl
\begin{eqalign}{
\int d\talpha d\tl \ensav{ \frac{1}{\MagBfield}\gyror{h_e' \tw_\perp}\dg\tpsi} = \\
		\quad \int d\talpha d\tl \ensav{ \gyror{\frac{h_e'}{\MagBfield(\bm{R}_e')} \tw_\perp} \dg\tpsi } 
		+ \int d\talpha d\tl \ensav{ \frac{w_\perp^2}{2\cycfreq[e]} \pd{\ln\MagBfield}{\talpha} h_e' }.
}\end{eqalign}
\label{close}
\end{equation}
We see that this term is just the $\nabla B$ drift in the exact magnetic field.

Now we handle the first term on the right-hand side of \eref{close}. Integrating by parts with respect to space, we have:
\begin{equation}
\fl
\begin{eqalign}{
\int d\talpha d\tl \ensav{ \gyror{\frac{h_e'}{\MagBfield(\bm{R}_e')} \tw_\perp}\dg\tpsi} = 
		\int d\talpha d\tl  \gyror{ \tw_\perp\left.\dgt\right|_{\energy[e],\magmom[e]}\ensav{\frac{h_e'}{\MagBfield(\bm{R}_e')}\tpsi } } \\
			\qquad	- \int d\talpha d\tl \ensav{\gyror{\tw_\perp\left.\dgt\right|_{\energy[e],\magmom[e]} \frac{h_e'}{\MagBfield(\bm{R}_e')}}\tpsi},
}\end{eqalign}
\end{equation}
where we have used the fact that $\nabla\tpsi = \widetilde{\nabla}_\perp\tpsi$ and the fact that we can then choose to take that derivative at constant $\energy[e]$ and $\magmom[e]$ rather than at constant $\tw$.

The crucial step in this derivation is to now write $\tpsi$ inside the first bracket as $\tpsi(\bm{R}_e') + (\bm{r} - \bm{R}_e')\dgt\tpsi$. Doing this, we obtain
\begin{equation}
\fl
\begin{eqalign}{
\int d\talpha d\tl \ensav{ \gyror{\frac{h_e'}{\MagBfield(\bm{R}_e')} \tw_\perp}\dg\tpsi} = \\
		\quad \int d\talpha d\tl  \gyror{ \tw_\perp\left.\dgt\right|_{\energy[e],\magmom[e]}\ensav{\frac{h_e'}{\MagBfield(\bm{R}_e')}\left[\tpsi(\bm{R}_e') + \frac{\bhat\times\tw}{\cycfreq[e]}\dg\tpsi\right] } } \\
		\quad - \int d\talpha d\tl\ensav{\gyror{\tw_\perp\left.\dgt\right|_{\energy[e],\magmom[e]} \frac{h_e'}{\MagBfield(\bm{R}_e')}}\tpsi}
}\end{eqalign}
\label{blahick}
\end{equation}
Now, for any function $H(\tpsi(\bm{R}_e'),\talpha(\bm{R}_e'),\tl,\energy[e],\magmom[e])$ we have that
\begin{equation}
\fl
\gyror{\bm{w}_\perp \left.\dgt\right|_{\energy[e],\magmom[e]} H} = \gyror{\cycfreq[e]\pd{\bm{R}_e'}{\gyr} \cdot\pd{H}{\bm{R}_e'}} = \cycfreq[e]\gyror{\pd{H}{\gyr}} = 0.
\end{equation}
Applying this result to \eref{blahick}, we obtain
\begin{equation}
\begin{eqalign}{
\int d\talpha d\tl \ensav{ \gyror{\frac{h_e'}{\MagBfield(\bm{R}_e')} \tw_\perp}\dg\tpsi} = \\
		\quad \int d\talpha d\tl  \gyror{ \tw_\perp\left.\dgt\right|_{\energy[e],\magmom[e]}\ensav{\frac{h_e'}{\MagBfield(\bm{R}_e')} \left(\frac{\bhat\times\tw}{\cycfreq[e]}\dg\tpsi\right) } }.
}\end{eqalign}
\end{equation}
We now work on writing this term as a perpendicular divergence. First, we convert the derivative back to one at constant $\tw$:
\begin{equation}
\begin{eqalign}{
\int d\talpha d\tl \ensav{ \gyror{\frac{h_e'}{\MagBfield(\bm{R}_e')} \tw_\perp}\dg\tpsi} = \\
		\quad \tvp\fluctfav{\MagBfield \gyror{ \tw\dgt\ensav{\frac{h_e'}{\MagBfield(\bm{R}_e')} \left(\frac{\bhat\times\tw}{\cycfreq[e]}\dg\tpsi\right) } }} \\
		\quad -\tvp\fluctfav{\ensav{ \frac{w_\perp^2}{2\cycfreq[e]} \bhat\times\nabla\tpsi\cdot\left(\pd{h_e}{\energy[e]} \nabla \energy[e] + \pd{h_e}{\magmom[e]} \nabla\magmom[e]\right)}}.
}\end{eqalign}
\end{equation}
Next, using the fact that $\partial\ensav{\cdot}/\partial\talpha \sim \gkeps\talpha$ and $({\partial\energy[e]}/{\partial\talpha}) \sim B (\partial\magmom[e] / \partial\talpha) \sim \gkeps T_e$ by axisymmetry, we have:
\begin{equation}
\begin{eqalign}{
\int d\talpha d\tl \ensav{ \gyror{\frac{h_e'}{\MagBfield(\bm{R}_e')} \tw_\perp}\dg\tpsi} = \\
		\qquad \tvp\fluctfav{\MagBfield  \widetilde{\nabla}_\perp\cdot\gyror{\tw\ensav{\frac{h_e'}{\MagBfield(\bm{R}_e')} \left(\frac{\bhat\times\tw}{\cycfreq[e]}\dg\tpsi\right) }} }.
}\end{eqalign}
\end{equation}
We can now perform the gyroaverage inside the divergence to find that
\begin{equation}
\begin{eqalign}{
\int d\talpha d\tl \ensav{ \gyror{\frac{h_e'}{\MagBfield(\bm{R}_e')} \tw_\perp}\dg\tpsi} = \\
		\qquad -\tvp\fluctfav{\MagBfield  \widetilde{\nabla}_\perp\cdot\gyror{\ensav{\frac{h_e'}{\MagBfield} \frac{w_\perp^2}{2\cycfreq[e]} \bhat\times\nabla\tpsi }} }.
}\end{eqalign}
\end{equation}
\begin{equation}
\fl
\begin{eqalign}{
\int d\talpha d\tl \ensav{ \gyror{\frac{h_e'}{\MagBfield(\bm{R}_e')} \tw_\perp}\dg\tpsi} = \Or(\massratio\gkeps^2\vth[i] f_e|\nabla\tpsi|B).
}\end{eqalign}
\end{equation}

Inserting this final result back into \eref{closer}, we obtain 
\begin{equation}
\begin{eqalign}{
\int d\talpha d\tl \ensav{ \frac{1}{\MagBfield}\gyror{h_e' \tw_\perp}\dg\tpsi} = 
		\int d\talpha d\tl \ensav{ \frac{w_\perp^2}{2\cycfreq[e]} \pd{\ln\MagBfield}{\talpha} h_e' }.
}\end{eqalign}
\label{result}
\end{equation}
Substituting for $h_e'$ from \eref{hTog} and using the fact that $\infrac{\partial \MagBfield}{\partial\talpha} = \partial \delBp / \partial \talpha$ we obtain \eref{heIdent} as required.
\section{The Electrostatic Limit}
\label{electrostaticLimit}
In this Appendix we provide the electrostatic limit of our equations.

\subsection{The Electrostatic Ordering}
Formally, this is an expansion in $\sqrt{\beta_i} \ll 1$ subsidiary to our expansions in $\massratio$ and $\gkeps$. In the main body of the paper the Alfv\'en velocity $\vA$ is assumed to be comparable to the ion thermal velocity $\vth[i]$; we must now distinguish the two.
We assume that the fluctuations we are interested in have frequencies comparable to those of sound waves, rather than Alfv\'en waves:
\begin{equation}
\left( \pd{}{t} + \bm{u}\dg\right) \sim k_\parallel \vth[i]
\end{equation}
and that typical fluctuating velocities scale with the ion thermal speed (rather than the Alfv\'en speed):
\begin{equation}
\wint h_s \bm{w} \sim \gkeps\vth[i]n_s,
\end{equation}
and that this scaling also applies to the fluctuating $\bm{E}\times\bm{B}$ velocity:
\begin{equation}
\frac{c}{B} \Meanb \times \nabla \delta\pot \sim \gkeps \vth[i].
\end{equation}

In order to consider electrostatic turbulence, we assume that $\delB_\perp$ and $\delAp$ are small:
\begin{equation}
\frac{\delB_\perp}{\MeanMagB} \sim \gkeps\sqrt{\beta} \mbox{\qquad,\qquad} \delAp \sim \frac{\vth}{c} \sqrt{\beta} \delpot.
\label{smallB}
\end{equation}
These orderings imply that $\tpsi$ and $\talpha$ differ from $\psi$ and $\alpha$ by amounts that are small in $\gkeps\sqrt{\beta}$.

Estimating the size of terms in \eref{final-bpar} we see that 
\begin{equation}
\frac{\delBp}{\MeanMagB} \sim \beta\gkeps,
	\label{noDelBp}
\end{equation}
and so we can drop all terms containing $\delBp$.

 As we wish to obtain the simplest equations possible, we will also neglect all finite-Mach-number effects. 
Thus, we combine this expansion with the low-Mach-number expansion of Section~\ref{P1-LowMach} of \cite{flowtome1}.
The primary result of the low-Mach expansion is that we can neglect $\pot_0$ and the poloidal variation of the mean density. In addition to this simplification, we can drop the distinction between $\gkupot$ and $\delpot$ -- from now on only working with the electrostatic potential $\delpot$.

In the subsequent sections we will apply this ordering to our collisionless and collisional equations.
One result is common to both the collisional and collisionless electrostatic limits -- the general solution for $\zeta$.
We obtain this by applying our electrostatic ordering to \eref{deltaApar}. The left-hand-side of \eref{deltaApar} is $\sqrt{\beta}$ smaller than the right; thus, to lowest order, we have
\begin{equation}
\Meanb\dg\left(\zeta - \delpot\right) = 0,
\end{equation}
where we have also used the fact that $\bhat = \Meanb + \Or(\sqrt{\beta})$.
Solving for $\zeta$, we obtain
\begin{equation}
\zeta = \delpot + \hat{\zeta}(\psi),
	\label{ESZetaSoln}
\end{equation}
with $\hat{\zeta}$ an arbitrary function to be determined. Using this result in the expression for the field-line velocity $\veff$, we see that $\veff$ is directed within the flux surfaces and merely moves one field line onto another without distorting them. Thus, this solution for $\zeta$ is consistent with our assumption that $\delB$ is small.

\subsection{The Collisional Electrostatic Limit}
In the collisional limit, we have already used the freedom in the definition of $\zeta$ to eliminate $\delta n_H$ and so cannot use this freedom to eliminate $\hat{\zeta}$. Instead, we use \eref{bibble} to determine $\hat{\zeta}$. Dropping all terms that are small in $\sqrt{\beta}$, \eref{bibble} becomes:
\begin{equation}
\fl
\begin{eqalign}{
\ddttwiddles \frac{e\zeta}{T_e} n_e + \MeanMagB \pd{ }{\tl} \frac{\upar}{\MeanMagB} = -\MaxVDrift \dg \frac{\delta T_e}{T_e}
	+ \MaxVDrift \dg\left(\delpot - \zeta\right) + c \pd{\zeta}{\talpha} \pd{ n_e }{\psi},
}\end{eqalign}
\end{equation}
where we have used the fact that $\tpsi - \psi \sim \gkeps \sqrt{\beta} \psi$.
Using \eref{ESZetaSoln} for $\zeta$ and averaging over a flux surface, we obtain
\begin{equation}
\ddttwiddles \left( \fav{\delpot} + \hat{\zeta} \right ) = 0,
\end{equation}
where we have used \eref{tmpG11} and \eref{tmpG12}.
Assuming no constant perturbations, we have 
\begin{equation}
\hat{\zeta} = -\fav{\delpot}.
\end{equation}

Applying the low-Mach-number limit of \cite{flowtome1} to \eref{dTH}, we obtain
\begin{equation}
\fl
\begin{eqalign}{
 &\pd{}{t} \left(\frac{\delta T_H}{T_e} + \fluctfav{\frac{2\delBp}{3B}}\right) n_e = 
 \frac{2}{3}\fluctfav{ \frac{1}{\MeanMagB}\pd{}{\tl}(\tpsi-\psi) } I n_e \frac{d\angvel}{d\psi},
}\end{eqalign}
\label{tmpdTH}
\end{equation}
where we have used the low-Mach-number result that $n_e$ is a flux function to remove it from the flux-surface averages whereupon we used \eref{tmpG11} and \eref{tmpG12} to eliminate most of the terms in \eref{dTH}.
Now, using \eref{noDelBp} and \eref{smallB} in \eref{tmpdTH} and again assuming that there are no constant perturbations, we have that $\delta T_H = 0$.

Substituting these results into \eref{tmpBlah} we have
\begin{equation}
\delta f_e = \frac{e \left(\delpot - \fav{\delpot}\right)}{T_e} F_{0e} + \Or(\gkeps \sqrt{\beta} F_{0e}),
\end{equation}
with $\delpot$ found from
\begin{equation}
\frac{1}{n_e} \left( \sum_{s=i} \frac{Z_s^2 e n_s}{T_s} \right) \delpot + \frac{e}{T_e} \left(\delpot - \fav{\delpot}\right) = \frac{1}{n_e} \sum_{s=i} Z_s \wint \gyror{h_s}.
\end{equation}
This is just the usual adiabatic electron model including the zero response to zonal perturbations~\cite{hammett1993developments}. This reassures us that we can recover known physics from our complex model and that there is a physically-relevant regime where the
adiabatic electron model is an exact limit of gyrokinetics and not an ad-hoc prescription.
\subsection{The Weakly-Collisional Electrostatic Limit}
\label{gangLimit}
In this limit, we can use the freedom in the definition of $\zeta$ to set $\hat{\zeta} = 0$ in \eref{ESZetaSoln}. Using \eref{uxi} we see that this implies that
\begin{equation}
\veff = \bm{u} + \Or(\gkeps\vth[i]\sqrt\beta).
\end{equation}
Equation \eref{trap0} for $g_{te}$ then becomes
\begin{equation}
\fl
\begin{eqalign}{
\left( \pd{ }{t} + \bm{u}\dg\right) g_{te} &+ \bav{\vdrift[e]\dg g_{te}} + \frac{c}{e\MeanMagB} \pb{\bav{\delpot}}{g_{te}} - \bav{\gyroR{\collop[g_{te}]}} \\
	&= - \left( \pd{ }{t} + \bm{u}\dg\right) \frac{ e \bav{\delpot} }{T_e} F_{0e} + c \pd{\bav{\delpot}}{\alpha} \pd{F_{0e}}{\psi},
}\end{eqalign}
\label{gangEq}
\end{equation}
where we have used \eref{smallB} and \eref{noDelBp} to drop terms involving $\delAp$ and $\delBp$.
This is the usual bounce-averaged kinetic equation originally derived in \cite{gang1990nonlinear}, again demonstrating that the preexisting exact models are contained in our model. However, we must still find a solution for the distribution function of passing electrons: $g_{pe}$ (which was not discussed in \cite{gang1990nonlinear}).
This is determined from the electrostatic limit of \eref{pass0}:
\begin{equation}
\fl
 \left(\pd{ }{t} + \bm{u}\dg\right)\left(\fav{\frac{\MeanMagB}{|w_\parallel|}}g_{pe}- \frac{eF_{0e}}{T_e}\fav{\frac{\MeanMagB\delpot}{|w_\parallel|}}\right) = \fav{\frac{\MeanMagB}{|w_\parallel|} \gyroR{\collop[g_{pe}]}} .
\label{passingElec}
\end{equation}
From this equation we see that $g_{pe}$ is driven by the competition between the shielding effects of the passing electrons that drive $g_{pe}$ towards the maximally-shielding solution
\begin{equation}
g_{pe} = - \frac{1}{\fav{\infrac{\MeanMagB}{|w_\parallel|}}} \fav{\frac{\MeanMagB \delpot}{|w_\parallel|}}\frac{e F_{0e}}{T_e}
\end{equation}
and collisions which drive $g_{pe}$ back towards a Maxwellian. The solution for $g_{pe}$ is non-zero and so the original discussion in \cite{gang1990nonlinear} is incomplete. Thus the solution for $\delta f_e$ is
\begin{equation}
\delta f_e = \frac{e\delpot}{T_e} F_{0e} + g_{e},
\end{equation}
with $g_e = g_{pe} + g_{te}$ found from \eref{gangEq} and \eref{passingElec} and $\delpot$ obtained from
\begin{equation}
\frac{1}{n_e} \left( \sum_{s=i} \frac{Z_s^2 n_s e}{T_s} + \frac{e}{T_e} \right) \delpot = \frac{1}{n_e} \wint g_e + \sum_{s=i} \frac{Z_s}{n_e} \wint \gyror{h_s}.
\end{equation}
\section*{References}
\bibliographystyle{unsrt}
\bibliography{references}
\end{document}